\documentclass[aps,pra,twocolumn,showpacs,superscriptaddress,bibliography]{revtex4-1}

\usepackage{graphicx}%
\usepackage{multirow}%
\usepackage{amsmath,amssymb,amsfonts}%
\usepackage{amsthm}%
\usepackage{mathrsfs}%
\usepackage[title]{appendix}%
\usepackage{xcolor}%
\usepackage{braket}
\begin{document}

\title{Scalable, high-fidelity all-electronic control of trapped-ion qubits}

\author{C. M. L{\"o}schnauer}
\affiliation{Oxford Ionics, Oxford, OX5 1PF, UK}
\author{J. Mosca Toba}
\affiliation{Oxford Ionics, Oxford, OX5 1PF, UK}
\author{A. C. Hughes}
\affiliation{Oxford Ionics, Oxford, OX5 1PF, UK}
\author{S. A. King}
\affiliation{Oxford Ionics, Oxford, OX5 1PF, UK}
\author{M. A. Weber}
\affiliation{Oxford Ionics, Oxford, OX5 1PF, UK}
\author{R. Srinivas}
\affiliation{Oxford Ionics, Oxford, OX5 1PF, UK}
\affiliation{Department of Physics, University of Oxford, Oxford, OX1 3PU, UK}
\author{R. Matt}
\affiliation{Oxford Ionics, Oxford, OX5 1PF, UK}
\author{R.~Nourshargh}
\affiliation{Oxford Ionics, Oxford, OX5 1PF, UK}
\author{D. T. C. Allcock}
\affiliation{Oxford Ionics, Oxford, OX5 1PF, UK}
\affiliation{Department of Physics, University of Oregon, Eugene, OR 97403, USA}
\author{C. J. Ballance}
\affiliation{Oxford Ionics, Oxford, OX5 1PF, UK}
\affiliation{Department of Physics, University of Oxford, Oxford, OX1 3PU, UK}
\author{C. Matthiesen}
\affiliation{Oxford Ionics, Oxford, OX5 1PF, UK}
\author{M. Malinowski}
\email{mm@oxionics.com}
\affiliation{Oxford Ionics, Oxford, OX5 1PF, UK}
\author{T. P. Harty}
\affiliation{Oxford Ionics, Oxford, OX5 1PF, UK}

\begin{abstract}
The central challenge of quantum computing is implementing high-fidelity quantum gates at scale. However, many existing approaches to qubit control suffer from a scale-performance trade-off, impeding progress towards the creation of useful devices. Here, we present a vision for an electronically controlled trapped-ion quantum computer that alleviates this bottleneck. Our architecture utilizes shared current-carrying traces and local tuning electrodes in a microfabricated chip to perform quantum gates with low noise and crosstalk regardless of device size. To verify our approach, we experimentally demonstrate low-noise site-selective single- and two-qubit gates in a seven-zone ion trap that can control up to 10 qubits. We implement electronic single-qubit gates with $99.99916(7) \%$ fidelity, and demonstrate consistent performance with low crosstalk across the device. We also electronically generate two-qubit maximally entangled states with $99.97(1) \%$ fidelity and long-term stable performance over continuous system operation. These state-of-the-art results validate the path to directly scaling these techniques to large-scale quantum computers based on electronically controlled trapped-ion qubits.

\end{abstract}
\maketitle

\section{Introduction}\label{sec:intro}

The last three decades have brought about impressive demonstrations of key building blocks of quantum computers (QCs).  Recent proof-of-principle experiments have shown that quantum gates on physical qubits can be implemented with high fidelity \cite{harty2014, clark2021, ding2023}, and that quantum error correction can be used to improve error rates \cite{acharya2023, dasilva2024}. Scaling to larger system sizes has been demonstrated for some platforms \cite{king2023, kim2023, bluvstein2024, manetsch2024}, as have methods for scaling using quantum interconnects \cite{Moehring2007, mills2019, Zhong2019, stephenson2020} and integrated control \cite{mehta2014, mehta2020, niffenegger2020, zwerver2022}. However, to build useful QCs, it is necessary to develop architectures that are \textit{simultaneously} low-noise and scalable. This requires co-optimizing the performance metrics - such as fidelity, connectivity, selectivity, and parallelism -- with the scalability metrics -- including manufacturability, qubit variability, footprint, signal delivery, and power consumption.

Chip-scale trapped-ion quantum computers (TIQCs) -- with atomic qubits suspended above a control chip \cite{chiaverini2005} -- have the potential to satisfy these criteria by combining the low-noise performance of atomic qubits \cite{sepiol2019} with the scalability of chip-integrated electronics. While TIQCs are, by many metrics, the best-performing QC systems today \cite{moses2023}, existing commercial systems are difficult to scale to useful sizes, as they rely on free-space light delivery to enable laser-based gates. Despite promising proof-of-principle demonstrations of laser-based gates with integrated photonics \cite{mehta2016, mehta2020}, significant performance, manufacturability, footprint, signal delivery, and power consumption challenges associated with integrated light generation, modulation, and distribution remain to be overcome \cite{mordini2024, kwon2024, hogle2023}.

An alternative method is to perform quantum gates using magnetic fields generated by chip-integrated traces \cite{ospelkaus2008}. This approach is immediately compatible with commercial microfabrication processes, opening the door to full device integration and scalability. Nevertheless, while previous work demonstrated high-quality single- and two-qubit laser-free gates \cite{harty2014, srinivas2021, weber2024}, the possibility of using this approach to build universal, integrated, multi-zone TIQCs remained unexplored so far.

In this paper, we propose an all-electronic TIQC architecture that combines laser-free gates with local electric-field tuning for high-performance integrated control. To validate our vision, we demonstrate the ability to perform all-electronic single- and two-qubit gates with state-of-the-art fidelities across a multi-zone microfabricated ion trap. We further show that electrical tuning allows for highly flexible, low-crosstalk coherent control. Our results validate the path to high-performance, integrated, large-scale QCs manufactured using established foundry processes.

\begin{figure*}[ht]
\centering
\includegraphics[width=1.0\textwidth]{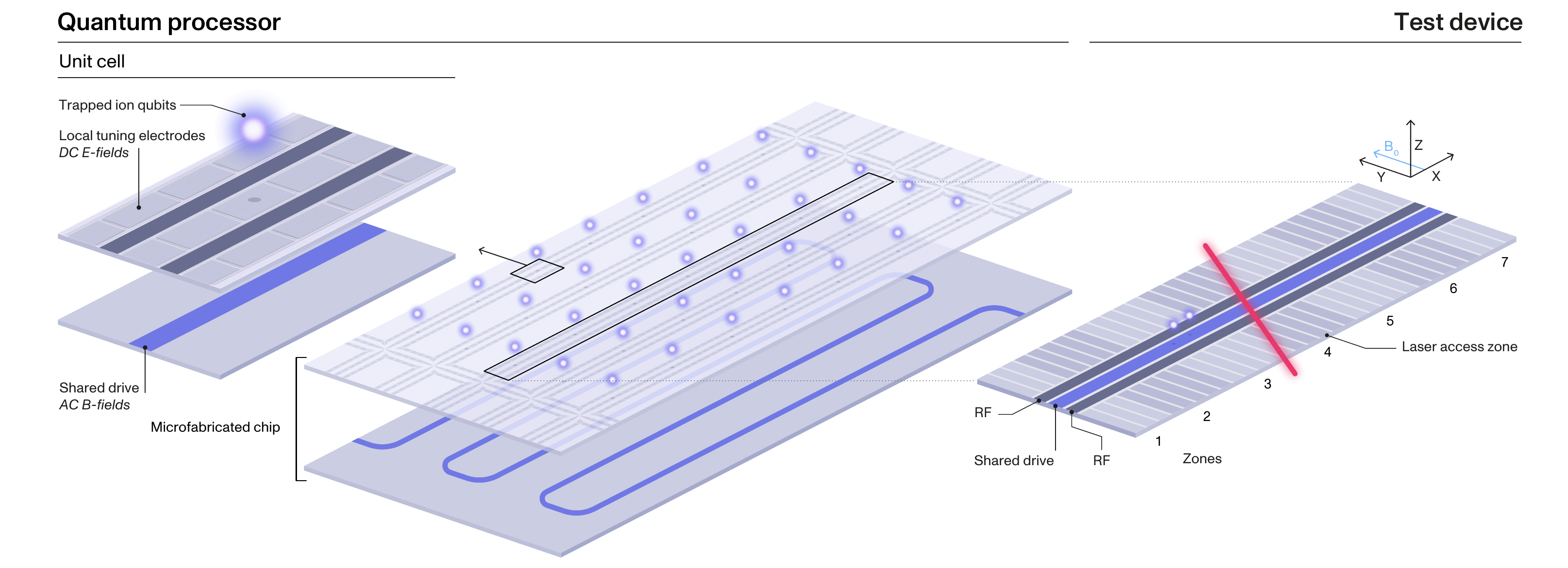}
\caption{Illustration of the all-electronic approach to the coherent control of trapped-ion qubits. The TIQC is a microfabricated chip (center) that contains a repeating 2D grid of unit cells (left), each composed of trapped-ion qubits stored near local tuning electrodes. A shared drive is routed to deliver AC magnetic fields to all unit cells at once. The strength of qubit coupling with the shared drive is controlled by applying DC electric fields to the tuning electrodes. Right: a linear seven-zone test device used in experiments in section~\ref{sec:experiments}. The tuning electrodes and a shared drive are placed on the same layer. The shared current-carrying trace is routed along $\hat{x}$, and the ions are trapped directly above at the height of $z \approx 40 \mathrm{\mu m}$. Laser light is delivered to zone 4 for ion loading, state preparation, and measurement, but is not used for coherent control. A static magnetic field $\mathrm{B_0} \approx 8.5$ mT is oriented along $\hat{y}$. }\label{fig:architecture}
\end{figure*}

\section{Architecture}\label{sec:architecture}

Our all-electronic architecture for coherent control of trapped-ion qubits is illustrated in Fig.~\ref{fig:architecture}. It employs a chip ion trap \cite{chiaverini2005}, where voltages applied to electrodes on the top metal layer hold ion qubits tens of microns above the surface. Building on previous proposals and demonstrations of addressed laser-free quantum gates \cite{leibfried2007, warring2013, lekitsch2017, srinivas2021}, two elements are incorporated into the chip to achieve site-selective control. First, a shared drive (Fig. \ref{fig:architecture}, purple) is used to deliver a control field globally to all qubits at once \cite{leibfried2007}. Second, local tuning electrodes (Fig. \ref{fig:architecture}, blue) are used to tune the control Hamiltonian on a per-zone basis, allowing for site-selective operations \cite{seck2020, srinivas2023, sutherland2023}.

The shared drive is implemented by passing oscillating (AC) currents through traces defined inside the chip. These AC currents generate AC magnetic fields at all sites, which form the basis of single- and multi-qubit quantum gates \cite{, warring2013a, sutherland2019}. To ensure efficient power and I/O use, traces generating magnetic fields in different zones are wired in series to a single source \cite{srinivas2021, enthoven2024}. Local tuning is implemented by applying additional DC voltages to the top-layer electrodes. These voltages generate electric fields that adjust the ion positions; as the on-chip currents generate large magnetic field gradients in the near field of the chip, even small changes in ion positions result in significant variations in the magnetic fields they experience~\cite{wineland1998, mintert2001}.  The voltages can also generate electric field curvatures that adjust ion motional frequencies. As shown experimentally in section~\ref{sec:experiments}, these degrees of freedom can be used to locally modulate the strength of single- and multi-qubit interactions.

The combination of the shared AC magnetic drive and local DC electric tuning allows coherent control with high fidelity, site selectivity, and parallelism. For example, to perform quantum operations globally (all qubits ``active''), the local electrodes are used to fine-tune small differences in the coupling rates between qubits and the magnetic drive such as to make the interaction Hamiltonians identical for all zones. Site-selectivity can then be achieved by using the local electrodes to ``hide'' qubits in certain zones such as to make their interaction Hamiltonian close to the identity \cite{warring2013, leu2023, lysne2024}. Finally, local electrodes can continuously tune the Hamiltonian between the ``active'' and the ``hidden'' values to perform different operations in different zones in parallel.

Compared to existing architectures for TIQC which use local laser and/or magnetic drives \cite{steane2006, lekitsch2017}, our architecture brings several performance and scalability benefits. First, all qubit tuning is achieved using DC voltages, which can be efficiently multiplexed using only standard chip-integrated electronics \cite{malinowski2023}. This alleviates the chip manufacturability and wiring challenges, allowing the use of standard microfabrication and packaging processes. Furthermore, DC voltages can be distributed across a chip with negligible crosstalk and power dissipation, and the electric-field patterns from multiple electrodes can be well localized to specific trap zones. Finally, our method does not require new dedicated structures, instead repurposing the electrodes already present in the TIQC for qubit shuttling and stray-field compensation. 

Second, all coherent control is achieved using electric and magnetic fields, which can be generated from electronically integrated, high-stability, low-cost sources with low phase and amplitude noise \cite{ball2016}. This is in contrast to laser-based gates, which typically rely on large, complex, free-space optical sources. Furthermore, while laser-based gates face stringent fidelity limits due to spontaneous emission \cite{ozeri2007, moore2023}, laser-free gates have a negligible fundamental noise floor \cite{sutherland2022, srinivas2021}, allowing significant fidelity improvements through hardware engineering alone. Finally, wiring the shared drive in series is highly power-efficient, allowing the number of qubits to be increased while keeping the power density constant.

To verify the performance of our architecture, we have designed and built a seven-zone microfabricated ion trap with local DC electrodes and a shared AC current-carrying trace. In the subsequent section, we demonstrate that all-electronic control can be used to perform single- and two-qubit gates in a site-selective fashion. Furthermore, we show the low-noise capabilities of our technology by benchmarking single-qubit gates with $99.99916(7) \%$ fidelity, and by creating two-qubit maximally entangled states with $99.97(1)\%$ fidelity -- to our knowledge, record-level performance for any QC platform to date. Finally, we discuss the path to scaling this performance to mid-scale QCs with thousands of qubits.

\section{Experiments}\label{sec:experiments}

\begin{figure*}[ht]
\centering
\includegraphics[width=0.95\textwidth]{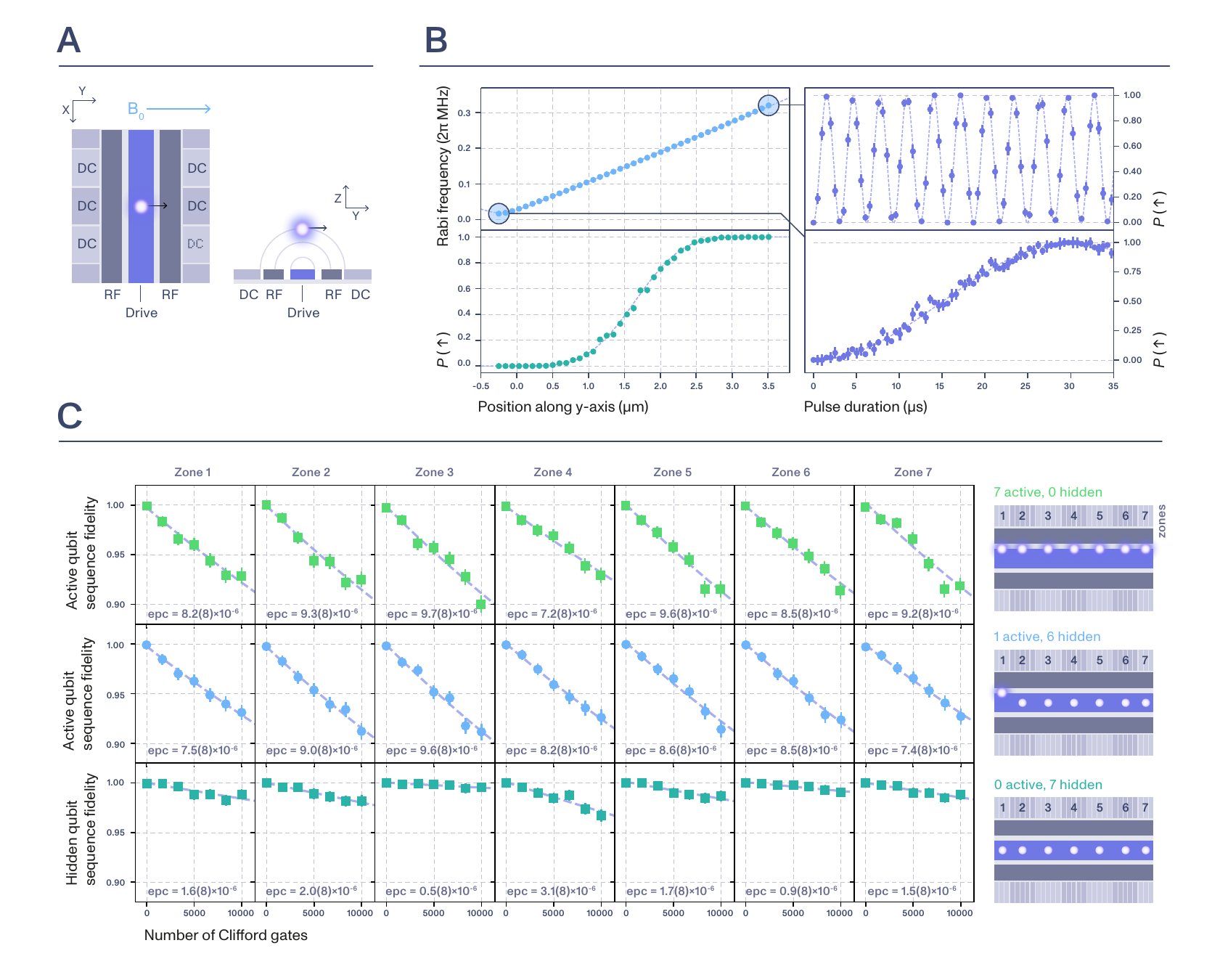}
\caption{Summary of single-qubit control experiments. A) Single-qubit control is achieved by using DC electrodes to adjust qubit position along $\hat{y}$. Directly above the shared trace ($y \approx 0$) the AC magnetic field is oriented in the $y$-direction. Since the static magnetic field is oriented along $y$ as well, we find that the single-qubit Rabi frequency $\Omega_1 \propto B_z \approx 0$, i.e. the qubit is ``hidden'' (appendix~\ref{secA1}). At the same time, $|\partial \Omega_1 / \partial y| \propto |\partial B_z / \partial y| > 0$, thus translating the qubit along $y$ causes interactions to be turned on. B) (Top left) Experimental measurements of the single-qubit Rabi frequency $\Omega_1$ vs $y$. (Right) Rabi oscillations in the ``active'' (top) and ``hidden'' (bottom) positions. (Bottom left) Spin-flip probability $P(\downarrow)$ vs $y$ for an SK1 sequence calibrated to drive a $\pi$-pulse in the ``active'' position. c) Results of single-qubit RB throughout the device. Each experiment uses one qubit to record the spin-flip probability $P(\downarrow)$ in a target zone in one of three configurations: all zones ``active'' (top), target zone ``active'' and other zones hidden (middle), and all zones ``hidden'' (bottom). The exponential fits (dashed lines) are used to estimate the error per Clifford (epc) in each configuration.
}
\label{fig:1q}
\end{figure*}
The architecture validation experiments are performed in a seven-zone single-layer linear microfabricated trap cooled to $\approx 5$ K (Fig.~\ref{fig:architecture}c). Each zone contains $6-10$ independent electrodes over $\approx 350$~{\textmu}m along $x$ which act as local tuners, and a current-carrying trace routed in the center of the chip acts as the shared drive. The qubit states $\ket{\uparrow}$ and $\ket{\downarrow}$ are encoded in the $4S_{1/2}$ Zeeman sublevels of $\mathrm{^{40}Ca^+}$ ions trapped $40 \ \mathrm{\mu m}$ above the chip surface. The qubit frequency $\omega_0 \approx 2 \pi \times 240$ MHz is set by a static magnetic field of $B_0 \approx 8.5$ mT, aligned in the plane of the chip perpendicular to the trap axis. Laser beams at 375\,nm, 397\,nm, 423\,nm, 729\,nm, 854\,nm, and 866\,nm are aligned above the central zone to facilitate ion loading, cooling, qubit initialization and measurement; in all other zones, all ion control is laser-free. 

Each experimental shot begins with Doppler and sideband cooling one or two “probe” ions to near the motional ground state \cite{metcalf2003}. The ions are then transported using time-dependent voltages on the DC electrodes to the center of the zone to be probed, and electrodes in all zones are set to initialize potential wells capable of holding ions. Following that, local electrode DC voltages are adjusted to selectively activate interactions in some zones only, and a sequence of AC current pulses, near-resonant with the qubit transition, is applied to the shared trace to drive the desired operation. Subsequently, the probe ion(s) are shuttled back to the central zone, state $\ket{\downarrow}$ is shelved using a sequence of 729 nm pulses, and state-dependent fluorescence is collected on a CMOS camera to measure the qubit state \cite{burrell2010}. In the experiments presented below, single-qubit gates are performed in all 7 trap zones, while two-qubit gates are only performed in zones 2--6 due to a reduced tuning electrode count in zones 1 and 7. The gates are benchmarked on stationary qubits, probing one zone at a time.

\subsection{High-fidelity tunable single-qubit gates}

Consider a qubit with frequency $\omega_0$ in zone $i$ (Fig.~\ref{fig:1q}a). Applying an oscillating current $I \cos(\omega_0 t + \phi)$ for duration $t_1$ through the shared trace results in a single-qubit rotation unitary $U_{1} = \exp(-i \sigma_\phi \theta_1/2)$, where $\sigma_\phi = \cos(\phi) \sigma_x + \sin(\phi) \sigma_y$, $\sigma_{x/y}$ are the Pauli spin matrices, and the gate angle $\theta_1= \Omega_1 t_1$, where the Rabi frequency $\Omega_1$ depends on the magnitude and polarization of the magnetic field from the shared trace (see appendix~\ref{secA1}). In our geometry, $\Omega_1$ is proportional to the $z$-polarized component of the magnetic field $B_z$. To achieve site-selective control, we leverage the fact that directly above the shared trace $B_z \approx 0$ while $\partial B_z / \partial y$ remains significant (Fig~\ref{fig:1q}a). Thus, zone-selective ion translation along $y$ results in zone-selective adjustment of $\theta_1$.

To validate the method, we measure the Rabi frequency for a range of ion positions $y$ in the central zone. The results are summarized in Fig.~\ref{fig:1q}b. At $I \approx 70$ mA, we find that we can tune the Rabi frequency between $\Omega_1 \approx 2 \pi \times 17$ kHz and $\Omega_1 \approx 2 \pi \times 330$ kHz by translating an ion from $y \approx -0.2$~{\textmu}m (qubit ``hidden'') to $y \approx 3.5$~{\textmu}m (qubit ``active''). This translation is performed by tuning voltages on local electrodes by $\approx 0.1$ V. We correct for the effect of the residual Rabi frequency at the ``hidden'' position (caused by $\approx 0.2$ deg misalignment of the static magnetic field) by employing a Solovay-Kitaev-1 gate sequence, which uses three resonant pulses to improve robustness to static Rabi frequency errors \cite{merrill2012}. As shown in Fig.~\ref{fig:1q}b for a total gate duration of $8 \ \mathrm{\mu}s$, we can switch between $\theta_1 \approx 0$ (qubit “hidden”) and $\theta_1 \approx \pi$ (qubit “active”) with high contrast and high degree of robustness against ion position errors.

We quantify the gate errors using single-qubit randomized benchmarking (RB) \cite{knill2008, magesan2011} with all zones set to ``active'', shown in Fig~\ref{fig:1q}c (top) and described further in appendix \ref{secA1}. Using all-electronic control, we achieve an average error of $8.4(7) \times 10^{-6}$ per Clifford, consistent across all 7 zones, and in agreement with independently measured noise in the signal delivery chain. This is, to our knowledge, the lowest single-qubit gate error ever recorded in a multi-zone quantum processor.

To estimate the crosstalk errors, we repeat RB with one qubit ``active'' at a time (Fig~\ref{fig:1q}c, middle). The resulting error per Clifford of $8.8(8) \times 10^{-6}$ is unchanged within measurement uncertainty, indicating that “hiding” or “activating” other qubits changes target qubit gate errors by $\lessapprox 1 \times 10^{-6}$. A more dominant source of crosstalk is the residual magnetic interaction between the shared drive and the “hidden” qubits. To estimate its magnitude, we record the effect of RB on “hidden” qubits (Fig~\ref{fig:1q}c, bottom), finding a spin flip probability of $1.6(8) \times 10^{-6}$ per Clifford, once again indicating very low crosstalk. The remaining error on the ``hidden'' qubits is dominated by coherent effects and can be suppressed with higher-order composite pulses \cite{merrill2012}.
% Upcoming work will serve to quantify crosstalk errors more rigorously \cite{sarovar2020, rudinger2021}, as well as to benchmark parallel operations, using techniques such as simultaneous RB \cite{gambetta2012, leu2023}.
\subsection{High-fidelity tunable two-qubit gates}

\begin{figure*}[ht]
\centering
\includegraphics[width=0.95\textwidth]{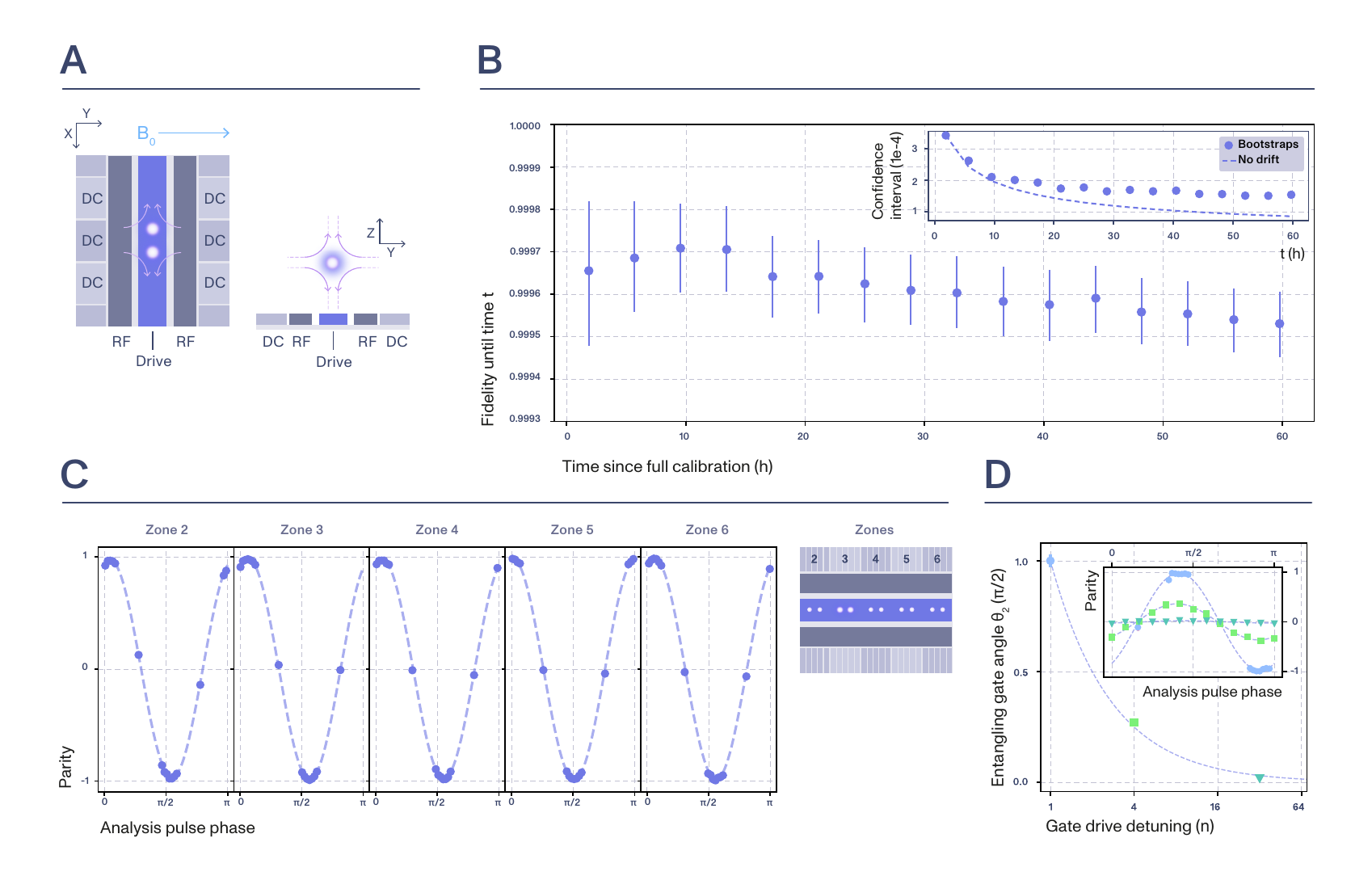}
\caption{Summary of two-qubit control experiments. A) Two-qubit Molmer-Sorensen interactions are implemented by driving currents through the shared trace. A state-dependent force on the in-plane radial mode is generated using the magnetic field gradient $\partial B_z / \partial y$, and the fact that $B_z \approx 0$ is used to minimize off-resonant qubit frequency shifts. Quadrupole potentials generated by the DC electrodes are used to tune the motional mode frequencies and orientations, adjusting the effective interaction strength. B) The entangled state fidelity in zone 4 following full system calibration. We record a fidelity of $0.9997(1)$ over the first 12 hours, and a fidelity of $0.9995(1)$ over the full 60 hours of data acquisition. The $68\%$ confidence intervals (CI) are calculated by bootstrapping with 10,000 resamples. After 12 hours, the CI deviates visibly from the $1/\sqrt{t}$ scaling (inset), indicating drifts in the underlying process.  C) Qubit parity oscillations in 5 zones (not corrected for state preparation and measurement errors), demonstrating the ability to prepare maximally entangled states $\theta_2 \approx \pi/2$ throughout the device. D) Experimental measurements of the entangling angle $\theta_2$ when the mode frequency is electrically adjusted to $n \times \delta$, where $\delta$ corresponds to a detuning of a maximally entangling operation. Insert shows parity oscillations used to infer the value of $\theta_2$.}
\label{fig:2q}
\end{figure*}

Consider a pair of qubits with frequency $\omega_0$ and a motional mode frequency $\omega_m$ in zone $i$ (Fig.~\ref{fig:2q}a). Applying a bichromatic oscillating current ${I_{\pm}(t) = I \cos((\omega_0 \pm (\omega_m + \delta)) t + \phi)}$ for duration $t_2$ to the shared trace results in a two-qubit entangling unitary $U_{2} = \exp(-i\sigma_\phi \otimes \sigma_\phi \theta_2/2)$ as long as $\delta = 2 \pi n/t_2$, where $n$ is an integer \cite{sorensen1999, ospelkaus2008} (see appendix \ref{secA2}). The entangling gate angle is then given by $\theta_2 = \Omega_2^2 t_2^2/(4 n \pi)$, where the gate Rabi frequency $\Omega_2$ depends on the magnitude of the magnetic field gradient along the direction of ion motion \cite{ospelkaus2008}.

To implement arbitrary quantum computation, it suffices to complement site-selective tunable single-qubit rotations with global maximally-entangling two-qubit gates ($\theta_2 = \pi/2$). We demonstrate this capability by preparing maximally entangled states as follows. We optically pump two ions to initial state $\ket{\uparrow \uparrow}$ and set the radial rocking mode to be at a frequency $\omega_m \approx 2 \pi \times 3.7$ MHz and aligned in the plane of the chip. We then apply a sequence of two bichromatic AC current pulses through the shared trace, each with $I \approx 0.7$ A and $\delta \approx 2\pi \times -16$ kHz, resulting in $\Omega_2 \approx 2 \pi \times 6$ kHz. Each pulse has a duration of $t_2/2 \approx 60\ \mathrm{\mu s}$, and we flip the phase difference between the currents by $\pi$ between the pulses to improve robustness to fluctuations in $\omega_m$ \cite{hayes2012}. This sequence ideally prepares a maximally entangled state $\left(\ket{\uparrow\uparrow} - i \ket{\downarrow\downarrow}\right) / \sqrt{2}$. We estimate the fidelity with that state by measuring the qubit populations at the end of the pulse, as well as parity contrast after an additional $\pi/2$ ``analysis'' pulse with a variable phase \cite{leibfried2003}. 

We begin the experiments by roughly tuning the two-qubit interactions throughout the trap. By electrically matching mode frequencies and orientations of all five two-qubit wells simultaneously, we generate maximally entangled states in every zone with $\lessapprox  10^{-2}$ error rate (Fig.~\ref{fig:2q}c), and we observe no statistically meaningful zone-to-zone variations. To characterize the gate performance with high precision, we then perform fine parameter calibration in zone 4 and repeat the measurement sequence over $\approx 12$ hours to estimate the fidelity with a statistical uncertainty of $\leq 1 \times 10^{-4}$. The results are shown in Fig.~\ref{fig:2q}b. Averaging over all $\approx 10^6$ data points, we find the error of preparing a maximally entangled state of $3(1) \times 10^{-4}$, a record for any QC platform. The uncertainty corresponds to a 68\% confidence interval estimated via bootstrapping~\cite{efron1994}, and the fidelity estimate has been corrected for the state preparation error of $1.4(2) \times 10^{-4}$ and the measurement error of $3.0(2) \times 10^{-4}$ which were independently monitored throughout the data acquisition period (see appendix \ref{secA2} for details). 

To benchmark the baseline passive stability of the system, we continued the data acquisition for a further 48 hours, bringing the total number of experimental runs to $\approx 5 \times 10^6$. We find that, beyond $t \approx 12$ hours, the uncertainty in the error deviates from the $t^{-1/2}$ scaling expected for purely statistical fluctuations (Fig.~\ref{fig:2q}d, inset), indicating a fidelity drift. Nonetheless, even when averaged over the full $t=60$ hours of data acquisition, we record an error of $4.7(7) \times 10^{-4}$, state-of-the-art for any QC platform. The average fidelity drift rate is at the level of $\approx 1 \times 10^{-4}$ per day over the full data run. During that period, only the motional frequency $\omega_m$ was regularly calibrated using an automated routine, with all other system parameters, including the qubit frequency, fixed at the initial calibration value. 

Based on independent memory benchmarking experiments \cite{omalley2015, sepiol2019}, we estimate that the majority of the remaining infidelity can be accounted for by qubit frequency fluctuations. At the same time, analysis of population errors in the experiments above, as well as independent measurements of mode heating rates ($< 1$ phonon per second on the gate motional mode), allow us to upper-bound the combined error of motional mode fluctuations and heating to $< 1 \times 10^{-4}$. These results suggest that the entangling operation error rate can be reduced to $< 1 \times 10^{-4}$ by employing standard techniques to protect from qubit dephasing, such as dynamical decoupling \cite{viola1998, bermudez2012, harty2016, weber2024} and/or field-insensitive encodings \cite{langer2005}. Simultaneously, the high level of stability in the fidelity measurements indicates that the errors caused by parameter drifts can be also kept to $< 1 \times 10^{-4}$ by adding only sparse calibration routines, e.g. one additional automated calibration per gate parameter per day. Together, these results show a direct path to reliable and robust quantum control with two-qubit gate error rates $< 1 \times 10^{-4}$.

% While the high-precision fidelity measurements above were conducted in zone 4, entangled states can be prepared in any zone by electrically matching the local mode frequencies and orientations, as shown in Fig.~\ref{fig:2q}c. We have verified through coarsely calibrated experiments ($\approx 0.99 - 0.999$ fidelity) that we can generate maximally entangled states in all zones with equal error rates, down to a statistical uncertainty of $3 \times 10^{-3}$, and upcoming upgrades to our control system will enable RB of two-qubit gates \cite{baldwin2020}, which will significantly speed up data acquisition, enabling gate fidelities to be measured with high precision throughout the processor.

Finally, we note that while the experiments above focussed on the maximally entangling operations, the method can be extended to implement tunable-angle two-qubit operations. We demonstrate this experimentally by electrically adjusting the potential experienced by the ions to offset the motional frequency by $n \times \delta$ from the shared drive. We then apply the same current pulse as above through the shared trace, and measure the contrast of qubit parity oscillations to estimate the entangling area $\theta_2$. In Fig.~\ref{fig:2q}d we show that this allows for the entangling area to be adjusted in discrete steps as $\theta_2 = \pi/(2n)$. We note that further continuous control over $\theta_2$ is possible by adjusting the mode orientation and that the two-qubit interactions can be fully switched off ($\theta_2 = 0$) by splitting the ion crystal before the shared drive is turned on. Such tunable-angle two-qubit gates could be used for further error reduction through more efficient unitary decomposition \cite{parra-rodriguez2020, foxen2020, lacroix2020}.

\section{Discussion}\label{sec:discussion}

The experiments presented in section~\ref{sec:experiments} demonstrate that the all-electronic control architecture discussed in section~\ref{sec:architecture} allows for low-noise, site-selective multi-zone control of trapped-ion qubits. Here, we outline how these primitives can be combined with established ion-trap and microfabrication technology to build next-generation high-performance QCs with many thousands of qubits. These mid-scale QCs are expected to provide early computational value \cite{clinton2024}, while simultaneously serving as architectural proof-points for large-scale fault-tolerant QCs.

We envision the all-electronic control architecture to be at the core of the TIQC built on a microfabricated multi-layer silicon chip \cite{guise2015}. Following the QCCD approach \cite{wineland1998}, the unit cells will be arranged on a 2D grid as in Fig.~\ref{fig:architecture}, with junctions placed throughout the device to facilitate all-to-all connectivity via ion shuttling \cite{burton2023}. To minimize qubit frequency variations across the chip, the device will be placed in a highly homogeneous magnetic field, and quantum information will be encoded into atomic states with reduced magnetic-field sensitivity \cite{langer2005}. 

Qubit state preparation and measurement (SPAM) will be achieved using laser light, delivered from off-chip sources using chip-integrated photonics \cite{mehta2020, niffenegger2020, ivory2021}. We propose to implement all SPAM using shared laser drives only, delivering light to all unit cells simultaneously, and once again using local electrodes to fine-tune the operations by moving ions in and out of laser beams. This alleviates the need for building optical systems for large-scale delivery of individually modulated laser beams, historically a major roadblock to scaling laser-based TIQCs \cite{wang2020, shih2021, pogorelov2021, binai-motlagh2023, sotirova2023}. Furthermore, SPAM requires orders of magnitude less optical power than high-fidelity coherent optical operations \cite{Ospelkaus2011}, relaxing constraints on optical losses and beam quality. Thus, our photonic delivery network can be built out of purely passive integrated components (waveguides, splitters, and grating couplers) \cite{mehta2017}, and using only processes and materials that are well known to the semiconductor industry \cite{mehta2016, west2019}. 

We envision that, using all-electronic control primitives, ever larger QCs can be built by repeating more and more unit cells. Assuming a zone spacing of $\approx 350$~{\textmu}m as in section~\ref{sec:experiments}, a 10,000-qubit device only requires a chip area of $\approx 3.5 \times 3.5 \ \mathrm{cm}^2$, comparable to a typical camera sensor. Power dissipation from the shared trace will be minimized by placing it on a dedicated low-loss metal layer and keeping the chip at cryogenic temperatures. For example, a shared trace of 50~{\textmu}m width, 2~{\textmu}m thickness, and $2 \times 10^{-10} \ \Omega$ m resistivity \cite{sambles1981} carrying a current of 1~A would lead to power dissipation of 0.35\,mW per cell, allowing a chip with tens of thousands of qubits to be operated in an off-the-shelf 4\,K cryostat, and with further order-of-magnitude gains possible by reducing the ion-electrode distance.

Brute-force device scaling could be limited by the practical wiring limits of purely passive devices. However, as argued in recent theoretical \cite{malinowski2023} and experimental \cite{delaney2024} work, most control electrodes can be co-wired across the device, such that a modest number ($\approx 50 - 100$) of analog voltage inputs suffices to achieve all-to-all connectivity regardless of device size. Furthermore, due to their low bandwidth, the local tuning electrodes can be efficiently multiplexed using only simple, cryogenically compatible, low-power chip-scale electronics \cite{malinowski2023}. This allows hundreds of electrodes to be controlled from a single analog input. By employing these two methods, a few hundred input lines -- a typical number of analog signals in today's TIQCs -- can already control devices with tens of thousands of qubits.

Practical QC manufacturing also requires careful design tolerancing and efficient device calibration, as cell-to-cell variations can have catastrophic effects on device performance and yield \cite{hertzberg2021, berke2022}. Compared to many QC modalities, the all-electronic TIQC architecture poses minimal new demands on the chip design, tolerancing, and manufacturing processes. In contrast to solid-state qubits, trapped-ion qubits leverage the physical separation between the qubit and the chip to suppress the detrimental effects of surface and material disorder on gate performance \cite{deleon2021}. The experiments presented in section III allow us to upper-bound the two-qubit gate error from surface-induced motional dephasing to less than $1 \times 10^{-4}$, and further gains can be achieved by employing coherent control \cite{hayes2012} and surface treatment techniques \cite{allcock2011, hite2012, brown2021, berlin-udi2022}.

Given that TIQCs allow for low surface noise and high qubit frequency uniformity, the primary goal of QC tolerancing and calibration is to equalize the control field parameters in different unit cells. Notably, the on-chip electronic control elements -- the shared trace and the local electrodes -- have typical minimum size scales of tens of microns, several orders of magnitude larger than the resolution of modern semiconductor processes. Thus, large-scale highly uniform fabrication of multi-layer qubit control structures is relatively straightforward. While integrated photonics requires sub-micron dielectric features \cite{beck2024}, their uniformity needs can be considerably relaxed by carrying out dissipative operation in the saturated steady-state regime \cite{lin2013, cole2022, malinowski2022}. Finally, the qubit coupling inhomogeneities caused by residual chip manufacturing and design imperfections can be shimmed out using local tuning electrodes. The combination of low calibration complexity, high parameter stability (section~\ref{sec:experiments}), and the ability to parallelize calibrations in different zones will allow continuous TIQC operation with minimal downtime. 

In summary, we have presented a vision for TIQCs based on all-electronic coherent control to alleviate the scale and performance challenges of today's QCs. We have experimentally tested the fundamental control primitives of our architecture -- site-selective single- and two-qubit gates -- and verified that both can be implemented with state-of-the-art fidelities. We have discussed the prospects for scaling these systems up, and argued that next-generation devices with thousands of qubits can be fabricated using only established microfabrication technology. This shows the path to increasing QC size without sacrificing performance, enabling mid- and large-scale QCs that can solve problems of commercial and scientific importance.

% % \bmhead{Acknowledgements}
\section*{Acknowledgements}
We thank the entire Oxford Ionics team for their contributions to this work. We also thank Daniel Slichter and Mario Gely for their helpful comments on an earlier version of the manuscript.
% \section*{Declarations}

% \begin{itemize}
% \item Funding. Nothing.
% \item Conflict of interest. Yes.
% \item Data availability. Available upon reasonable request.
% \item Author contribution. Low detail.
% \end{itemize}

\bibliography{mylib}

\begin{thebibliography}{89}%
\makeatletter
\providecommand \@ifxundefined [1]{%
 \@ifx{#1\undefined}
}%
\providecommand \@ifnum [1]{%
 \ifnum #1\expandafter \@firstoftwo
 \else \expandafter \@secondoftwo
 \fi
}%
\providecommand \@ifx [1]{%
 \ifx #1\expandafter \@firstoftwo
 \else \expandafter \@secondoftwo
 \fi
}%
\providecommand \natexlab [1]{#1}%
\providecommand \enquote  [1]{``#1''}%
\providecommand \bibnamefont  [1]{#1}%
\providecommand \bibfnamefont [1]{#1}%
\providecommand \citenamefont [1]{#1}%
\providecommand \href@noop [0]{\@secondoftwo}%
\providecommand \href [0]{\begingroup \@sanitize@url \@href}%
\providecommand \@href[1]{\@@startlink{#1}\@@href}%
\providecommand \@@href[1]{\endgroup#1\@@endlink}%
\providecommand \@sanitize@url [0]{\catcode `\\12\catcode `\$12\catcode `\&12\catcode `\#12\catcode `\^12\catcode `\_12\catcode `\%12\relax}%
\providecommand \@@startlink[1]{}%
\providecommand \@@endlink[0]{}%
\providecommand \url  [0]{\begingroup\@sanitize@url \@url }%
\providecommand \@url [1]{\endgroup\@href {#1}{\urlprefix }}%
\providecommand \urlprefix  [0]{URL }%
\providecommand \Eprint [0]{\href }%
\providecommand \doibase [0]{http://dx.doi.org/}%
\providecommand \selectlanguage [0]{\@gobble}%
\providecommand \bibinfo  [0]{\@secondoftwo}%
\providecommand \bibfield  [0]{\@secondoftwo}%
\providecommand \translation [1]{[#1]}%
\providecommand \BibitemOpen [0]{}%
\providecommand \bibitemStop [0]{}%
\providecommand \bibitemNoStop [0]{.\EOS\space}%
\providecommand \EOS [0]{\spacefactor3000\relax}%
\providecommand \BibitemShut  [1]{\csname bibitem#1\endcsname}%
\let\auto@bib@innerbib\@empty
%</preamble>
\bibitem [{\citenamefont {Harty}\ \emph {et~al.}(2014)\citenamefont {Harty}, \citenamefont {Allcock}, \citenamefont {Ballance}, \citenamefont {Guidoni}, \citenamefont {Janacek}, \citenamefont {Linke}, \citenamefont {Stacey},\ and\ \citenamefont {Lucas}}]{harty2014}%
  \BibitemOpen
  \bibfield  {author} {\bibinfo {author} {\bibfnamefont {T.~P.}\ \bibnamefont {Harty}}, \bibinfo {author} {\bibfnamefont {D.~T.~C.}\ \bibnamefont {Allcock}}, \bibinfo {author} {\bibfnamefont {C.~J.}\ \bibnamefont {Ballance}}, \bibinfo {author} {\bibfnamefont {L.}~\bibnamefont {Guidoni}}, \bibinfo {author} {\bibfnamefont {H.~A.}\ \bibnamefont {Janacek}}, \bibinfo {author} {\bibfnamefont {N.~M.}\ \bibnamefont {Linke}}, \bibinfo {author} {\bibfnamefont {D.~N.}\ \bibnamefont {Stacey}}, \ and\ \bibinfo {author} {\bibfnamefont {D.~M.}\ \bibnamefont {Lucas}},\ }\href {\doibase 10.1103/PhysRevLett.113.220501} {\bibfield  {journal} {\bibinfo  {journal} {Physical Review Letters}\ }\textbf {\bibinfo {volume} {113}},\ \bibinfo {pages} {220501} (\bibinfo {year} {2014})}\BibitemShut {NoStop}%
\bibitem [{\citenamefont {Clark}\ \emph {et~al.}(2021)\citenamefont {Clark}, \citenamefont {Tinkey}, \citenamefont {Sawyer}, \citenamefont {Meier}, \citenamefont {Burkhardt}, \citenamefont {Seck}, \citenamefont {Shappert}, \citenamefont {Guise}, \citenamefont {Volin}, \citenamefont {Fallek}, \citenamefont {Hayden}, \citenamefont {Rellergert},\ and\ \citenamefont {Brown}}]{clark2021}%
  \BibitemOpen
  \bibfield  {author} {\bibinfo {author} {\bibfnamefont {C.~R.}\ \bibnamefont {Clark}}, \bibinfo {author} {\bibfnamefont {H.~N.}\ \bibnamefont {Tinkey}}, \bibinfo {author} {\bibfnamefont {B.~C.}\ \bibnamefont {Sawyer}}, \bibinfo {author} {\bibfnamefont {A.~M.}\ \bibnamefont {Meier}}, \bibinfo {author} {\bibfnamefont {K.~A.}\ \bibnamefont {Burkhardt}}, \bibinfo {author} {\bibfnamefont {C.~M.}\ \bibnamefont {Seck}}, \bibinfo {author} {\bibfnamefont {C.~M.}\ \bibnamefont {Shappert}}, \bibinfo {author} {\bibfnamefont {N.~D.}\ \bibnamefont {Guise}}, \bibinfo {author} {\bibfnamefont {C.~E.}\ \bibnamefont {Volin}}, \bibinfo {author} {\bibfnamefont {S.~D.}\ \bibnamefont {Fallek}}, \bibinfo {author} {\bibfnamefont {H.~T.}\ \bibnamefont {Hayden}}, \bibinfo {author} {\bibfnamefont {W.~G.}\ \bibnamefont {Rellergert}}, \ and\ \bibinfo {author} {\bibfnamefont {K.~R.}\ \bibnamefont {Brown}},\ }\href {\doibase 10.1103/PhysRevLett.127.130505} {\bibfield  {journal} {\bibinfo  {journal} {Physical Review Letters}\ }\textbf
  {\bibinfo {volume} {127}},\ \bibinfo {pages} {130505} (\bibinfo {year} {2021})}\BibitemShut {NoStop}%
\bibitem [{\citenamefont {Ding}\ \emph {et~al.}(2023)\citenamefont {Ding}, \citenamefont {Hays}, \citenamefont {Sung}, \citenamefont {Kannan}, \citenamefont {An}, \citenamefont {Di~Paolo}, \citenamefont {Karamlou}, \citenamefont {Hazard}, \citenamefont {Azar}, \citenamefont {Kim}, \citenamefont {Niedzielski}, \citenamefont {Melville}, \citenamefont {Schwartz}, \citenamefont {Yoder}, \citenamefont {Orlando}, \citenamefont {Gustavsson}, \citenamefont {Grover}, \citenamefont {Serniak},\ and\ \citenamefont {Oliver}}]{ding2023}%
  \BibitemOpen
  \bibfield  {author} {\bibinfo {author} {\bibfnamefont {L.}~\bibnamefont {Ding}}, \bibinfo {author} {\bibfnamefont {M.}~\bibnamefont {Hays}}, \bibinfo {author} {\bibfnamefont {Y.}~\bibnamefont {Sung}}, \bibinfo {author} {\bibfnamefont {B.}~\bibnamefont {Kannan}}, \bibinfo {author} {\bibfnamefont {J.}~\bibnamefont {An}}, \bibinfo {author} {\bibfnamefont {A.}~\bibnamefont {Di~Paolo}}, \bibinfo {author} {\bibfnamefont {A.~H.}\ \bibnamefont {Karamlou}}, \bibinfo {author} {\bibfnamefont {T.~M.}\ \bibnamefont {Hazard}}, \bibinfo {author} {\bibfnamefont {K.}~\bibnamefont {Azar}}, \bibinfo {author} {\bibfnamefont {D.~K.}\ \bibnamefont {Kim}}, \bibinfo {author} {\bibfnamefont {B.~M.}\ \bibnamefont {Niedzielski}}, \bibinfo {author} {\bibfnamefont {A.}~\bibnamefont {Melville}}, \bibinfo {author} {\bibfnamefont {M.~E.}\ \bibnamefont {Schwartz}}, \bibinfo {author} {\bibfnamefont {J.~L.}\ \bibnamefont {Yoder}}, \bibinfo {author} {\bibfnamefont {T.~P.}\ \bibnamefont {Orlando}}, \bibinfo {author} {\bibfnamefont
  {S.}~\bibnamefont {Gustavsson}}, \bibinfo {author} {\bibfnamefont {J.~A.}\ \bibnamefont {Grover}}, \bibinfo {author} {\bibfnamefont {K.}~\bibnamefont {Serniak}}, \ and\ \bibinfo {author} {\bibfnamefont {W.~D.}\ \bibnamefont {Oliver}},\ }\href {\doibase 10.1103/PhysRevX.13.031035} {\bibfield  {journal} {\bibinfo  {journal} {Physical Review X}\ }\textbf {\bibinfo {volume} {13}},\ \bibinfo {pages} {031035} (\bibinfo {year} {2023})}\BibitemShut {NoStop}%
\bibitem [{\citenamefont {Acharya}\ \emph {et~al.}(2023)\citenamefont {Acharya}, \citenamefont {Aleiner}, \citenamefont {Allen}, \citenamefont {Andersen}, \citenamefont {Ansmann}, \citenamefont {Arute}, \citenamefont {Arya}, \citenamefont {Asfaw}, \citenamefont {Atalaya}, \citenamefont {Babbush}, \citenamefont {Bacon}, \citenamefont {Bardin}, \citenamefont {Basso}, \citenamefont {Bengtsson}, \citenamefont {Boixo}, \citenamefont {Bortoli}, \citenamefont {Bourassa}, \citenamefont {Bovaird}, \citenamefont {Brill}, \citenamefont {Broughton}, \citenamefont {Buckley}, \citenamefont {Buell}, \citenamefont {Burger}, \citenamefont {Burkett}, \citenamefont {Bushnell}, \citenamefont {Chen}, \citenamefont {Chen}, \citenamefont {Chiaro}, \citenamefont {Cogan}, \citenamefont {Collins}, \citenamefont {Conner}, \citenamefont {Courtney}, \citenamefont {Crook}, \citenamefont {Curtin}, \citenamefont {Debroy}, \citenamefont {Del Toro~Barba}, \citenamefont {Demura}, \citenamefont {Dunsworth}, \citenamefont {Eppens}, \citenamefont
  {Erickson}, \citenamefont {Faoro}, \citenamefont {Farhi}, \citenamefont {Fatemi}, \citenamefont {Flores~Burgos}, \citenamefont {Forati}, \citenamefont {Fowler}, \citenamefont {Foxen}, \citenamefont {Giang}, \citenamefont {Gidney}, \citenamefont {Gilboa}, \citenamefont {Giustina}, \citenamefont {Grajales~Dau}, \citenamefont {Gross}, \citenamefont {Habegger}, \citenamefont {Hamilton}, \citenamefont {Harrigan}, \citenamefont {Harrington}, \citenamefont {Higgott}, \citenamefont {Hilton}, \citenamefont {Hoffmann}, \citenamefont {Hong}, \citenamefont {Huang}, \citenamefont {Huff}, \citenamefont {Huggins}, \citenamefont {Ioffe}, \citenamefont {Isakov}, \citenamefont {Iveland}, \citenamefont {Jeffrey}, \citenamefont {Jiang}, \citenamefont {Jones}, \citenamefont {Juhas}, \citenamefont {Kafri}, \citenamefont {Kechedzhi}, \citenamefont {Kelly}, \citenamefont {Khattar}, \citenamefont {Khezri}, \citenamefont {Kieferov{\'a}}, \citenamefont {Kim}, \citenamefont {Kitaev}, \citenamefont {Klimov}, \citenamefont {Klots},
  \citenamefont {Korotkov}, \citenamefont {Kostritsa}, \citenamefont {Kreikebaum}, \citenamefont {Landhuis}, \citenamefont {Laptev}, \citenamefont {Lau}, \citenamefont {Laws}, \citenamefont {Lee}, \citenamefont {Lee}, \citenamefont {Lester}, \citenamefont {Lill}, \citenamefont {Liu}, \citenamefont {Locharla}, \citenamefont {Lucero}, \citenamefont {Malone}, \citenamefont {Marshall}, \citenamefont {Martin}, \citenamefont {McClean}, \citenamefont {McCourt}, \citenamefont {McEwen}, \citenamefont {Megrant}, \citenamefont {Meurer~Costa}, \citenamefont {Mi}, \citenamefont {Miao}, \citenamefont {Mohseni}, \citenamefont {Montazeri}, \citenamefont {Morvan}, \citenamefont {Mount}, \citenamefont {Mruczkiewicz}, \citenamefont {Naaman}, \citenamefont {Neeley}, \citenamefont {Neill}, \citenamefont {Nersisyan}, \citenamefont {Neven}, \citenamefont {Newman}, \citenamefont {Ng}, \citenamefont {Nguyen}, \citenamefont {Nguyen}, \citenamefont {Niu}, \citenamefont {O'Brien}, \citenamefont {Opremcak}, \citenamefont {Platt},
  \citenamefont {Petukhov}, \citenamefont {Potter}, \citenamefont {Pryadko}, \citenamefont {Quintana}, \citenamefont {Roushan}, \citenamefont {Rubin}, \citenamefont {Saei}, \citenamefont {Sank}, \citenamefont {Sankaragomathi}, \citenamefont {Satzinger}, \citenamefont {Schurkus}, \citenamefont {Schuster}, \citenamefont {Shearn}, \citenamefont {Shorter}, \citenamefont {Shvarts}, \citenamefont {Skruzny}, \citenamefont {Smelyanskiy}, \citenamefont {Smith}, \citenamefont {Sterling}, \citenamefont {Strain}, \citenamefont {Szalay}, \citenamefont {Torres}, \citenamefont {Vidal}, \citenamefont {Villalonga}, \citenamefont {Vollgraff~Heidweiller}, \citenamefont {White}, \citenamefont {Xing}, \citenamefont {Yao}, \citenamefont {Yeh}, \citenamefont {Yoo}, \citenamefont {Young}, \citenamefont {Zalcman}, \citenamefont {Zhang}, \citenamefont {Zhu},\ and\ \citenamefont {{Google Quantum AI}}}]{acharya2023}%
  \BibitemOpen
  \bibfield  {author} {\bibinfo {author} {\bibfnamefont {R.}~\bibnamefont {Acharya}}, \bibinfo {author} {\bibfnamefont {I.}~\bibnamefont {Aleiner}}, \bibinfo {author} {\bibfnamefont {R.}~\bibnamefont {Allen}}, \bibinfo {author} {\bibfnamefont {T.~I.}\ \bibnamefont {Andersen}}, \bibinfo {author} {\bibfnamefont {M.}~\bibnamefont {Ansmann}}, \bibinfo {author} {\bibfnamefont {F.}~\bibnamefont {Arute}}, \bibinfo {author} {\bibfnamefont {K.}~\bibnamefont {Arya}}, \bibinfo {author} {\bibfnamefont {A.}~\bibnamefont {Asfaw}}, \bibinfo {author} {\bibfnamefont {J.}~\bibnamefont {Atalaya}}, \bibinfo {author} {\bibfnamefont {R.}~\bibnamefont {Babbush}}, \bibinfo {author} {\bibfnamefont {D.}~\bibnamefont {Bacon}}, \bibinfo {author} {\bibfnamefont {J.~C.}\ \bibnamefont {Bardin}}, \bibinfo {author} {\bibfnamefont {J.}~\bibnamefont {Basso}}, \bibinfo {author} {\bibfnamefont {A.}~\bibnamefont {Bengtsson}}, \bibinfo {author} {\bibfnamefont {S.}~\bibnamefont {Boixo}}, \bibinfo {author} {\bibfnamefont {G.}~\bibnamefont
  {Bortoli}}, \bibinfo {author} {\bibfnamefont {A.}~\bibnamefont {Bourassa}}, \bibinfo {author} {\bibfnamefont {J.}~\bibnamefont {Bovaird}}, \bibinfo {author} {\bibfnamefont {L.}~\bibnamefont {Brill}}, \bibinfo {author} {\bibfnamefont {M.}~\bibnamefont {Broughton}}, \bibinfo {author} {\bibfnamefont {B.~B.}\ \bibnamefont {Buckley}}, \bibinfo {author} {\bibfnamefont {D.~A.}\ \bibnamefont {Buell}}, \bibinfo {author} {\bibfnamefont {T.}~\bibnamefont {Burger}}, \bibinfo {author} {\bibfnamefont {B.}~\bibnamefont {Burkett}}, \bibinfo {author} {\bibfnamefont {N.}~\bibnamefont {Bushnell}}, \bibinfo {author} {\bibfnamefont {Y.}~\bibnamefont {Chen}}, \bibinfo {author} {\bibfnamefont {Z.}~\bibnamefont {Chen}}, \bibinfo {author} {\bibfnamefont {B.}~\bibnamefont {Chiaro}}, \bibinfo {author} {\bibfnamefont {J.}~\bibnamefont {Cogan}}, \bibinfo {author} {\bibfnamefont {R.}~\bibnamefont {Collins}}, \bibinfo {author} {\bibfnamefont {P.}~\bibnamefont {Conner}}, \bibinfo {author} {\bibfnamefont {W.}~\bibnamefont {Courtney}},
  \bibinfo {author} {\bibfnamefont {A.~L.}\ \bibnamefont {Crook}}, \bibinfo {author} {\bibfnamefont {B.}~\bibnamefont {Curtin}}, \bibinfo {author} {\bibfnamefont {D.~M.}\ \bibnamefont {Debroy}}, \bibinfo {author} {\bibfnamefont {A.}~\bibnamefont {Del Toro~Barba}}, \bibinfo {author} {\bibfnamefont {S.}~\bibnamefont {Demura}}, \bibinfo {author} {\bibfnamefont {A.}~\bibnamefont {Dunsworth}}, \bibinfo {author} {\bibfnamefont {D.}~\bibnamefont {Eppens}}, \bibinfo {author} {\bibfnamefont {C.}~\bibnamefont {Erickson}}, \bibinfo {author} {\bibfnamefont {L.}~\bibnamefont {Faoro}}, \bibinfo {author} {\bibfnamefont {E.}~\bibnamefont {Farhi}}, \bibinfo {author} {\bibfnamefont {R.}~\bibnamefont {Fatemi}}, \bibinfo {author} {\bibfnamefont {L.}~\bibnamefont {Flores~Burgos}}, \bibinfo {author} {\bibfnamefont {E.}~\bibnamefont {Forati}}, \bibinfo {author} {\bibfnamefont {A.~G.}\ \bibnamefont {Fowler}}, \bibinfo {author} {\bibfnamefont {B.}~\bibnamefont {Foxen}}, \bibinfo {author} {\bibfnamefont {W.}~\bibnamefont {Giang}},
  \bibinfo {author} {\bibfnamefont {C.}~\bibnamefont {Gidney}}, \bibinfo {author} {\bibfnamefont {D.}~\bibnamefont {Gilboa}}, \bibinfo {author} {\bibfnamefont {M.}~\bibnamefont {Giustina}}, \bibinfo {author} {\bibfnamefont {A.}~\bibnamefont {Grajales~Dau}}, \bibinfo {author} {\bibfnamefont {J.~A.}\ \bibnamefont {Gross}}, \bibinfo {author} {\bibfnamefont {S.}~\bibnamefont {Habegger}}, \bibinfo {author} {\bibfnamefont {M.~C.}\ \bibnamefont {Hamilton}}, \bibinfo {author} {\bibfnamefont {M.~P.}\ \bibnamefont {Harrigan}}, \bibinfo {author} {\bibfnamefont {S.~D.}\ \bibnamefont {Harrington}}, \bibinfo {author} {\bibfnamefont {O.}~\bibnamefont {Higgott}}, \bibinfo {author} {\bibfnamefont {J.}~\bibnamefont {Hilton}}, \bibinfo {author} {\bibfnamefont {M.}~\bibnamefont {Hoffmann}}, \bibinfo {author} {\bibfnamefont {S.}~\bibnamefont {Hong}}, \bibinfo {author} {\bibfnamefont {T.}~\bibnamefont {Huang}}, \bibinfo {author} {\bibfnamefont {A.}~\bibnamefont {Huff}}, \bibinfo {author} {\bibfnamefont {W.~J.}\ \bibnamefont
  {Huggins}}, \bibinfo {author} {\bibfnamefont {L.~B.}\ \bibnamefont {Ioffe}}, \bibinfo {author} {\bibfnamefont {S.~V.}\ \bibnamefont {Isakov}}, \bibinfo {author} {\bibfnamefont {J.}~\bibnamefont {Iveland}}, \bibinfo {author} {\bibfnamefont {E.}~\bibnamefont {Jeffrey}}, \bibinfo {author} {\bibfnamefont {Z.}~\bibnamefont {Jiang}}, \bibinfo {author} {\bibfnamefont {C.}~\bibnamefont {Jones}}, \bibinfo {author} {\bibfnamefont {P.}~\bibnamefont {Juhas}}, \bibinfo {author} {\bibfnamefont {D.}~\bibnamefont {Kafri}}, \bibinfo {author} {\bibfnamefont {K.}~\bibnamefont {Kechedzhi}}, \bibinfo {author} {\bibfnamefont {J.}~\bibnamefont {Kelly}}, \bibinfo {author} {\bibfnamefont {T.}~\bibnamefont {Khattar}}, \bibinfo {author} {\bibfnamefont {M.}~\bibnamefont {Khezri}}, \bibinfo {author} {\bibfnamefont {M.}~\bibnamefont {Kieferov{\'a}}}, \bibinfo {author} {\bibfnamefont {S.}~\bibnamefont {Kim}}, \bibinfo {author} {\bibfnamefont {A.}~\bibnamefont {Kitaev}}, \bibinfo {author} {\bibfnamefont {P.~V.}\ \bibnamefont {Klimov}},
  \bibinfo {author} {\bibfnamefont {A.~R.}\ \bibnamefont {Klots}}, \bibinfo {author} {\bibfnamefont {A.~N.}\ \bibnamefont {Korotkov}}, \bibinfo {author} {\bibfnamefont {F.}~\bibnamefont {Kostritsa}}, \bibinfo {author} {\bibfnamefont {J.~M.}\ \bibnamefont {Kreikebaum}}, \bibinfo {author} {\bibfnamefont {D.}~\bibnamefont {Landhuis}}, \bibinfo {author} {\bibfnamefont {P.}~\bibnamefont {Laptev}}, \bibinfo {author} {\bibfnamefont {K.-M.}\ \bibnamefont {Lau}}, \bibinfo {author} {\bibfnamefont {L.}~\bibnamefont {Laws}}, \bibinfo {author} {\bibfnamefont {J.}~\bibnamefont {Lee}}, \bibinfo {author} {\bibfnamefont {K.}~\bibnamefont {Lee}}, \bibinfo {author} {\bibfnamefont {B.~J.}\ \bibnamefont {Lester}}, \bibinfo {author} {\bibfnamefont {A.}~\bibnamefont {Lill}}, \bibinfo {author} {\bibfnamefont {W.}~\bibnamefont {Liu}}, \bibinfo {author} {\bibfnamefont {A.}~\bibnamefont {Locharla}}, \bibinfo {author} {\bibfnamefont {E.}~\bibnamefont {Lucero}}, \bibinfo {author} {\bibfnamefont {F.~D.}\ \bibnamefont {Malone}}, \bibinfo
  {author} {\bibfnamefont {J.}~\bibnamefont {Marshall}}, \bibinfo {author} {\bibfnamefont {O.}~\bibnamefont {Martin}}, \bibinfo {author} {\bibfnamefont {J.~R.}\ \bibnamefont {McClean}}, \bibinfo {author} {\bibfnamefont {T.}~\bibnamefont {McCourt}}, \bibinfo {author} {\bibfnamefont {M.}~\bibnamefont {McEwen}}, \bibinfo {author} {\bibfnamefont {A.}~\bibnamefont {Megrant}}, \bibinfo {author} {\bibfnamefont {B.}~\bibnamefont {Meurer~Costa}}, \bibinfo {author} {\bibfnamefont {X.}~\bibnamefont {Mi}}, \bibinfo {author} {\bibfnamefont {K.~C.}\ \bibnamefont {Miao}}, \bibinfo {author} {\bibfnamefont {M.}~\bibnamefont {Mohseni}}, \bibinfo {author} {\bibfnamefont {S.}~\bibnamefont {Montazeri}}, \bibinfo {author} {\bibfnamefont {A.}~\bibnamefont {Morvan}}, \bibinfo {author} {\bibfnamefont {E.}~\bibnamefont {Mount}}, \bibinfo {author} {\bibfnamefont {W.}~\bibnamefont {Mruczkiewicz}}, \bibinfo {author} {\bibfnamefont {O.}~\bibnamefont {Naaman}}, \bibinfo {author} {\bibfnamefont {M.}~\bibnamefont {Neeley}}, \bibinfo {author}
  {\bibfnamefont {C.}~\bibnamefont {Neill}}, \bibinfo {author} {\bibfnamefont {A.}~\bibnamefont {Nersisyan}}, \bibinfo {author} {\bibfnamefont {H.}~\bibnamefont {Neven}}, \bibinfo {author} {\bibfnamefont {M.}~\bibnamefont {Newman}}, \bibinfo {author} {\bibfnamefont {J.~H.}\ \bibnamefont {Ng}}, \bibinfo {author} {\bibfnamefont {A.}~\bibnamefont {Nguyen}}, \bibinfo {author} {\bibfnamefont {M.}~\bibnamefont {Nguyen}}, \bibinfo {author} {\bibfnamefont {M.~Y.}\ \bibnamefont {Niu}}, \bibinfo {author} {\bibfnamefont {T.~E.}\ \bibnamefont {O'Brien}}, \bibinfo {author} {\bibfnamefont {A.}~\bibnamefont {Opremcak}}, \bibinfo {author} {\bibfnamefont {J.}~\bibnamefont {Platt}}, \bibinfo {author} {\bibfnamefont {A.}~\bibnamefont {Petukhov}}, \bibinfo {author} {\bibfnamefont {R.}~\bibnamefont {Potter}}, \bibinfo {author} {\bibfnamefont {L.~P.}\ \bibnamefont {Pryadko}}, \bibinfo {author} {\bibfnamefont {C.}~\bibnamefont {Quintana}}, \bibinfo {author} {\bibfnamefont {P.}~\bibnamefont {Roushan}}, \bibinfo {author}
  {\bibfnamefont {N.~C.}\ \bibnamefont {Rubin}}, \bibinfo {author} {\bibfnamefont {N.}~\bibnamefont {Saei}}, \bibinfo {author} {\bibfnamefont {D.}~\bibnamefont {Sank}}, \bibinfo {author} {\bibfnamefont {K.}~\bibnamefont {Sankaragomathi}}, \bibinfo {author} {\bibfnamefont {K.~J.}\ \bibnamefont {Satzinger}}, \bibinfo {author} {\bibfnamefont {H.~F.}\ \bibnamefont {Schurkus}}, \bibinfo {author} {\bibfnamefont {C.}~\bibnamefont {Schuster}}, \bibinfo {author} {\bibfnamefont {M.~J.}\ \bibnamefont {Shearn}}, \bibinfo {author} {\bibfnamefont {A.}~\bibnamefont {Shorter}}, \bibinfo {author} {\bibfnamefont {V.}~\bibnamefont {Shvarts}}, \bibinfo {author} {\bibfnamefont {J.}~\bibnamefont {Skruzny}}, \bibinfo {author} {\bibfnamefont {V.}~\bibnamefont {Smelyanskiy}}, \bibinfo {author} {\bibfnamefont {W.~C.}\ \bibnamefont {Smith}}, \bibinfo {author} {\bibfnamefont {G.}~\bibnamefont {Sterling}}, \bibinfo {author} {\bibfnamefont {D.}~\bibnamefont {Strain}}, \bibinfo {author} {\bibfnamefont {M.}~\bibnamefont {Szalay}}, \bibinfo
  {author} {\bibfnamefont {A.}~\bibnamefont {Torres}}, \bibinfo {author} {\bibfnamefont {G.}~\bibnamefont {Vidal}}, \bibinfo {author} {\bibfnamefont {B.}~\bibnamefont {Villalonga}}, \bibinfo {author} {\bibfnamefont {C.}~\bibnamefont {Vollgraff~Heidweiller}}, \bibinfo {author} {\bibfnamefont {T.}~\bibnamefont {White}}, \bibinfo {author} {\bibfnamefont {C.}~\bibnamefont {Xing}}, \bibinfo {author} {\bibfnamefont {Z.~J.}\ \bibnamefont {Yao}}, \bibinfo {author} {\bibfnamefont {P.}~\bibnamefont {Yeh}}, \bibinfo {author} {\bibfnamefont {J.}~\bibnamefont {Yoo}}, \bibinfo {author} {\bibfnamefont {G.}~\bibnamefont {Young}}, \bibinfo {author} {\bibfnamefont {A.}~\bibnamefont {Zalcman}}, \bibinfo {author} {\bibfnamefont {Y.}~\bibnamefont {Zhang}}, \bibinfo {author} {\bibfnamefont {N.}~\bibnamefont {Zhu}}, \ and\ \bibinfo {author} {\bibnamefont {{Google Quantum AI}}},\ }\href {\doibase 10.1038/s41586-022-05434-1} {\bibfield  {journal} {\bibinfo  {journal} {Nature}\ }\textbf {\bibinfo {volume} {614}},\ \bibinfo {pages}
  {676} (\bibinfo {year} {2023})}\BibitemShut {NoStop}%
\bibitem [{\citenamefont {{da Silva}}\ \emph {et~al.}(2024)\citenamefont {{da Silva}}, \citenamefont {{Ryan-Anderson}}, \citenamefont {{Bello-Rivas}}, \citenamefont {Chernoguzov}, \citenamefont {Dreiling}, \citenamefont {Foltz}, \citenamefont {Frachon}, \citenamefont {Gaebler}, \citenamefont {Gatterman}, \citenamefont {{Grans-Samuelsson}}, \citenamefont {Hayes}, \citenamefont {Hewitt}, \citenamefont {Johansen}, \citenamefont {Lucchetti}, \citenamefont {Mills}, \citenamefont {Moses}, \citenamefont {Neyenhuis}, \citenamefont {Paz}, \citenamefont {Pino}, \citenamefont {Siegfried}, \citenamefont {Strabley}, \citenamefont {Sundaram}, \citenamefont {Tom}, \citenamefont {Wernli}, \citenamefont {Zanner}, \citenamefont {Stutz},\ and\ \citenamefont {Svore}}]{dasilva2024}%
  \BibitemOpen
  \bibfield  {author} {\bibinfo {author} {\bibfnamefont {M.~P.}\ \bibnamefont {{da Silva}}}, \bibinfo {author} {\bibfnamefont {C.}~\bibnamefont {{Ryan-Anderson}}}, \bibinfo {author} {\bibfnamefont {J.~M.}\ \bibnamefont {{Bello-Rivas}}}, \bibinfo {author} {\bibfnamefont {A.}~\bibnamefont {Chernoguzov}}, \bibinfo {author} {\bibfnamefont {J.~M.}\ \bibnamefont {Dreiling}}, \bibinfo {author} {\bibfnamefont {C.}~\bibnamefont {Foltz}}, \bibinfo {author} {\bibfnamefont {F.}~\bibnamefont {Frachon}}, \bibinfo {author} {\bibfnamefont {J.~P.}\ \bibnamefont {Gaebler}}, \bibinfo {author} {\bibfnamefont {T.~M.}\ \bibnamefont {Gatterman}}, \bibinfo {author} {\bibfnamefont {L.}~\bibnamefont {{Grans-Samuelsson}}}, \bibinfo {author} {\bibfnamefont {D.}~\bibnamefont {Hayes}}, \bibinfo {author} {\bibfnamefont {N.}~\bibnamefont {Hewitt}}, \bibinfo {author} {\bibfnamefont {J.}~\bibnamefont {Johansen}}, \bibinfo {author} {\bibfnamefont {D.}~\bibnamefont {Lucchetti}}, \bibinfo {author} {\bibfnamefont {M.}~\bibnamefont {Mills}},
  \bibinfo {author} {\bibfnamefont {S.~A.}\ \bibnamefont {Moses}}, \bibinfo {author} {\bibfnamefont {B.}~\bibnamefont {Neyenhuis}}, \bibinfo {author} {\bibfnamefont {A.}~\bibnamefont {Paz}}, \bibinfo {author} {\bibfnamefont {J.}~\bibnamefont {Pino}}, \bibinfo {author} {\bibfnamefont {P.}~\bibnamefont {Siegfried}}, \bibinfo {author} {\bibfnamefont {J.}~\bibnamefont {Strabley}}, \bibinfo {author} {\bibfnamefont {A.}~\bibnamefont {Sundaram}}, \bibinfo {author} {\bibfnamefont {D.}~\bibnamefont {Tom}}, \bibinfo {author} {\bibfnamefont {S.~J.}\ \bibnamefont {Wernli}}, \bibinfo {author} {\bibfnamefont {M.}~\bibnamefont {Zanner}}, \bibinfo {author} {\bibfnamefont {R.~P.}\ \bibnamefont {Stutz}}, \ and\ \bibinfo {author} {\bibfnamefont {K.~M.}\ \bibnamefont {Svore}},\ }\href {\doibase 10.48550/arXiv.2404.02280} {\enquote {\bibinfo {title} {Demonstration of logical qubits and repeated error correction with better-than-physical error rates},}\ } (\bibinfo {year} {2024}),\ \Eprint {http://arxiv.org/abs/2404.02280}
  {arXiv:2404.02280 [quant-ph]} \BibitemShut {NoStop}%
\bibitem [{\citenamefont {King}\ \emph {et~al.}(2023)\citenamefont {King}, \citenamefont {Raymond}, \citenamefont {Lanting}, \citenamefont {Harris}, \citenamefont {Zucca}, \citenamefont {Altomare}, \citenamefont {Berkley}, \citenamefont {Boothby}, \citenamefont {Ejtemaee}, \citenamefont {Enderud}, \citenamefont {Hoskinson}, \citenamefont {Huang}, \citenamefont {Ladizinsky}, \citenamefont {MacDonald}, \citenamefont {Marsden}, \citenamefont {Molavi}, \citenamefont {Oh}, \citenamefont {{Poulin-Lamarre}}, \citenamefont {Reis}, \citenamefont {Rich}, \citenamefont {Sato}, \citenamefont {Tsai}, \citenamefont {Volkmann}, \citenamefont {Whittaker}, \citenamefont {Yao}, \citenamefont {Sandvik},\ and\ \citenamefont {Amin}}]{king2023}%
  \BibitemOpen
  \bibfield  {author} {\bibinfo {author} {\bibfnamefont {A.~D.}\ \bibnamefont {King}}, \bibinfo {author} {\bibfnamefont {J.}~\bibnamefont {Raymond}}, \bibinfo {author} {\bibfnamefont {T.}~\bibnamefont {Lanting}}, \bibinfo {author} {\bibfnamefont {R.}~\bibnamefont {Harris}}, \bibinfo {author} {\bibfnamefont {A.}~\bibnamefont {Zucca}}, \bibinfo {author} {\bibfnamefont {F.}~\bibnamefont {Altomare}}, \bibinfo {author} {\bibfnamefont {A.~J.}\ \bibnamefont {Berkley}}, \bibinfo {author} {\bibfnamefont {K.}~\bibnamefont {Boothby}}, \bibinfo {author} {\bibfnamefont {S.}~\bibnamefont {Ejtemaee}}, \bibinfo {author} {\bibfnamefont {C.}~\bibnamefont {Enderud}}, \bibinfo {author} {\bibfnamefont {E.}~\bibnamefont {Hoskinson}}, \bibinfo {author} {\bibfnamefont {S.}~\bibnamefont {Huang}}, \bibinfo {author} {\bibfnamefont {E.}~\bibnamefont {Ladizinsky}}, \bibinfo {author} {\bibfnamefont {A.~J.~R.}\ \bibnamefont {MacDonald}}, \bibinfo {author} {\bibfnamefont {G.}~\bibnamefont {Marsden}}, \bibinfo {author} {\bibfnamefont
  {R.}~\bibnamefont {Molavi}}, \bibinfo {author} {\bibfnamefont {T.}~\bibnamefont {Oh}}, \bibinfo {author} {\bibfnamefont {G.}~\bibnamefont {{Poulin-Lamarre}}}, \bibinfo {author} {\bibfnamefont {M.}~\bibnamefont {Reis}}, \bibinfo {author} {\bibfnamefont {C.}~\bibnamefont {Rich}}, \bibinfo {author} {\bibfnamefont {Y.}~\bibnamefont {Sato}}, \bibinfo {author} {\bibfnamefont {N.}~\bibnamefont {Tsai}}, \bibinfo {author} {\bibfnamefont {M.}~\bibnamefont {Volkmann}}, \bibinfo {author} {\bibfnamefont {J.~D.}\ \bibnamefont {Whittaker}}, \bibinfo {author} {\bibfnamefont {J.}~\bibnamefont {Yao}}, \bibinfo {author} {\bibfnamefont {A.~W.}\ \bibnamefont {Sandvik}}, \ and\ \bibinfo {author} {\bibfnamefont {M.~H.}\ \bibnamefont {Amin}},\ }\href {\doibase 10.1038/s41586-023-05867-2} {\bibfield  {journal} {\bibinfo  {journal} {Nature}\ }\textbf {\bibinfo {volume} {617}},\ \bibinfo {pages} {61} (\bibinfo {year} {2023})}\BibitemShut {NoStop}%
\bibitem [{\citenamefont {Kim}\ \emph {et~al.}(2023)\citenamefont {Kim}, \citenamefont {Eddins}, \citenamefont {Anand}, \citenamefont {Wei}, \citenamefont {{van den Berg}}, \citenamefont {Rosenblatt}, \citenamefont {Nayfeh}, \citenamefont {Wu}, \citenamefont {Zaletel}, \citenamefont {Temme},\ and\ \citenamefont {Kandala}}]{kim2023}%
  \BibitemOpen
  \bibfield  {author} {\bibinfo {author} {\bibfnamefont {Y.}~\bibnamefont {Kim}}, \bibinfo {author} {\bibfnamefont {A.}~\bibnamefont {Eddins}}, \bibinfo {author} {\bibfnamefont {S.}~\bibnamefont {Anand}}, \bibinfo {author} {\bibfnamefont {K.~X.}\ \bibnamefont {Wei}}, \bibinfo {author} {\bibfnamefont {E.}~\bibnamefont {{van den Berg}}}, \bibinfo {author} {\bibfnamefont {S.}~\bibnamefont {Rosenblatt}}, \bibinfo {author} {\bibfnamefont {H.}~\bibnamefont {Nayfeh}}, \bibinfo {author} {\bibfnamefont {Y.}~\bibnamefont {Wu}}, \bibinfo {author} {\bibfnamefont {M.}~\bibnamefont {Zaletel}}, \bibinfo {author} {\bibfnamefont {K.}~\bibnamefont {Temme}}, \ and\ \bibinfo {author} {\bibfnamefont {A.}~\bibnamefont {Kandala}},\ }\href {\doibase 10.1038/s41586-023-06096-3} {\bibfield  {journal} {\bibinfo  {journal} {Nature}\ }\textbf {\bibinfo {volume} {618}},\ \bibinfo {pages} {500} (\bibinfo {year} {2023})}\BibitemShut {NoStop}%
\bibitem [{\citenamefont {Bluvstein}\ \emph {et~al.}(2024)\citenamefont {Bluvstein}, \citenamefont {Evered}, \citenamefont {Geim}, \citenamefont {Li}, \citenamefont {Zhou}, \citenamefont {Manovitz}, \citenamefont {Ebadi}, \citenamefont {Cain}, \citenamefont {Kalinowski}, \citenamefont {Hangleiter}, \citenamefont {Bonilla~Ataides}, \citenamefont {Maskara}, \citenamefont {Cong}, \citenamefont {Gao}, \citenamefont {Sales~Rodriguez}, \citenamefont {Karolyshyn}, \citenamefont {Semeghini}, \citenamefont {Gullans}, \citenamefont {Greiner}, \citenamefont {Vuleti{\'c}},\ and\ \citenamefont {Lukin}}]{bluvstein2024}%
  \BibitemOpen
  \bibfield  {author} {\bibinfo {author} {\bibfnamefont {D.}~\bibnamefont {Bluvstein}}, \bibinfo {author} {\bibfnamefont {S.~J.}\ \bibnamefont {Evered}}, \bibinfo {author} {\bibfnamefont {A.~A.}\ \bibnamefont {Geim}}, \bibinfo {author} {\bibfnamefont {S.~H.}\ \bibnamefont {Li}}, \bibinfo {author} {\bibfnamefont {H.}~\bibnamefont {Zhou}}, \bibinfo {author} {\bibfnamefont {T.}~\bibnamefont {Manovitz}}, \bibinfo {author} {\bibfnamefont {S.}~\bibnamefont {Ebadi}}, \bibinfo {author} {\bibfnamefont {M.}~\bibnamefont {Cain}}, \bibinfo {author} {\bibfnamefont {M.}~\bibnamefont {Kalinowski}}, \bibinfo {author} {\bibfnamefont {D.}~\bibnamefont {Hangleiter}}, \bibinfo {author} {\bibfnamefont {J.~P.}\ \bibnamefont {Bonilla~Ataides}}, \bibinfo {author} {\bibfnamefont {N.}~\bibnamefont {Maskara}}, \bibinfo {author} {\bibfnamefont {I.}~\bibnamefont {Cong}}, \bibinfo {author} {\bibfnamefont {X.}~\bibnamefont {Gao}}, \bibinfo {author} {\bibfnamefont {P.}~\bibnamefont {Sales~Rodriguez}}, \bibinfo {author} {\bibfnamefont
  {T.}~\bibnamefont {Karolyshyn}}, \bibinfo {author} {\bibfnamefont {G.}~\bibnamefont {Semeghini}}, \bibinfo {author} {\bibfnamefont {M.~J.}\ \bibnamefont {Gullans}}, \bibinfo {author} {\bibfnamefont {M.}~\bibnamefont {Greiner}}, \bibinfo {author} {\bibfnamefont {V.}~\bibnamefont {Vuleti{\'c}}}, \ and\ \bibinfo {author} {\bibfnamefont {M.~D.}\ \bibnamefont {Lukin}},\ }\href {\doibase 10.1038/s41586-023-06927-3} {\bibfield  {journal} {\bibinfo  {journal} {Nature}\ }\textbf {\bibinfo {volume} {626}},\ \bibinfo {pages} {58} (\bibinfo {year} {2024})}\BibitemShut {NoStop}%
\bibitem [{\citenamefont {Manetsch}\ \emph {et~al.}(2024)\citenamefont {Manetsch}, \citenamefont {Nomura}, \citenamefont {Bataille}, \citenamefont {Leung}, \citenamefont {Lv},\ and\ \citenamefont {Endres}}]{manetsch2024}%
  \BibitemOpen
  \bibfield  {author} {\bibinfo {author} {\bibfnamefont {H.~J.}\ \bibnamefont {Manetsch}}, \bibinfo {author} {\bibfnamefont {G.}~\bibnamefont {Nomura}}, \bibinfo {author} {\bibfnamefont {E.}~\bibnamefont {Bataille}}, \bibinfo {author} {\bibfnamefont {K.~H.}\ \bibnamefont {Leung}}, \bibinfo {author} {\bibfnamefont {X.}~\bibnamefont {Lv}}, \ and\ \bibinfo {author} {\bibfnamefont {M.}~\bibnamefont {Endres}},\ }\href {\doibase 10.48550/arXiv.2403.12021} {\enquote {\bibinfo {title} {A tweezer array with 6100 highly coherent atomic qubits},}\ } (\bibinfo {year} {2024}),\ \Eprint {http://arxiv.org/abs/2403.12021} {arXiv:2403.12021 [cond-mat, physics:physics, physics:quant-ph]} \BibitemShut {NoStop}%
\bibitem [{\citenamefont {Moehring}\ \emph {et~al.}(2007)\citenamefont {Moehring}, \citenamefont {Maunz}, \citenamefont {Olmschenk}, \citenamefont {Younge}, \citenamefont {Matsukevich}, \citenamefont {Duan},\ and\ \citenamefont {Monroe}}]{Moehring2007}%
  \BibitemOpen
  \bibfield  {author} {\bibinfo {author} {\bibfnamefont {D.~L.}\ \bibnamefont {Moehring}}, \bibinfo {author} {\bibfnamefont {P.}~\bibnamefont {Maunz}}, \bibinfo {author} {\bibfnamefont {S.}~\bibnamefont {Olmschenk}}, \bibinfo {author} {\bibfnamefont {K.~C.}\ \bibnamefont {Younge}}, \bibinfo {author} {\bibfnamefont {D.~N.}\ \bibnamefont {Matsukevich}}, \bibinfo {author} {\bibfnamefont {L.-M.}\ \bibnamefont {Duan}}, \ and\ \bibinfo {author} {\bibfnamefont {C.}~\bibnamefont {Monroe}},\ }\href {\doibase 10.1038/nature06118} {\bibfield  {journal} {\bibinfo  {journal} {Nature}\ }\textbf {\bibinfo {volume} {449}},\ \bibinfo {pages} {68} (\bibinfo {year} {2007})}\BibitemShut {NoStop}%
\bibitem [{\citenamefont {Mills}\ \emph {et~al.}(2019)\citenamefont {Mills}, \citenamefont {Zajac}, \citenamefont {Gullans}, \citenamefont {Schupp}, \citenamefont {Hazard},\ and\ \citenamefont {Petta}}]{mills2019}%
  \BibitemOpen
  \bibfield  {author} {\bibinfo {author} {\bibfnamefont {A.~R.}\ \bibnamefont {Mills}}, \bibinfo {author} {\bibfnamefont {D.~M.}\ \bibnamefont {Zajac}}, \bibinfo {author} {\bibfnamefont {M.~J.}\ \bibnamefont {Gullans}}, \bibinfo {author} {\bibfnamefont {F.~J.}\ \bibnamefont {Schupp}}, \bibinfo {author} {\bibfnamefont {T.~M.}\ \bibnamefont {Hazard}}, \ and\ \bibinfo {author} {\bibfnamefont {J.~R.}\ \bibnamefont {Petta}},\ }\href {\doibase 10.1038/s41467-019-08970-z} {\bibfield  {journal} {\bibinfo  {journal} {Nature Communications}\ }\textbf {\bibinfo {volume} {10}},\ \bibinfo {pages} {1063} (\bibinfo {year} {2019})}\BibitemShut {NoStop}%
\bibitem [{\citenamefont {Zhong}\ \emph {et~al.}(2019)\citenamefont {Zhong}, \citenamefont {Chang}, \citenamefont {Satzinger}, \citenamefont {Chou}, \citenamefont {Bienfait}, \citenamefont {Conner}, \citenamefont {Dumur}, \citenamefont {Grebel}, \citenamefont {Peairs}, \citenamefont {Povey}, \citenamefont {Schuster},\ and\ \citenamefont {Cleland}}]{Zhong2019}%
  \BibitemOpen
  \bibfield  {author} {\bibinfo {author} {\bibfnamefont {Y.~P.}\ \bibnamefont {Zhong}}, \bibinfo {author} {\bibfnamefont {H.-S.}\ \bibnamefont {Chang}}, \bibinfo {author} {\bibfnamefont {K.~J.}\ \bibnamefont {Satzinger}}, \bibinfo {author} {\bibfnamefont {M.-H.}\ \bibnamefont {Chou}}, \bibinfo {author} {\bibfnamefont {A.}~\bibnamefont {Bienfait}}, \bibinfo {author} {\bibfnamefont {C.~R.}\ \bibnamefont {Conner}}, \bibinfo {author} {\bibfnamefont {{\'E}.}~\bibnamefont {Dumur}}, \bibinfo {author} {\bibfnamefont {J.}~\bibnamefont {Grebel}}, \bibinfo {author} {\bibfnamefont {G.~A.}\ \bibnamefont {Peairs}}, \bibinfo {author} {\bibfnamefont {R.~G.}\ \bibnamefont {Povey}}, \bibinfo {author} {\bibfnamefont {D.~I.}\ \bibnamefont {Schuster}}, \ and\ \bibinfo {author} {\bibfnamefont {A.~N.}\ \bibnamefont {Cleland}},\ }\href {\doibase 10.1038/s41567-019-0507-7} {\bibfield  {journal} {\bibinfo  {journal} {Nature Physics}\ }\textbf {\bibinfo {volume} {15}},\ \bibinfo {pages} {741} (\bibinfo {year} {2019})}\BibitemShut
  {NoStop}%
\bibitem [{\citenamefont {Stephenson}\ \emph {et~al.}(2020)\citenamefont {Stephenson}, \citenamefont {Nadlinger}, \citenamefont {Nichol}, \citenamefont {An}, \citenamefont {Drmota}, \citenamefont {Ballance}, \citenamefont {Thirumalai}, \citenamefont {Goodwin}, \citenamefont {Lucas},\ and\ \citenamefont {Ballance}}]{stephenson2020}%
  \BibitemOpen
  \bibfield  {author} {\bibinfo {author} {\bibfnamefont {L.~J.}\ \bibnamefont {Stephenson}}, \bibinfo {author} {\bibfnamefont {D.~P.}\ \bibnamefont {Nadlinger}}, \bibinfo {author} {\bibfnamefont {B.~C.}\ \bibnamefont {Nichol}}, \bibinfo {author} {\bibfnamefont {S.}~\bibnamefont {An}}, \bibinfo {author} {\bibfnamefont {P.}~\bibnamefont {Drmota}}, \bibinfo {author} {\bibfnamefont {T.~G.}\ \bibnamefont {Ballance}}, \bibinfo {author} {\bibfnamefont {K.}~\bibnamefont {Thirumalai}}, \bibinfo {author} {\bibfnamefont {J.~F.}\ \bibnamefont {Goodwin}}, \bibinfo {author} {\bibfnamefont {D.~M.}\ \bibnamefont {Lucas}}, \ and\ \bibinfo {author} {\bibfnamefont {C.~J.}\ \bibnamefont {Ballance}},\ }\href {\doibase 10.1103/PhysRevLett.124.110501} {\bibfield  {journal} {\bibinfo  {journal} {Physical Review Letters}\ }\textbf {\bibinfo {volume} {124}},\ \bibinfo {pages} {110501} (\bibinfo {year} {2020})}\BibitemShut {NoStop}%
\bibitem [{\citenamefont {Mehta}\ \emph {et~al.}(2014)\citenamefont {Mehta}, \citenamefont {Eltony}, \citenamefont {Bruzewicz}, \citenamefont {Chuang}, \citenamefont {Ram}, \citenamefont {Sage},\ and\ \citenamefont {Chiaverini}}]{mehta2014}%
  \BibitemOpen
  \bibfield  {author} {\bibinfo {author} {\bibfnamefont {K.~K.}\ \bibnamefont {Mehta}}, \bibinfo {author} {\bibfnamefont {A.~M.}\ \bibnamefont {Eltony}}, \bibinfo {author} {\bibfnamefont {C.~D.}\ \bibnamefont {Bruzewicz}}, \bibinfo {author} {\bibfnamefont {I.~L.}\ \bibnamefont {Chuang}}, \bibinfo {author} {\bibfnamefont {R.~J.}\ \bibnamefont {Ram}}, \bibinfo {author} {\bibfnamefont {J.~M.}\ \bibnamefont {Sage}}, \ and\ \bibinfo {author} {\bibfnamefont {J.}~\bibnamefont {Chiaverini}},\ }\href {\doibase 10.1063/1.4892061} {\bibfield  {journal} {\bibinfo  {journal} {Applied Physics Letters}\ }\textbf {\bibinfo {volume} {105}},\ \bibinfo {pages} {044103} (\bibinfo {year} {2014})}\BibitemShut {NoStop}%
\bibitem [{\citenamefont {Mehta}\ \emph {et~al.}(2020)\citenamefont {Mehta}, \citenamefont {Zhang}, \citenamefont {Malinowski}, \citenamefont {Nguyen}, \citenamefont {Stadler},\ and\ \citenamefont {Home}}]{mehta2020}%
  \BibitemOpen
  \bibfield  {author} {\bibinfo {author} {\bibfnamefont {K.~K.}\ \bibnamefont {Mehta}}, \bibinfo {author} {\bibfnamefont {C.}~\bibnamefont {Zhang}}, \bibinfo {author} {\bibfnamefont {M.}~\bibnamefont {Malinowski}}, \bibinfo {author} {\bibfnamefont {T.-L.}\ \bibnamefont {Nguyen}}, \bibinfo {author} {\bibfnamefont {M.}~\bibnamefont {Stadler}}, \ and\ \bibinfo {author} {\bibfnamefont {J.~P.}\ \bibnamefont {Home}},\ }\href {\doibase 10.1038/s41586-020-2823-6} {\bibfield  {journal} {\bibinfo  {journal} {Nature}\ }\textbf {\bibinfo {volume} {586}},\ \bibinfo {pages} {533} (\bibinfo {year} {2020})}\BibitemShut {NoStop}%
\bibitem [{\citenamefont {Niffenegger}\ \emph {et~al.}(2020)\citenamefont {Niffenegger}, \citenamefont {Stuart}, \citenamefont {{Sorace-Agaskar}}, \citenamefont {Kharas}, \citenamefont {Bramhavar}, \citenamefont {Bruzewicz}, \citenamefont {Loh}, \citenamefont {Maxson}, \citenamefont {McConnell}, \citenamefont {Reens}, \citenamefont {West}, \citenamefont {Sage},\ and\ \citenamefont {Chiaverini}}]{niffenegger2020}%
  \BibitemOpen
  \bibfield  {author} {\bibinfo {author} {\bibfnamefont {R.~J.}\ \bibnamefont {Niffenegger}}, \bibinfo {author} {\bibfnamefont {J.}~\bibnamefont {Stuart}}, \bibinfo {author} {\bibfnamefont {C.}~\bibnamefont {{Sorace-Agaskar}}}, \bibinfo {author} {\bibfnamefont {D.}~\bibnamefont {Kharas}}, \bibinfo {author} {\bibfnamefont {S.}~\bibnamefont {Bramhavar}}, \bibinfo {author} {\bibfnamefont {C.~D.}\ \bibnamefont {Bruzewicz}}, \bibinfo {author} {\bibfnamefont {W.}~\bibnamefont {Loh}}, \bibinfo {author} {\bibfnamefont {R.~T.}\ \bibnamefont {Maxson}}, \bibinfo {author} {\bibfnamefont {R.}~\bibnamefont {McConnell}}, \bibinfo {author} {\bibfnamefont {D.}~\bibnamefont {Reens}}, \bibinfo {author} {\bibfnamefont {G.~N.}\ \bibnamefont {West}}, \bibinfo {author} {\bibfnamefont {J.~M.}\ \bibnamefont {Sage}}, \ and\ \bibinfo {author} {\bibfnamefont {J.}~\bibnamefont {Chiaverini}},\ }\href {\doibase 10.1038/s41586-020-2811-x} {\bibfield  {journal} {\bibinfo  {journal} {Nature}\ }\textbf {\bibinfo {volume} {586}},\ \bibinfo
  {pages} {538} (\bibinfo {year} {2020})}\BibitemShut {NoStop}%
\bibitem [{\citenamefont {Zwerver}\ \emph {et~al.}(2022)\citenamefont {Zwerver}, \citenamefont {Kr{\"a}henmann}, \citenamefont {Watson}, \citenamefont {Lampert}, \citenamefont {George}, \citenamefont {Pillarisetty}, \citenamefont {Bojarski}, \citenamefont {Amin}, \citenamefont {Amitonov}, \citenamefont {Boter}, \citenamefont {Caudillo}, \citenamefont {{Correas-Serrano}}, \citenamefont {Dehollain}, \citenamefont {Droulers}, \citenamefont {Henry}, \citenamefont {Kotlyar}, \citenamefont {Lodari}, \citenamefont {L{\"u}thi}, \citenamefont {Michalak}, \citenamefont {Mueller}, \citenamefont {Neyens}, \citenamefont {Roberts}, \citenamefont {Samkharadze}, \citenamefont {Zheng}, \citenamefont {Zietz}, \citenamefont {Scappucci}, \citenamefont {Veldhorst}, \citenamefont {Vandersypen},\ and\ \citenamefont {Clarke}}]{zwerver2022}%
  \BibitemOpen
  \bibfield  {author} {\bibinfo {author} {\bibfnamefont {A.~M.~J.}\ \bibnamefont {Zwerver}}, \bibinfo {author} {\bibfnamefont {T.}~\bibnamefont {Kr{\"a}henmann}}, \bibinfo {author} {\bibfnamefont {T.~F.}\ \bibnamefont {Watson}}, \bibinfo {author} {\bibfnamefont {L.}~\bibnamefont {Lampert}}, \bibinfo {author} {\bibfnamefont {H.~C.}\ \bibnamefont {George}}, \bibinfo {author} {\bibfnamefont {R.}~\bibnamefont {Pillarisetty}}, \bibinfo {author} {\bibfnamefont {S.~A.}\ \bibnamefont {Bojarski}}, \bibinfo {author} {\bibfnamefont {P.}~\bibnamefont {Amin}}, \bibinfo {author} {\bibfnamefont {S.~V.}\ \bibnamefont {Amitonov}}, \bibinfo {author} {\bibfnamefont {J.~M.}\ \bibnamefont {Boter}}, \bibinfo {author} {\bibfnamefont {R.}~\bibnamefont {Caudillo}}, \bibinfo {author} {\bibfnamefont {D.}~\bibnamefont {{Correas-Serrano}}}, \bibinfo {author} {\bibfnamefont {J.~P.}\ \bibnamefont {Dehollain}}, \bibinfo {author} {\bibfnamefont {G.}~\bibnamefont {Droulers}}, \bibinfo {author} {\bibfnamefont {E.~M.}\ \bibnamefont {Henry}},
  \bibinfo {author} {\bibfnamefont {R.}~\bibnamefont {Kotlyar}}, \bibinfo {author} {\bibfnamefont {M.}~\bibnamefont {Lodari}}, \bibinfo {author} {\bibfnamefont {F.}~\bibnamefont {L{\"u}thi}}, \bibinfo {author} {\bibfnamefont {D.~J.}\ \bibnamefont {Michalak}}, \bibinfo {author} {\bibfnamefont {B.~K.}\ \bibnamefont {Mueller}}, \bibinfo {author} {\bibfnamefont {S.}~\bibnamefont {Neyens}}, \bibinfo {author} {\bibfnamefont {J.}~\bibnamefont {Roberts}}, \bibinfo {author} {\bibfnamefont {N.}~\bibnamefont {Samkharadze}}, \bibinfo {author} {\bibfnamefont {G.}~\bibnamefont {Zheng}}, \bibinfo {author} {\bibfnamefont {O.~K.}\ \bibnamefont {Zietz}}, \bibinfo {author} {\bibfnamefont {G.}~\bibnamefont {Scappucci}}, \bibinfo {author} {\bibfnamefont {M.}~\bibnamefont {Veldhorst}}, \bibinfo {author} {\bibfnamefont {L.~M.~K.}\ \bibnamefont {Vandersypen}}, \ and\ \bibinfo {author} {\bibfnamefont {J.~S.}\ \bibnamefont {Clarke}},\ }\href {\doibase 10.1038/s41928-022-00727-9} {\bibfield  {journal} {\bibinfo  {journal} {Nature
  Electronics}\ }\textbf {\bibinfo {volume} {5}},\ \bibinfo {pages} {184} (\bibinfo {year} {2022})}\BibitemShut {NoStop}%
\bibitem [{\citenamefont {Chiaverini}\ \emph {et~al.}(2005)\citenamefont {Chiaverini}, \citenamefont {Blakestad}, \citenamefont {Britton}, \citenamefont {Jost}, \citenamefont {Langer}, \citenamefont {Leibfried}, \citenamefont {Ozeri},\ and\ \citenamefont {Wineland}}]{chiaverini2005}%
  \BibitemOpen
  \bibfield  {author} {\bibinfo {author} {\bibfnamefont {J.}~\bibnamefont {Chiaverini}}, \bibinfo {author} {\bibfnamefont {R.~B.}\ \bibnamefont {Blakestad}}, \bibinfo {author} {\bibfnamefont {J.}~\bibnamefont {Britton}}, \bibinfo {author} {\bibfnamefont {J.~D.}\ \bibnamefont {Jost}}, \bibinfo {author} {\bibfnamefont {C.}~\bibnamefont {Langer}}, \bibinfo {author} {\bibfnamefont {D.}~\bibnamefont {Leibfried}}, \bibinfo {author} {\bibfnamefont {R.}~\bibnamefont {Ozeri}}, \ and\ \bibinfo {author} {\bibfnamefont {D.~J.}\ \bibnamefont {Wineland}},\ }\href {\doibase 10.48550/arXiv.quant-ph/0501147} {\enquote {\bibinfo {title} {Surface-{{Electrode Architecture}} for {{Ion-Trap Quantum Information Processing}}},}\ } (\bibinfo {year} {2005}),\ \Eprint {http://arxiv.org/abs/quant-ph/0501147} {arXiv:quant-ph/0501147} \BibitemShut {NoStop}%
\bibitem [{\citenamefont {Sepiol}\ \emph {et~al.}(2019)\citenamefont {Sepiol}, \citenamefont {Hughes}, \citenamefont {Tarlton}, \citenamefont {Nadlinger}, \citenamefont {Ballance}, \citenamefont {Ballance}, \citenamefont {Harty}, \citenamefont {Steane}, \citenamefont {Goodwin},\ and\ \citenamefont {Lucas}}]{sepiol2019}%
  \BibitemOpen
  \bibfield  {author} {\bibinfo {author} {\bibfnamefont {M.~A.}\ \bibnamefont {Sepiol}}, \bibinfo {author} {\bibfnamefont {A.~C.}\ \bibnamefont {Hughes}}, \bibinfo {author} {\bibfnamefont {J.~E.}\ \bibnamefont {Tarlton}}, \bibinfo {author} {\bibfnamefont {D.~P.}\ \bibnamefont {Nadlinger}}, \bibinfo {author} {\bibfnamefont {T.~G.}\ \bibnamefont {Ballance}}, \bibinfo {author} {\bibfnamefont {C.~J.}\ \bibnamefont {Ballance}}, \bibinfo {author} {\bibfnamefont {T.~P.}\ \bibnamefont {Harty}}, \bibinfo {author} {\bibfnamefont {A.~M.}\ \bibnamefont {Steane}}, \bibinfo {author} {\bibfnamefont {J.~F.}\ \bibnamefont {Goodwin}}, \ and\ \bibinfo {author} {\bibfnamefont {D.~M.}\ \bibnamefont {Lucas}},\ }\href {\doibase 10.1103/PhysRevLett.123.110503} {\bibfield  {journal} {\bibinfo  {journal} {Physical Review Letters}\ }\textbf {\bibinfo {volume} {123}},\ \bibinfo {pages} {110503} (\bibinfo {year} {2019})}\BibitemShut {NoStop}%
\bibitem [{\citenamefont {Moses}\ \emph {et~al.}(2023)\citenamefont {Moses}, \citenamefont {Baldwin}, \citenamefont {Allman}, \citenamefont {Ancona}, \citenamefont {Ascarrunz}, \citenamefont {Barnes}, \citenamefont {Bartolotta}, \citenamefont {Bjork}, \citenamefont {Blanchard}, \citenamefont {Bohn}, \citenamefont {Bohnet}, \citenamefont {Brown}, \citenamefont {Burdick}, \citenamefont {Burton}, \citenamefont {Campbell}, \citenamefont {Campora}, \citenamefont {Carron}, \citenamefont {Chambers}, \citenamefont {Chan}, \citenamefont {Chen}, \citenamefont {Chernoguzov}, \citenamefont {Chertkov}, \citenamefont {Colina}, \citenamefont {Curtis}, \citenamefont {Daniel}, \citenamefont {DeCross}, \citenamefont {Deen}, \citenamefont {Delaney}, \citenamefont {Dreiling}, \citenamefont {Ertsgaard}, \citenamefont {Esposito}, \citenamefont {Estey}, \citenamefont {Fabrikant}, \citenamefont {Figgatt}, \citenamefont {Foltz}, \citenamefont {{Foss-Feig}}, \citenamefont {Francois}, \citenamefont {Gaebler}, \citenamefont {Gatterman},
  \citenamefont {Gilbreth}, \citenamefont {Giles}, \citenamefont {Glynn}, \citenamefont {Hall}, \citenamefont {Hankin}, \citenamefont {Hansen}, \citenamefont {Hayes}, \citenamefont {Higashi}, \citenamefont {Hoffman}, \citenamefont {Horning}, \citenamefont {Hout}, \citenamefont {Jacobs}, \citenamefont {Johansen}, \citenamefont {Jones}, \citenamefont {Karcz}, \citenamefont {Klein}, \citenamefont {Lauria}, \citenamefont {Lee}, \citenamefont {Liefer}, \citenamefont {Lu}, \citenamefont {Lucchetti}, \citenamefont {Lytle}, \citenamefont {Malm}, \citenamefont {Matheny}, \citenamefont {Mathewson}, \citenamefont {Mayer}, \citenamefont {Miller}, \citenamefont {Mills}, \citenamefont {Neyenhuis}, \citenamefont {Nugent}, \citenamefont {Olson}, \citenamefont {Parks}, \citenamefont {Price}, \citenamefont {Price}, \citenamefont {Pugh}, \citenamefont {Ransford}, \citenamefont {Reed}, \citenamefont {Roman}, \citenamefont {Rowe}, \citenamefont {{Ryan-Anderson}}, \citenamefont {Sanders}, \citenamefont {Sedlacek}, \citenamefont
  {Shevchuk}, \citenamefont {Siegfried}, \citenamefont {Skripka}, \citenamefont {Spaun}, \citenamefont {Sprenkle}, \citenamefont {Stutz}, \citenamefont {Swallows}, \citenamefont {Tobey}, \citenamefont {Tran}, \citenamefont {Tran}, \citenamefont {Vogt}, \citenamefont {Volin}, \citenamefont {Walker}, \citenamefont {Zolot},\ and\ \citenamefont {Pino}}]{moses2023}%
  \BibitemOpen
  \bibfield  {author} {\bibinfo {author} {\bibfnamefont {S.~A.}\ \bibnamefont {Moses}}, \bibinfo {author} {\bibfnamefont {C.~H.}\ \bibnamefont {Baldwin}}, \bibinfo {author} {\bibfnamefont {M.~S.}\ \bibnamefont {Allman}}, \bibinfo {author} {\bibfnamefont {R.}~\bibnamefont {Ancona}}, \bibinfo {author} {\bibfnamefont {L.}~\bibnamefont {Ascarrunz}}, \bibinfo {author} {\bibfnamefont {C.}~\bibnamefont {Barnes}}, \bibinfo {author} {\bibfnamefont {J.}~\bibnamefont {Bartolotta}}, \bibinfo {author} {\bibfnamefont {B.}~\bibnamefont {Bjork}}, \bibinfo {author} {\bibfnamefont {P.}~\bibnamefont {Blanchard}}, \bibinfo {author} {\bibfnamefont {M.}~\bibnamefont {Bohn}}, \bibinfo {author} {\bibfnamefont {J.~G.}\ \bibnamefont {Bohnet}}, \bibinfo {author} {\bibfnamefont {N.~C.}\ \bibnamefont {Brown}}, \bibinfo {author} {\bibfnamefont {N.~Q.}\ \bibnamefont {Burdick}}, \bibinfo {author} {\bibfnamefont {W.~C.}\ \bibnamefont {Burton}}, \bibinfo {author} {\bibfnamefont {S.~L.}\ \bibnamefont {Campbell}}, \bibinfo {author}
  {\bibfnamefont {J.~P.}\ \bibnamefont {Campora}}, \bibinfo {author} {\bibfnamefont {C.}~\bibnamefont {Carron}}, \bibinfo {author} {\bibfnamefont {J.}~\bibnamefont {Chambers}}, \bibinfo {author} {\bibfnamefont {J.~W.}\ \bibnamefont {Chan}}, \bibinfo {author} {\bibfnamefont {Y.~H.}\ \bibnamefont {Chen}}, \bibinfo {author} {\bibfnamefont {A.}~\bibnamefont {Chernoguzov}}, \bibinfo {author} {\bibfnamefont {E.}~\bibnamefont {Chertkov}}, \bibinfo {author} {\bibfnamefont {J.}~\bibnamefont {Colina}}, \bibinfo {author} {\bibfnamefont {J.~P.}\ \bibnamefont {Curtis}}, \bibinfo {author} {\bibfnamefont {R.}~\bibnamefont {Daniel}}, \bibinfo {author} {\bibfnamefont {M.}~\bibnamefont {DeCross}}, \bibinfo {author} {\bibfnamefont {D.}~\bibnamefont {Deen}}, \bibinfo {author} {\bibfnamefont {C.}~\bibnamefont {Delaney}}, \bibinfo {author} {\bibfnamefont {J.~M.}\ \bibnamefont {Dreiling}}, \bibinfo {author} {\bibfnamefont {C.~T.}\ \bibnamefont {Ertsgaard}}, \bibinfo {author} {\bibfnamefont {J.}~\bibnamefont {Esposito}}, \bibinfo
  {author} {\bibfnamefont {B.}~\bibnamefont {Estey}}, \bibinfo {author} {\bibfnamefont {M.}~\bibnamefont {Fabrikant}}, \bibinfo {author} {\bibfnamefont {C.}~\bibnamefont {Figgatt}}, \bibinfo {author} {\bibfnamefont {C.}~\bibnamefont {Foltz}}, \bibinfo {author} {\bibfnamefont {M.}~\bibnamefont {{Foss-Feig}}}, \bibinfo {author} {\bibfnamefont {D.}~\bibnamefont {Francois}}, \bibinfo {author} {\bibfnamefont {J.~P.}\ \bibnamefont {Gaebler}}, \bibinfo {author} {\bibfnamefont {T.~M.}\ \bibnamefont {Gatterman}}, \bibinfo {author} {\bibfnamefont {C.~N.}\ \bibnamefont {Gilbreth}}, \bibinfo {author} {\bibfnamefont {J.}~\bibnamefont {Giles}}, \bibinfo {author} {\bibfnamefont {E.}~\bibnamefont {Glynn}}, \bibinfo {author} {\bibfnamefont {A.}~\bibnamefont {Hall}}, \bibinfo {author} {\bibfnamefont {A.~M.}\ \bibnamefont {Hankin}}, \bibinfo {author} {\bibfnamefont {A.}~\bibnamefont {Hansen}}, \bibinfo {author} {\bibfnamefont {D.}~\bibnamefont {Hayes}}, \bibinfo {author} {\bibfnamefont {B.}~\bibnamefont {Higashi}}, \bibinfo
  {author} {\bibfnamefont {I.~M.}\ \bibnamefont {Hoffman}}, \bibinfo {author} {\bibfnamefont {B.}~\bibnamefont {Horning}}, \bibinfo {author} {\bibfnamefont {J.~J.}\ \bibnamefont {Hout}}, \bibinfo {author} {\bibfnamefont {R.}~\bibnamefont {Jacobs}}, \bibinfo {author} {\bibfnamefont {J.}~\bibnamefont {Johansen}}, \bibinfo {author} {\bibfnamefont {L.}~\bibnamefont {Jones}}, \bibinfo {author} {\bibfnamefont {J.}~\bibnamefont {Karcz}}, \bibinfo {author} {\bibfnamefont {T.}~\bibnamefont {Klein}}, \bibinfo {author} {\bibfnamefont {P.}~\bibnamefont {Lauria}}, \bibinfo {author} {\bibfnamefont {P.}~\bibnamefont {Lee}}, \bibinfo {author} {\bibfnamefont {D.}~\bibnamefont {Liefer}}, \bibinfo {author} {\bibfnamefont {S.~T.}\ \bibnamefont {Lu}}, \bibinfo {author} {\bibfnamefont {D.}~\bibnamefont {Lucchetti}}, \bibinfo {author} {\bibfnamefont {C.}~\bibnamefont {Lytle}}, \bibinfo {author} {\bibfnamefont {A.}~\bibnamefont {Malm}}, \bibinfo {author} {\bibfnamefont {M.}~\bibnamefont {Matheny}}, \bibinfo {author} {\bibfnamefont
  {B.}~\bibnamefont {Mathewson}}, \bibinfo {author} {\bibfnamefont {K.}~\bibnamefont {Mayer}}, \bibinfo {author} {\bibfnamefont {D.~B.}\ \bibnamefont {Miller}}, \bibinfo {author} {\bibfnamefont {M.}~\bibnamefont {Mills}}, \bibinfo {author} {\bibfnamefont {B.}~\bibnamefont {Neyenhuis}}, \bibinfo {author} {\bibfnamefont {L.}~\bibnamefont {Nugent}}, \bibinfo {author} {\bibfnamefont {S.}~\bibnamefont {Olson}}, \bibinfo {author} {\bibfnamefont {J.}~\bibnamefont {Parks}}, \bibinfo {author} {\bibfnamefont {G.~N.}\ \bibnamefont {Price}}, \bibinfo {author} {\bibfnamefont {Z.}~\bibnamefont {Price}}, \bibinfo {author} {\bibfnamefont {M.}~\bibnamefont {Pugh}}, \bibinfo {author} {\bibfnamefont {A.}~\bibnamefont {Ransford}}, \bibinfo {author} {\bibfnamefont {A.~P.}\ \bibnamefont {Reed}}, \bibinfo {author} {\bibfnamefont {C.}~\bibnamefont {Roman}}, \bibinfo {author} {\bibfnamefont {M.}~\bibnamefont {Rowe}}, \bibinfo {author} {\bibfnamefont {C.}~\bibnamefont {{Ryan-Anderson}}}, \bibinfo {author} {\bibfnamefont
  {S.}~\bibnamefont {Sanders}}, \bibinfo {author} {\bibfnamefont {J.}~\bibnamefont {Sedlacek}}, \bibinfo {author} {\bibfnamefont {P.}~\bibnamefont {Shevchuk}}, \bibinfo {author} {\bibfnamefont {P.}~\bibnamefont {Siegfried}}, \bibinfo {author} {\bibfnamefont {T.}~\bibnamefont {Skripka}}, \bibinfo {author} {\bibfnamefont {B.}~\bibnamefont {Spaun}}, \bibinfo {author} {\bibfnamefont {R.~T.}\ \bibnamefont {Sprenkle}}, \bibinfo {author} {\bibfnamefont {R.~P.}\ \bibnamefont {Stutz}}, \bibinfo {author} {\bibfnamefont {M.}~\bibnamefont {Swallows}}, \bibinfo {author} {\bibfnamefont {R.~I.}\ \bibnamefont {Tobey}}, \bibinfo {author} {\bibfnamefont {A.}~\bibnamefont {Tran}}, \bibinfo {author} {\bibfnamefont {T.}~\bibnamefont {Tran}}, \bibinfo {author} {\bibfnamefont {E.}~\bibnamefont {Vogt}}, \bibinfo {author} {\bibfnamefont {C.}~\bibnamefont {Volin}}, \bibinfo {author} {\bibfnamefont {J.}~\bibnamefont {Walker}}, \bibinfo {author} {\bibfnamefont {A.~M.}\ \bibnamefont {Zolot}}, \ and\ \bibinfo {author} {\bibfnamefont
  {J.~M.}\ \bibnamefont {Pino}},\ }\href {\doibase 10.1103/PhysRevX.13.041052} {\bibfield  {journal} {\bibinfo  {journal} {Physical Review X}\ }\textbf {\bibinfo {volume} {13}},\ \bibinfo {pages} {041052} (\bibinfo {year} {2023})}\BibitemShut {NoStop}%
\bibitem [{\citenamefont {Mehta}\ \emph {et~al.}(2016)\citenamefont {Mehta}, \citenamefont {Bruzewicz}, \citenamefont {McConnell}, \citenamefont {Ram}, \citenamefont {Sage},\ and\ \citenamefont {Chiaverini}}]{mehta2016}%
  \BibitemOpen
  \bibfield  {author} {\bibinfo {author} {\bibfnamefont {K.~K.}\ \bibnamefont {Mehta}}, \bibinfo {author} {\bibfnamefont {C.~D.}\ \bibnamefont {Bruzewicz}}, \bibinfo {author} {\bibfnamefont {R.}~\bibnamefont {McConnell}}, \bibinfo {author} {\bibfnamefont {R.~J.}\ \bibnamefont {Ram}}, \bibinfo {author} {\bibfnamefont {J.~M.}\ \bibnamefont {Sage}}, \ and\ \bibinfo {author} {\bibfnamefont {J.}~\bibnamefont {Chiaverini}},\ }\href {\doibase 10.1038/nnano.2016.139} {\bibfield  {journal} {\bibinfo  {journal} {Nature Nanotechnology}\ }\textbf {\bibinfo {volume} {11}},\ \bibinfo {pages} {1066} (\bibinfo {year} {2016})}\BibitemShut {NoStop}%
\bibitem [{\citenamefont {Mordini}\ \emph {et~al.}(2024)\citenamefont {Mordini}, \citenamefont {Vasquez}, \citenamefont {Motohashi}, \citenamefont {M{\"u}ller}, \citenamefont {Malinowski}, \citenamefont {Zhang}, \citenamefont {Mehta}, \citenamefont {Kienzler},\ and\ \citenamefont {Home}}]{mordini2024}%
  \BibitemOpen
  \bibfield  {author} {\bibinfo {author} {\bibfnamefont {C.}~\bibnamefont {Mordini}}, \bibinfo {author} {\bibfnamefont {A.~R.}\ \bibnamefont {Vasquez}}, \bibinfo {author} {\bibfnamefont {Y.}~\bibnamefont {Motohashi}}, \bibinfo {author} {\bibfnamefont {M.}~\bibnamefont {M{\"u}ller}}, \bibinfo {author} {\bibfnamefont {M.}~\bibnamefont {Malinowski}}, \bibinfo {author} {\bibfnamefont {C.}~\bibnamefont {Zhang}}, \bibinfo {author} {\bibfnamefont {K.~K.}\ \bibnamefont {Mehta}}, \bibinfo {author} {\bibfnamefont {D.}~\bibnamefont {Kienzler}}, \ and\ \bibinfo {author} {\bibfnamefont {J.~P.}\ \bibnamefont {Home}},\ }\href {\doibase 10.48550/arXiv.2401.18056} {\enquote {\bibinfo {title} {Multi-zone trapped-ion qubit control in an integrated photonics {{QCCD}} device},}\ } (\bibinfo {year} {2024}),\ \Eprint {http://arxiv.org/abs/2401.18056} {arXiv:2401.18056 [physics, physics:quant-ph]} \BibitemShut {NoStop}%
\bibitem [{\citenamefont {Kwon}\ \emph {et~al.}(2024)\citenamefont {Kwon}, \citenamefont {Setzer}, \citenamefont {Gehl}, \citenamefont {Karl}, \citenamefont {Van Der~Wall}, \citenamefont {Law}, \citenamefont {Blain}, \citenamefont {Stick},\ and\ \citenamefont {McGuinness}}]{kwon2024}%
  \BibitemOpen
  \bibfield  {author} {\bibinfo {author} {\bibfnamefont {J.}~\bibnamefont {Kwon}}, \bibinfo {author} {\bibfnamefont {W.~J.}\ \bibnamefont {Setzer}}, \bibinfo {author} {\bibfnamefont {M.}~\bibnamefont {Gehl}}, \bibinfo {author} {\bibfnamefont {N.}~\bibnamefont {Karl}}, \bibinfo {author} {\bibfnamefont {J.}~\bibnamefont {Van Der~Wall}}, \bibinfo {author} {\bibfnamefont {R.}~\bibnamefont {Law}}, \bibinfo {author} {\bibfnamefont {M.~G.}\ \bibnamefont {Blain}}, \bibinfo {author} {\bibfnamefont {D.}~\bibnamefont {Stick}}, \ and\ \bibinfo {author} {\bibfnamefont {H.~J.}\ \bibnamefont {McGuinness}},\ }\href {\doibase 10.48550/arXiv.2308.14918} {\enquote {\bibinfo {title} {Multi-site {{Integrated Optical Addressing}} of {{Trapped Ions}}},}\ } (\bibinfo {year} {2024}),\ \Eprint {http://arxiv.org/abs/2308.14918} {arXiv:2308.14918 [quant-ph]} \BibitemShut {NoStop}%
\bibitem [{\citenamefont {Hogle}\ \emph {et~al.}(2023)\citenamefont {Hogle}, \citenamefont {Dominguez}, \citenamefont {Dong}, \citenamefont {Leenheer}, \citenamefont {McGuinness}, \citenamefont {Ruzic}, \citenamefont {Eichenfield},\ and\ \citenamefont {Stick}}]{hogle2023}%
  \BibitemOpen
  \bibfield  {author} {\bibinfo {author} {\bibfnamefont {C.~W.}\ \bibnamefont {Hogle}}, \bibinfo {author} {\bibfnamefont {D.}~\bibnamefont {Dominguez}}, \bibinfo {author} {\bibfnamefont {M.}~\bibnamefont {Dong}}, \bibinfo {author} {\bibfnamefont {A.}~\bibnamefont {Leenheer}}, \bibinfo {author} {\bibfnamefont {H.~J.}\ \bibnamefont {McGuinness}}, \bibinfo {author} {\bibfnamefont {B.~P.}\ \bibnamefont {Ruzic}}, \bibinfo {author} {\bibfnamefont {M.}~\bibnamefont {Eichenfield}}, \ and\ \bibinfo {author} {\bibfnamefont {D.}~\bibnamefont {Stick}},\ }\href {\doibase 10.1038/s41534-023-00737-1} {\bibfield  {journal} {\bibinfo  {journal} {npj Quantum Information}\ }\textbf {\bibinfo {volume} {9}} (\bibinfo {year} {2023}),\ 10.1038/s41534-023-00737-1}\BibitemShut {NoStop}%
\bibitem [{\citenamefont {Ospelkaus}\ \emph {et~al.}(2008)\citenamefont {Ospelkaus}, \citenamefont {Langer}, \citenamefont {Amini}, \citenamefont {Brown}, \citenamefont {Leibfried},\ and\ \citenamefont {Wineland}}]{ospelkaus2008}%
  \BibitemOpen
  \bibfield  {author} {\bibinfo {author} {\bibfnamefont {C.}~\bibnamefont {Ospelkaus}}, \bibinfo {author} {\bibfnamefont {C.~E.}\ \bibnamefont {Langer}}, \bibinfo {author} {\bibfnamefont {J.~M.}\ \bibnamefont {Amini}}, \bibinfo {author} {\bibfnamefont {K.~R.}\ \bibnamefont {Brown}}, \bibinfo {author} {\bibfnamefont {D.}~\bibnamefont {Leibfried}}, \ and\ \bibinfo {author} {\bibfnamefont {D.~J.}\ \bibnamefont {Wineland}},\ }\href {\doibase 10.1103/PhysRevLett.101.090502} {\bibfield  {journal} {\bibinfo  {journal} {Physical Review Letters}\ }\textbf {\bibinfo {volume} {101}},\ \bibinfo {pages} {090502} (\bibinfo {year} {2008})}\BibitemShut {NoStop}%
\bibitem [{\citenamefont {Srinivas}\ \emph {et~al.}(2021)\citenamefont {Srinivas}, \citenamefont {Burd}, \citenamefont {Knaack}, \citenamefont {Sutherland}, \citenamefont {Kwiatkowski}, \citenamefont {Glancy}, \citenamefont {Knill}, \citenamefont {Wineland}, \citenamefont {Leibfried}, \citenamefont {Wilson}, \citenamefont {Allcock},\ and\ \citenamefont {Slichter}}]{srinivas2021}%
  \BibitemOpen
  \bibfield  {author} {\bibinfo {author} {\bibfnamefont {R.}~\bibnamefont {Srinivas}}, \bibinfo {author} {\bibfnamefont {S.~C.}\ \bibnamefont {Burd}}, \bibinfo {author} {\bibfnamefont {H.~M.}\ \bibnamefont {Knaack}}, \bibinfo {author} {\bibfnamefont {R.~T.}\ \bibnamefont {Sutherland}}, \bibinfo {author} {\bibfnamefont {A.}~\bibnamefont {Kwiatkowski}}, \bibinfo {author} {\bibfnamefont {S.}~\bibnamefont {Glancy}}, \bibinfo {author} {\bibfnamefont {E.}~\bibnamefont {Knill}}, \bibinfo {author} {\bibfnamefont {D.~J.}\ \bibnamefont {Wineland}}, \bibinfo {author} {\bibfnamefont {D.}~\bibnamefont {Leibfried}}, \bibinfo {author} {\bibfnamefont {A.~C.}\ \bibnamefont {Wilson}}, \bibinfo {author} {\bibfnamefont {D.~T.~C.}\ \bibnamefont {Allcock}}, \ and\ \bibinfo {author} {\bibfnamefont {D.~H.}\ \bibnamefont {Slichter}},\ }\href {\doibase 10.1038/s41586-021-03809-4} {\bibfield  {journal} {\bibinfo  {journal} {Nature}\ }\textbf {\bibinfo {volume} {597}},\ \bibinfo {pages} {209} (\bibinfo {year} {2021})}\BibitemShut
  {NoStop}%
\bibitem [{\citenamefont {Weber}\ \emph {et~al.}(2024)\citenamefont {Weber}, \citenamefont {Gely}, \citenamefont {Hanley}, \citenamefont {Harty}, \citenamefont {Leu}, \citenamefont {L{\"o}schnauer}, \citenamefont {Nadlinger},\ and\ \citenamefont {Lucas}}]{weber2024}%
  \BibitemOpen
  \bibfield  {author} {\bibinfo {author} {\bibfnamefont {M.~A.}\ \bibnamefont {Weber}}, \bibinfo {author} {\bibfnamefont {M.~F.}\ \bibnamefont {Gely}}, \bibinfo {author} {\bibfnamefont {R.~K.}\ \bibnamefont {Hanley}}, \bibinfo {author} {\bibfnamefont {T.~P.}\ \bibnamefont {Harty}}, \bibinfo {author} {\bibfnamefont {A.~D.}\ \bibnamefont {Leu}}, \bibinfo {author} {\bibfnamefont {C.~M.}\ \bibnamefont {L{\"o}schnauer}}, \bibinfo {author} {\bibfnamefont {D.~P.}\ \bibnamefont {Nadlinger}}, \ and\ \bibinfo {author} {\bibfnamefont {D.~M.}\ \bibnamefont {Lucas}},\ }\href {\doibase 10.48550/arXiv.2402.12955} {\enquote {\bibinfo {title} {Robust and fast microwave-driven quantum logic for trapped-ion qubits},}\ } (\bibinfo {year} {2024}),\ \Eprint {http://arxiv.org/abs/2402.12955} {arXiv:2402.12955 [quant-ph]} \BibitemShut {NoStop}%
\bibitem [{\citenamefont {Leibfried}\ \emph {et~al.}(2007)\citenamefont {Leibfried}, \citenamefont {Knill}, \citenamefont {Ospelkaus},\ and\ \citenamefont {Wineland}}]{leibfried2007}%
  \BibitemOpen
  \bibfield  {author} {\bibinfo {author} {\bibfnamefont {D.}~\bibnamefont {Leibfried}}, \bibinfo {author} {\bibfnamefont {E.}~\bibnamefont {Knill}}, \bibinfo {author} {\bibfnamefont {C.}~\bibnamefont {Ospelkaus}}, \ and\ \bibinfo {author} {\bibfnamefont {D.~J.}\ \bibnamefont {Wineland}},\ }\href {\doibase 10.1103/PhysRevA.76.032324} {\bibfield  {journal} {\bibinfo  {journal} {Physical Review A}\ }\textbf {\bibinfo {volume} {76}},\ \bibinfo {pages} {032324} (\bibinfo {year} {2007})}\BibitemShut {NoStop}%
\bibitem [{\citenamefont {Warring}\ \emph {et~al.}(2013{\natexlab{a}})\citenamefont {Warring}, \citenamefont {Ospelkaus}, \citenamefont {Colombe}, \citenamefont {J{\"o}rdens}, \citenamefont {Leibfried},\ and\ \citenamefont {Wineland}}]{warring2013}%
  \BibitemOpen
  \bibfield  {author} {\bibinfo {author} {\bibfnamefont {U.}~\bibnamefont {Warring}}, \bibinfo {author} {\bibfnamefont {C.}~\bibnamefont {Ospelkaus}}, \bibinfo {author} {\bibfnamefont {Y.}~\bibnamefont {Colombe}}, \bibinfo {author} {\bibfnamefont {R.}~\bibnamefont {J{\"o}rdens}}, \bibinfo {author} {\bibfnamefont {D.}~\bibnamefont {Leibfried}}, \ and\ \bibinfo {author} {\bibfnamefont {D.~J.}\ \bibnamefont {Wineland}},\ }\href {\doibase 10.1103/PhysRevLett.110.173002} {\bibfield  {journal} {\bibinfo  {journal} {Physical Review Letters}\ }\textbf {\bibinfo {volume} {110}},\ \bibinfo {pages} {173002} (\bibinfo {year} {2013}{\natexlab{a}})}\BibitemShut {NoStop}%
\bibitem [{\citenamefont {Lekitsch}\ \emph {et~al.}(2017)\citenamefont {Lekitsch}, \citenamefont {Weidt}, \citenamefont {Fowler}, \citenamefont {M{\o}lmer}, \citenamefont {Devitt}, \citenamefont {Wunderlich},\ and\ \citenamefont {Hensinger}}]{lekitsch2017}%
  \BibitemOpen
  \bibfield  {author} {\bibinfo {author} {\bibfnamefont {B.}~\bibnamefont {Lekitsch}}, \bibinfo {author} {\bibfnamefont {S.}~\bibnamefont {Weidt}}, \bibinfo {author} {\bibfnamefont {A.~G.}\ \bibnamefont {Fowler}}, \bibinfo {author} {\bibfnamefont {K.}~\bibnamefont {M{\o}lmer}}, \bibinfo {author} {\bibfnamefont {S.~J.}\ \bibnamefont {Devitt}}, \bibinfo {author} {\bibfnamefont {C.}~\bibnamefont {Wunderlich}}, \ and\ \bibinfo {author} {\bibfnamefont {W.~K.}\ \bibnamefont {Hensinger}},\ }\href {\doibase 10.1126/sciadv.1601540} {\bibfield  {journal} {\bibinfo  {journal} {Science Advances}\ }\textbf {\bibinfo {volume} {3}},\ \bibinfo {pages} {e1601540} (\bibinfo {year} {2017})}\BibitemShut {NoStop}%
\bibitem [{\citenamefont {Seck}\ \emph {et~al.}(2020)\citenamefont {Seck}, \citenamefont {Meier}, \citenamefont {Merrill}, \citenamefont {Hayden}, \citenamefont {Sawyer}, \citenamefont {Volin},\ and\ \citenamefont {Brown}}]{seck2020}%
  \BibitemOpen
  \bibfield  {author} {\bibinfo {author} {\bibfnamefont {C.~M.}\ \bibnamefont {Seck}}, \bibinfo {author} {\bibfnamefont {A.~M.}\ \bibnamefont {Meier}}, \bibinfo {author} {\bibfnamefont {J.~T.}\ \bibnamefont {Merrill}}, \bibinfo {author} {\bibfnamefont {H.~T.}\ \bibnamefont {Hayden}}, \bibinfo {author} {\bibfnamefont {B.~C.}\ \bibnamefont {Sawyer}}, \bibinfo {author} {\bibfnamefont {C.~E.}\ \bibnamefont {Volin}}, \ and\ \bibinfo {author} {\bibfnamefont {K.~R.}\ \bibnamefont {Brown}},\ }\href {\doibase 10.1088/1367-2630/ab8046} {\bibfield  {journal} {\bibinfo  {journal} {New Journal of Physics}\ }\textbf {\bibinfo {volume} {22}},\ \bibinfo {pages} {053024} (\bibinfo {year} {2020})}\BibitemShut {NoStop}%
\bibitem [{\citenamefont {Srinivas}\ \emph {et~al.}(2023)\citenamefont {Srinivas}, \citenamefont {L{\"o}schnauer}, \citenamefont {Malinowski}, \citenamefont {Hughes}, \citenamefont {Nourshargh}, \citenamefont {Negnevitsky}, \citenamefont {Allcock}, \citenamefont {King}, \citenamefont {Matthiesen}, \citenamefont {Harty},\ and\ \citenamefont {Ballance}}]{srinivas2023}%
  \BibitemOpen
  \bibfield  {author} {\bibinfo {author} {\bibfnamefont {R.}~\bibnamefont {Srinivas}}, \bibinfo {author} {\bibfnamefont {C.~M.}\ \bibnamefont {L{\"o}schnauer}}, \bibinfo {author} {\bibfnamefont {M.}~\bibnamefont {Malinowski}}, \bibinfo {author} {\bibfnamefont {A.~C.}\ \bibnamefont {Hughes}}, \bibinfo {author} {\bibfnamefont {R.}~\bibnamefont {Nourshargh}}, \bibinfo {author} {\bibfnamefont {V.}~\bibnamefont {Negnevitsky}}, \bibinfo {author} {\bibfnamefont {D.~T.~C.}\ \bibnamefont {Allcock}}, \bibinfo {author} {\bibfnamefont {S.~A.}\ \bibnamefont {King}}, \bibinfo {author} {\bibfnamefont {C.}~\bibnamefont {Matthiesen}}, \bibinfo {author} {\bibfnamefont {T.~P.}\ \bibnamefont {Harty}}, \ and\ \bibinfo {author} {\bibfnamefont {C.~J.}\ \bibnamefont {Ballance}},\ }\href {\doibase 10.1103/PhysRevLett.131.020601} {\bibfield  {journal} {\bibinfo  {journal} {Physical Review Letters}\ }\textbf {\bibinfo {volume} {131}},\ \bibinfo {pages} {020601} (\bibinfo {year} {2023})}\BibitemShut {NoStop}%
\bibitem [{\citenamefont {Sutherland}\ \emph {et~al.}(2023)\citenamefont {Sutherland}, \citenamefont {Srinivas},\ and\ \citenamefont {Allcock}}]{sutherland2023}%
  \BibitemOpen
  \bibfield  {author} {\bibinfo {author} {\bibfnamefont {R.~T.}\ \bibnamefont {Sutherland}}, \bibinfo {author} {\bibfnamefont {R.}~\bibnamefont {Srinivas}}, \ and\ \bibinfo {author} {\bibfnamefont {D.~T.~C.}\ \bibnamefont {Allcock}},\ }\href {\doibase 10.1103/PhysRevA.107.032604} {\bibfield  {journal} {\bibinfo  {journal} {Physical Review A}\ }\textbf {\bibinfo {volume} {107}},\ \bibinfo {pages} {032604} (\bibinfo {year} {2023})}\BibitemShut {NoStop}%
\bibitem [{\citenamefont {Warring}\ \emph {et~al.}(2013{\natexlab{b}})\citenamefont {Warring}, \citenamefont {Ospelkaus}, \citenamefont {Colombe}, \citenamefont {Brown}, \citenamefont {Amini}, \citenamefont {Carsjens}, \citenamefont {Leibfried},\ and\ \citenamefont {Wineland}}]{warring2013a}%
  \BibitemOpen
  \bibfield  {author} {\bibinfo {author} {\bibfnamefont {U.}~\bibnamefont {Warring}}, \bibinfo {author} {\bibfnamefont {C.}~\bibnamefont {Ospelkaus}}, \bibinfo {author} {\bibfnamefont {Y.}~\bibnamefont {Colombe}}, \bibinfo {author} {\bibfnamefont {K.~R.}\ \bibnamefont {Brown}}, \bibinfo {author} {\bibfnamefont {J.~M.}\ \bibnamefont {Amini}}, \bibinfo {author} {\bibfnamefont {M.}~\bibnamefont {Carsjens}}, \bibinfo {author} {\bibfnamefont {D.}~\bibnamefont {Leibfried}}, \ and\ \bibinfo {author} {\bibfnamefont {D.~J.}\ \bibnamefont {Wineland}},\ }\href {\doibase 10.1103/PhysRevA.87.013437} {\bibfield  {journal} {\bibinfo  {journal} {Physical Review A}\ }\textbf {\bibinfo {volume} {87}},\ \bibinfo {pages} {013437} (\bibinfo {year} {2013}{\natexlab{b}})}\BibitemShut {NoStop}%
\bibitem [{\citenamefont {Sutherland}\ \emph {et~al.}(2019)\citenamefont {Sutherland}, \citenamefont {Srinivas}, \citenamefont {Burd}, \citenamefont {Leibfried}, \citenamefont {Wilson}, \citenamefont {Wineland}, \citenamefont {Allcock}, \citenamefont {Slichter},\ and\ \citenamefont {Libby}}]{sutherland2019}%
  \BibitemOpen
  \bibfield  {author} {\bibinfo {author} {\bibfnamefont {R.~T.}\ \bibnamefont {Sutherland}}, \bibinfo {author} {\bibfnamefont {R.}~\bibnamefont {Srinivas}}, \bibinfo {author} {\bibfnamefont {S.~C.}\ \bibnamefont {Burd}}, \bibinfo {author} {\bibfnamefont {D.}~\bibnamefont {Leibfried}}, \bibinfo {author} {\bibfnamefont {A.~C.}\ \bibnamefont {Wilson}}, \bibinfo {author} {\bibfnamefont {D.~J.}\ \bibnamefont {Wineland}}, \bibinfo {author} {\bibfnamefont {D.~T.~C.}\ \bibnamefont {Allcock}}, \bibinfo {author} {\bibfnamefont {D.~H.}\ \bibnamefont {Slichter}}, \ and\ \bibinfo {author} {\bibfnamefont {S.~B.}\ \bibnamefont {Libby}},\ }\href {\doibase 10.1088/1367-2630/ab0be5} {\bibfield  {journal} {\bibinfo  {journal} {New Journal of Physics}\ }\textbf {\bibinfo {volume} {21}},\ \bibinfo {pages} {033033} (\bibinfo {year} {2019})}\BibitemShut {NoStop}%
\bibitem [{\citenamefont {Enthoven}\ \emph {et~al.}(2024)\citenamefont {Enthoven}, \citenamefont {Babaie},\ and\ \citenamefont {Sebastiano}}]{enthoven2024}%
  \BibitemOpen
  \bibfield  {author} {\bibinfo {author} {\bibfnamefont {L.}~\bibnamefont {Enthoven}}, \bibinfo {author} {\bibfnamefont {M.}~\bibnamefont {Babaie}}, \ and\ \bibinfo {author} {\bibfnamefont {F.}~\bibnamefont {Sebastiano}},\ }\href {\doibase 10.48550/arXiv.2403.09526} {\enquote {\bibinfo {title} {Optimizing the {{Electrical Interface}} for {{Large-Scale Color-Center Quantum Processors}}},}\ } (\bibinfo {year} {2024}),\ \Eprint {http://arxiv.org/abs/2403.09526} {arXiv:2403.09526 [quant-ph]} \BibitemShut {NoStop}%
\bibitem [{\citenamefont {Wineland}\ \emph {et~al.}(1998)\citenamefont {Wineland}, \citenamefont {Monroe}, \citenamefont {Itano}, \citenamefont {Leibfried}, \citenamefont {King},\ and\ \citenamefont {Meekhof}}]{wineland1998}%
  \BibitemOpen
  \bibfield  {author} {\bibinfo {author} {\bibfnamefont {D.~J.}\ \bibnamefont {Wineland}}, \bibinfo {author} {\bibfnamefont {C.}~\bibnamefont {Monroe}}, \bibinfo {author} {\bibfnamefont {W.~M.}\ \bibnamefont {Itano}}, \bibinfo {author} {\bibfnamefont {D.}~\bibnamefont {Leibfried}}, \bibinfo {author} {\bibfnamefont {B.~E.}\ \bibnamefont {King}}, \ and\ \bibinfo {author} {\bibfnamefont {D.~M.}\ \bibnamefont {Meekhof}},\ }\href {\doibase 10.6028/jres.103.019} {\bibfield  {journal} {\bibinfo  {journal} {Journal of Research of the National Institute of Standards and Technology}\ }\textbf {\bibinfo {volume} {103}},\ \bibinfo {pages} {259} (\bibinfo {year} {1998})}\BibitemShut {NoStop}%
\bibitem [{\citenamefont {Mintert}\ and\ \citenamefont {Wunderlich}(2001)}]{mintert2001}%
  \BibitemOpen
  \bibfield  {author} {\bibinfo {author} {\bibfnamefont {F.}~\bibnamefont {Mintert}}\ and\ \bibinfo {author} {\bibfnamefont {C.}~\bibnamefont {Wunderlich}},\ }\href {\doibase 10.1103/PhysRevLett.87.257904} {\bibfield  {journal} {\bibinfo  {journal} {Physical Review Letters}\ }\textbf {\bibinfo {volume} {87}},\ \bibinfo {pages} {257904} (\bibinfo {year} {2001})}\BibitemShut {NoStop}%
\bibitem [{\citenamefont {Leu}\ \emph {et~al.}(2023)\citenamefont {Leu}, \citenamefont {Gely}, \citenamefont {Weber}, \citenamefont {Smith}, \citenamefont {Nadlinger},\ and\ \citenamefont {Lucas}}]{leu2023}%
  \BibitemOpen
  \bibfield  {author} {\bibinfo {author} {\bibfnamefont {A.~D.}\ \bibnamefont {Leu}}, \bibinfo {author} {\bibfnamefont {M.~F.}\ \bibnamefont {Gely}}, \bibinfo {author} {\bibfnamefont {M.~A.}\ \bibnamefont {Weber}}, \bibinfo {author} {\bibfnamefont {M.~C.}\ \bibnamefont {Smith}}, \bibinfo {author} {\bibfnamefont {D.~P.}\ \bibnamefont {Nadlinger}}, \ and\ \bibinfo {author} {\bibfnamefont {D.~M.}\ \bibnamefont {Lucas}},\ }\href {\doibase 10.1103/PhysRevLett.131.120601} {\bibfield  {journal} {\bibinfo  {journal} {Physical Review Letters}\ }\textbf {\bibinfo {volume} {131}},\ \bibinfo {pages} {120601} (\bibinfo {year} {2023})}\BibitemShut {NoStop}%
\bibitem [{\citenamefont {Lysne}\ \emph {et~al.}(2024)\citenamefont {Lysne}, \citenamefont {Niedermeyer}, \citenamefont {Wilson}, \citenamefont {Slichter},\ and\ \citenamefont {Leibfried}}]{lysne2024}%
  \BibitemOpen
  \bibfield  {author} {\bibinfo {author} {\bibfnamefont {N.~K.}\ \bibnamefont {Lysne}}, \bibinfo {author} {\bibfnamefont {J.~F.}\ \bibnamefont {Niedermeyer}}, \bibinfo {author} {\bibfnamefont {A.~C.}\ \bibnamefont {Wilson}}, \bibinfo {author} {\bibfnamefont {D.~H.}\ \bibnamefont {Slichter}}, \ and\ \bibinfo {author} {\bibfnamefont {D.}~\bibnamefont {Leibfried}},\ }\href {\doibase 10.48550/arXiv.2402.05857} {\enquote {\bibinfo {title} {Individual addressing and state readout of trapped ions utilizing rf micromotion},}\ } (\bibinfo {year} {2024}),\ \Eprint {http://arxiv.org/abs/2402.05857} {arXiv:2402.05857 [physics, physics:quant-ph]} \BibitemShut {NoStop}%
\bibitem [{\citenamefont {Steane}(2006)}]{steane2006}%
  \BibitemOpen
  \bibfield  {author} {\bibinfo {author} {\bibfnamefont {A.~M.}\ \bibnamefont {Steane}},\ }\href {\doibase 10.48550/arXiv.quant-ph/0412165} {\enquote {\bibinfo {title} {How to build a 300 bit, 1 {{Giga-operation}} quantum computer},}\ } (\bibinfo {year} {2006}),\ \Eprint {http://arxiv.org/abs/quant-ph/0412165} {arXiv:quant-ph/0412165} \BibitemShut {NoStop}%
\bibitem [{\citenamefont {Malinowski}\ \emph {et~al.}(2023)\citenamefont {Malinowski}, \citenamefont {Allcock},\ and\ \citenamefont {Ballance}}]{malinowski2023}%
  \BibitemOpen
  \bibfield  {author} {\bibinfo {author} {\bibfnamefont {M.}~\bibnamefont {Malinowski}}, \bibinfo {author} {\bibfnamefont {D.}~\bibnamefont {Allcock}}, \ and\ \bibinfo {author} {\bibfnamefont {C.}~\bibnamefont {Ballance}},\ }\href {\doibase 10.1103/PRXQuantum.4.040313} {\bibfield  {journal} {\bibinfo  {journal} {PRX Quantum}\ }\textbf {\bibinfo {volume} {4}},\ \bibinfo {pages} {040313} (\bibinfo {year} {2023})}\BibitemShut {NoStop}%
\bibitem [{\citenamefont {Ball}\ \emph {et~al.}(2016)\citenamefont {Ball}, \citenamefont {Oliver},\ and\ \citenamefont {Biercuk}}]{ball2016}%
  \BibitemOpen
  \bibfield  {author} {\bibinfo {author} {\bibfnamefont {H.}~\bibnamefont {Ball}}, \bibinfo {author} {\bibfnamefont {W.~D.}\ \bibnamefont {Oliver}}, \ and\ \bibinfo {author} {\bibfnamefont {M.~J.}\ \bibnamefont {Biercuk}},\ }\href {\doibase 10.1038/npjqi.2016.33} {\bibfield  {journal} {\bibinfo  {journal} {npj Quantum Information}\ }\textbf {\bibinfo {volume} {2}} (\bibinfo {year} {2016}),\ 10.1038/npjqi.2016.33}\BibitemShut {NoStop}%
\bibitem [{\citenamefont {Ozeri}\ \emph {et~al.}(2007)\citenamefont {Ozeri}, \citenamefont {Itano}, \citenamefont {Blakestad}, \citenamefont {Britton}, \citenamefont {Chiaverini}, \citenamefont {Jost}, \citenamefont {Langer}, \citenamefont {Leibfried}, \citenamefont {Reichle}, \citenamefont {Seidelin}, \citenamefont {Wesenberg},\ and\ \citenamefont {Wineland}}]{ozeri2007}%
  \BibitemOpen
  \bibfield  {author} {\bibinfo {author} {\bibfnamefont {R.}~\bibnamefont {Ozeri}}, \bibinfo {author} {\bibfnamefont {W.~M.}\ \bibnamefont {Itano}}, \bibinfo {author} {\bibfnamefont {R.~B.}\ \bibnamefont {Blakestad}}, \bibinfo {author} {\bibfnamefont {J.}~\bibnamefont {Britton}}, \bibinfo {author} {\bibfnamefont {J.}~\bibnamefont {Chiaverini}}, \bibinfo {author} {\bibfnamefont {J.~D.}\ \bibnamefont {Jost}}, \bibinfo {author} {\bibfnamefont {C.}~\bibnamefont {Langer}}, \bibinfo {author} {\bibfnamefont {D.}~\bibnamefont {Leibfried}}, \bibinfo {author} {\bibfnamefont {R.}~\bibnamefont {Reichle}}, \bibinfo {author} {\bibfnamefont {S.}~\bibnamefont {Seidelin}}, \bibinfo {author} {\bibfnamefont {J.~H.}\ \bibnamefont {Wesenberg}}, \ and\ \bibinfo {author} {\bibfnamefont {D.~J.}\ \bibnamefont {Wineland}},\ }\href {\doibase 10.1103/PhysRevA.75.042329} {\bibfield  {journal} {\bibinfo  {journal} {Physical Review A}\ }\textbf {\bibinfo {volume} {75}},\ \bibinfo {pages} {042329} (\bibinfo {year} {2007})}\BibitemShut
  {NoStop}%
\bibitem [{\citenamefont {Moore}\ \emph {et~al.}(2023)\citenamefont {Moore}, \citenamefont {Campbell}, \citenamefont {Hudson}, \citenamefont {Boguslawski}, \citenamefont {Wineland},\ and\ \citenamefont {Allcock}}]{moore2023}%
  \BibitemOpen
  \bibfield  {author} {\bibinfo {author} {\bibfnamefont {I.~D.}\ \bibnamefont {Moore}}, \bibinfo {author} {\bibfnamefont {W.~C.}\ \bibnamefont {Campbell}}, \bibinfo {author} {\bibfnamefont {E.~R.}\ \bibnamefont {Hudson}}, \bibinfo {author} {\bibfnamefont {M.~J.}\ \bibnamefont {Boguslawski}}, \bibinfo {author} {\bibfnamefont {D.~J.}\ \bibnamefont {Wineland}}, \ and\ \bibinfo {author} {\bibfnamefont {D.~T.~C.}\ \bibnamefont {Allcock}},\ }\href {\doibase 10.1103/PhysRevA.107.032413} {\bibfield  {journal} {\bibinfo  {journal} {Physical Review A}\ }\textbf {\bibinfo {volume} {107}},\ \bibinfo {pages} {032413} (\bibinfo {year} {2023})}\BibitemShut {NoStop}%
\bibitem [{\citenamefont {Sutherland}\ \emph {et~al.}(2022)\citenamefont {Sutherland}, \citenamefont {Yu}, \citenamefont {Beck},\ and\ \citenamefont {H{\"a}ffner}}]{sutherland2022}%
  \BibitemOpen
  \bibfield  {author} {\bibinfo {author} {\bibfnamefont {R.~T.}\ \bibnamefont {Sutherland}}, \bibinfo {author} {\bibfnamefont {Q.}~\bibnamefont {Yu}}, \bibinfo {author} {\bibfnamefont {K.~M.}\ \bibnamefont {Beck}}, \ and\ \bibinfo {author} {\bibfnamefont {H.}~\bibnamefont {H{\"a}ffner}},\ }\href {\doibase 10.1103/PhysRevA.105.022437} {\bibfield  {journal} {\bibinfo  {journal} {Physical Review A}\ }\textbf {\bibinfo {volume} {105}},\ \bibinfo {pages} {022437} (\bibinfo {year} {2022})}\BibitemShut {NoStop}%
\bibitem [{\citenamefont {Metcalf}\ and\ \citenamefont {van~der Straten}(2003)}]{metcalf2003}%
  \BibitemOpen
  \bibfield  {author} {\bibinfo {author} {\bibfnamefont {H.~J.}\ \bibnamefont {Metcalf}}\ and\ \bibinfo {author} {\bibfnamefont {P.}~\bibnamefont {van~der Straten}},\ }\href {\doibase 10.1364/JOSAB.20.000887} {\bibfield  {journal} {\bibinfo  {journal} {JOSA B}\ }\textbf {\bibinfo {volume} {20}},\ \bibinfo {pages} {887} (\bibinfo {year} {2003})}\BibitemShut {NoStop}%
\bibitem [{\citenamefont {Burrell}\ \emph {et~al.}(2010)\citenamefont {Burrell}, \citenamefont {Szwer}, \citenamefont {Webster},\ and\ \citenamefont {Lucas}}]{burrell2010}%
  \BibitemOpen
  \bibfield  {author} {\bibinfo {author} {\bibfnamefont {A.~H.}\ \bibnamefont {Burrell}}, \bibinfo {author} {\bibfnamefont {D.~J.}\ \bibnamefont {Szwer}}, \bibinfo {author} {\bibfnamefont {S.~C.}\ \bibnamefont {Webster}}, \ and\ \bibinfo {author} {\bibfnamefont {D.~M.}\ \bibnamefont {Lucas}},\ }\href {\doibase 10.1103/PhysRevA.81.040302} {\bibfield  {journal} {\bibinfo  {journal} {Physical Review A}\ }\textbf {\bibinfo {volume} {81}},\ \bibinfo {pages} {040302} (\bibinfo {year} {2010})}\BibitemShut {NoStop}%
\bibitem [{\citenamefont {Merrill}\ and\ \citenamefont {Brown}(2012)}]{merrill2012}%
  \BibitemOpen
  \bibfield  {author} {\bibinfo {author} {\bibfnamefont {J.~T.}\ \bibnamefont {Merrill}}\ and\ \bibinfo {author} {\bibfnamefont {K.~R.}\ \bibnamefont {Brown}},\ }\href {\doibase 10.48550/arXiv.1203.6392} {\enquote {\bibinfo {title} {Progress in compensating pulse sequences for quantum computation},}\ } (\bibinfo {year} {2012}),\ \Eprint {http://arxiv.org/abs/1203.6392} {arXiv:1203.6392} \BibitemShut {NoStop}%
\bibitem [{\citenamefont {Knill}\ \emph {et~al.}(2008)\citenamefont {Knill}, \citenamefont {Leibfried}, \citenamefont {Reichle}, \citenamefont {Britton}, \citenamefont {Blakestad}, \citenamefont {Jost}, \citenamefont {Langer}, \citenamefont {Ozeri}, \citenamefont {Seidelin},\ and\ \citenamefont {Wineland}}]{knill2008}%
  \BibitemOpen
  \bibfield  {author} {\bibinfo {author} {\bibfnamefont {E.}~\bibnamefont {Knill}}, \bibinfo {author} {\bibfnamefont {D.}~\bibnamefont {Leibfried}}, \bibinfo {author} {\bibfnamefont {R.}~\bibnamefont {Reichle}}, \bibinfo {author} {\bibfnamefont {J.}~\bibnamefont {Britton}}, \bibinfo {author} {\bibfnamefont {R.~B.}\ \bibnamefont {Blakestad}}, \bibinfo {author} {\bibfnamefont {J.~D.}\ \bibnamefont {Jost}}, \bibinfo {author} {\bibfnamefont {C.}~\bibnamefont {Langer}}, \bibinfo {author} {\bibfnamefont {R.}~\bibnamefont {Ozeri}}, \bibinfo {author} {\bibfnamefont {S.}~\bibnamefont {Seidelin}}, \ and\ \bibinfo {author} {\bibfnamefont {D.~J.}\ \bibnamefont {Wineland}},\ }\href {\doibase 10.1103/PhysRevA.77.012307} {\bibfield  {journal} {\bibinfo  {journal} {Physical Review A}\ }\textbf {\bibinfo {volume} {77}},\ \bibinfo {pages} {012307} (\bibinfo {year} {2008})}\BibitemShut {NoStop}%
\bibitem [{\citenamefont {Magesan}\ \emph {et~al.}(2011)\citenamefont {Magesan}, \citenamefont {Gambetta},\ and\ \citenamefont {Emerson}}]{magesan2011}%
  \BibitemOpen
  \bibfield  {author} {\bibinfo {author} {\bibfnamefont {E.}~\bibnamefont {Magesan}}, \bibinfo {author} {\bibfnamefont {J.~M.}\ \bibnamefont {Gambetta}}, \ and\ \bibinfo {author} {\bibfnamefont {J.}~\bibnamefont {Emerson}},\ }\href {\doibase 10.1103/PhysRevLett.106.180504} {\bibfield  {journal} {\bibinfo  {journal} {Physical Review Letters}\ }\textbf {\bibinfo {volume} {106}},\ \bibinfo {pages} {180504} (\bibinfo {year} {2011})}\BibitemShut {NoStop}%
\bibitem [{\citenamefont {S{\o}rensen}\ and\ \citenamefont {M{\o}lmer}(1999)}]{sorensen1999}%
  \BibitemOpen
  \bibfield  {author} {\bibinfo {author} {\bibfnamefont {A.}~\bibnamefont {S{\o}rensen}}\ and\ \bibinfo {author} {\bibfnamefont {K.}~\bibnamefont {M{\o}lmer}},\ }\href {\doibase 10.1103/PhysRevLett.82.1971} {\bibfield  {journal} {\bibinfo  {journal} {Physical Review Letters}\ }\textbf {\bibinfo {volume} {82}},\ \bibinfo {pages} {1971} (\bibinfo {year} {1999})}\BibitemShut {NoStop}%
\bibitem [{\citenamefont {Hayes}\ \emph {et~al.}(2012)\citenamefont {Hayes}, \citenamefont {Clark}, \citenamefont {Debnath}, \citenamefont {Hucul}, \citenamefont {Inlek}, \citenamefont {Lee}, \citenamefont {Quraishi},\ and\ \citenamefont {Monroe}}]{hayes2012}%
  \BibitemOpen
  \bibfield  {author} {\bibinfo {author} {\bibfnamefont {D.}~\bibnamefont {Hayes}}, \bibinfo {author} {\bibfnamefont {S.~M.}\ \bibnamefont {Clark}}, \bibinfo {author} {\bibfnamefont {S.}~\bibnamefont {Debnath}}, \bibinfo {author} {\bibfnamefont {D.}~\bibnamefont {Hucul}}, \bibinfo {author} {\bibfnamefont {I.~V.}\ \bibnamefont {Inlek}}, \bibinfo {author} {\bibfnamefont {K.~W.}\ \bibnamefont {Lee}}, \bibinfo {author} {\bibfnamefont {Q.}~\bibnamefont {Quraishi}}, \ and\ \bibinfo {author} {\bibfnamefont {C.}~\bibnamefont {Monroe}},\ }\href {\doibase 10.1103/PhysRevLett.109.020503} {\bibfield  {journal} {\bibinfo  {journal} {Physical Review Letters}\ }\textbf {\bibinfo {volume} {109}},\ \bibinfo {pages} {020503} (\bibinfo {year} {2012})}\BibitemShut {NoStop}%
\bibitem [{\citenamefont {Leibfried}\ \emph {et~al.}(2003)\citenamefont {Leibfried}, \citenamefont {DeMarco}, \citenamefont {Meyer}, \citenamefont {Lucas}, \citenamefont {Barrett}, \citenamefont {Britton}, \citenamefont {Itano}, \citenamefont {Jelenkovi{\'c}}, \citenamefont {Langer}, \citenamefont {Rosenband},\ and\ \citenamefont {Wineland}}]{leibfried2003}%
  \BibitemOpen
  \bibfield  {author} {\bibinfo {author} {\bibfnamefont {D.}~\bibnamefont {Leibfried}}, \bibinfo {author} {\bibfnamefont {B.}~\bibnamefont {DeMarco}}, \bibinfo {author} {\bibfnamefont {V.}~\bibnamefont {Meyer}}, \bibinfo {author} {\bibfnamefont {D.}~\bibnamefont {Lucas}}, \bibinfo {author} {\bibfnamefont {M.}~\bibnamefont {Barrett}}, \bibinfo {author} {\bibfnamefont {J.}~\bibnamefont {Britton}}, \bibinfo {author} {\bibfnamefont {W.~M.}\ \bibnamefont {Itano}}, \bibinfo {author} {\bibfnamefont {B.}~\bibnamefont {Jelenkovi{\'c}}}, \bibinfo {author} {\bibfnamefont {C.}~\bibnamefont {Langer}}, \bibinfo {author} {\bibfnamefont {T.}~\bibnamefont {Rosenband}}, \ and\ \bibinfo {author} {\bibfnamefont {D.~J.}\ \bibnamefont {Wineland}},\ }\href {\doibase 10.1038/nature01492} {\bibfield  {journal} {\bibinfo  {journal} {Nature}\ }\textbf {\bibinfo {volume} {422}},\ \bibinfo {pages} {412} (\bibinfo {year} {2003})}\BibitemShut {NoStop}%
\bibitem [{\citenamefont {Efron}\ and\ \citenamefont {Tibshirani}(1994)}]{efron1994}%
  \BibitemOpen
  \bibfield  {author} {\bibinfo {author} {\bibfnamefont {B.}~\bibnamefont {Efron}}\ and\ \bibinfo {author} {\bibfnamefont {R.~J.}\ \bibnamefont {Tibshirani}},\ }\href@noop {} {\emph {\bibinfo {title} {An Introduction to the Bootstrap}}}\ (\bibinfo  {publisher} {CRC Press},\ \bibinfo {year} {1994})\BibitemShut {NoStop}%
\bibitem [{\citenamefont {O'Malley}\ \emph {et~al.}(2015)\citenamefont {O'Malley}, \citenamefont {Kelly}, \citenamefont {Barends}, \citenamefont {Campbell}, \citenamefont {Chen}, \citenamefont {Chen}, \citenamefont {Chiaro}, \citenamefont {Dunsworth}, \citenamefont {Fowler}, \citenamefont {Hoi}, \citenamefont {Jeffrey}, \citenamefont {Megrant}, \citenamefont {Mutus}, \citenamefont {Neill}, \citenamefont {Quintana}, \citenamefont {Roushan}, \citenamefont {Sank}, \citenamefont {Vainsencher}, \citenamefont {Wenner}, \citenamefont {White}, \citenamefont {Korotkov}, \citenamefont {Cleland},\ and\ \citenamefont {Martinis}}]{omalley2015}%
  \BibitemOpen
  \bibfield  {author} {\bibinfo {author} {\bibfnamefont {P.~J.~J.}\ \bibnamefont {O'Malley}}, \bibinfo {author} {\bibfnamefont {J.}~\bibnamefont {Kelly}}, \bibinfo {author} {\bibfnamefont {R.}~\bibnamefont {Barends}}, \bibinfo {author} {\bibfnamefont {B.}~\bibnamefont {Campbell}}, \bibinfo {author} {\bibfnamefont {Y.}~\bibnamefont {Chen}}, \bibinfo {author} {\bibfnamefont {Z.}~\bibnamefont {Chen}}, \bibinfo {author} {\bibfnamefont {B.}~\bibnamefont {Chiaro}}, \bibinfo {author} {\bibfnamefont {A.}~\bibnamefont {Dunsworth}}, \bibinfo {author} {\bibfnamefont {A.~G.}\ \bibnamefont {Fowler}}, \bibinfo {author} {\bibfnamefont {I.-C.}\ \bibnamefont {Hoi}}, \bibinfo {author} {\bibfnamefont {E.}~\bibnamefont {Jeffrey}}, \bibinfo {author} {\bibfnamefont {A.}~\bibnamefont {Megrant}}, \bibinfo {author} {\bibfnamefont {J.}~\bibnamefont {Mutus}}, \bibinfo {author} {\bibfnamefont {C.}~\bibnamefont {Neill}}, \bibinfo {author} {\bibfnamefont {C.}~\bibnamefont {Quintana}}, \bibinfo {author} {\bibfnamefont {P.}~\bibnamefont
  {Roushan}}, \bibinfo {author} {\bibfnamefont {D.}~\bibnamefont {Sank}}, \bibinfo {author} {\bibfnamefont {A.}~\bibnamefont {Vainsencher}}, \bibinfo {author} {\bibfnamefont {J.}~\bibnamefont {Wenner}}, \bibinfo {author} {\bibfnamefont {T.~C.}\ \bibnamefont {White}}, \bibinfo {author} {\bibfnamefont {A.~N.}\ \bibnamefont {Korotkov}}, \bibinfo {author} {\bibfnamefont {A.~N.}\ \bibnamefont {Cleland}}, \ and\ \bibinfo {author} {\bibfnamefont {J.~M.}\ \bibnamefont {Martinis}},\ }\href {\doibase 10.1103/PhysRevApplied.3.044009} {\bibfield  {journal} {\bibinfo  {journal} {Physical Review Applied}\ }\textbf {\bibinfo {volume} {3}},\ \bibinfo {pages} {044009} (\bibinfo {year} {2015})}\BibitemShut {NoStop}%
\bibitem [{\citenamefont {Viola}\ and\ \citenamefont {Lloyd}(1998)}]{viola1998}%
  \BibitemOpen
  \bibfield  {author} {\bibinfo {author} {\bibfnamefont {L.}~\bibnamefont {Viola}}\ and\ \bibinfo {author} {\bibfnamefont {S.}~\bibnamefont {Lloyd}},\ }\href {\doibase 10.1103/PhysRevA.58.2733} {\bibfield  {journal} {\bibinfo  {journal} {Physical Review A}\ }\textbf {\bibinfo {volume} {58}},\ \bibinfo {pages} {2733} (\bibinfo {year} {1998})}\BibitemShut {NoStop}%
\bibitem [{\citenamefont {Bermudez}\ \emph {et~al.}(2012)\citenamefont {Bermudez}, \citenamefont {Schmidt}, \citenamefont {Plenio},\ and\ \citenamefont {Retzker}}]{bermudez2012}%
  \BibitemOpen
  \bibfield  {author} {\bibinfo {author} {\bibfnamefont {A.}~\bibnamefont {Bermudez}}, \bibinfo {author} {\bibfnamefont {P.~O.}\ \bibnamefont {Schmidt}}, \bibinfo {author} {\bibfnamefont {M.~B.}\ \bibnamefont {Plenio}}, \ and\ \bibinfo {author} {\bibfnamefont {A.}~\bibnamefont {Retzker}},\ }\href {\doibase 10.1103/PhysRevA.85.040302} {\bibfield  {journal} {\bibinfo  {journal} {Physical Review A}\ }\textbf {\bibinfo {volume} {85}},\ \bibinfo {pages} {040302} (\bibinfo {year} {2012})}\BibitemShut {NoStop}%
\bibitem [{\citenamefont {Harty}\ \emph {et~al.}(2016)\citenamefont {Harty}, \citenamefont {Sepiol}, \citenamefont {Allcock}, \citenamefont {Ballance}, \citenamefont {Tarlton},\ and\ \citenamefont {Lucas}}]{harty2016}%
  \BibitemOpen
  \bibfield  {author} {\bibinfo {author} {\bibfnamefont {T.~P.}\ \bibnamefont {Harty}}, \bibinfo {author} {\bibfnamefont {M.~A.}\ \bibnamefont {Sepiol}}, \bibinfo {author} {\bibfnamefont {D.~T.~C.}\ \bibnamefont {Allcock}}, \bibinfo {author} {\bibfnamefont {C.~J.}\ \bibnamefont {Ballance}}, \bibinfo {author} {\bibfnamefont {J.~E.}\ \bibnamefont {Tarlton}}, \ and\ \bibinfo {author} {\bibfnamefont {D.~M.}\ \bibnamefont {Lucas}},\ }\href {\doibase 10.1103/PhysRevLett.117.140501} {\bibfield  {journal} {\bibinfo  {journal} {Physical Review Letters}\ }\textbf {\bibinfo {volume} {117}},\ \bibinfo {pages} {140501} (\bibinfo {year} {2016})}\BibitemShut {NoStop}%
\bibitem [{\citenamefont {Langer}\ \emph {et~al.}(2005)\citenamefont {Langer}, \citenamefont {Ozeri}, \citenamefont {Jost}, \citenamefont {Chiaverini}, \citenamefont {DeMarco}, \citenamefont {{Ben-Kish}}, \citenamefont {Blakestad}, \citenamefont {Britton}, \citenamefont {Hume}, \citenamefont {Itano}, \citenamefont {Leibfried}, \citenamefont {Reichle}, \citenamefont {Rosenband}, \citenamefont {Schaetz}, \citenamefont {Schmidt},\ and\ \citenamefont {Wineland}}]{langer2005}%
  \BibitemOpen
  \bibfield  {author} {\bibinfo {author} {\bibfnamefont {C.}~\bibnamefont {Langer}}, \bibinfo {author} {\bibfnamefont {R.}~\bibnamefont {Ozeri}}, \bibinfo {author} {\bibfnamefont {J.~D.}\ \bibnamefont {Jost}}, \bibinfo {author} {\bibfnamefont {J.}~\bibnamefont {Chiaverini}}, \bibinfo {author} {\bibfnamefont {B.}~\bibnamefont {DeMarco}}, \bibinfo {author} {\bibfnamefont {A.}~\bibnamefont {{Ben-Kish}}}, \bibinfo {author} {\bibfnamefont {R.~B.}\ \bibnamefont {Blakestad}}, \bibinfo {author} {\bibfnamefont {J.}~\bibnamefont {Britton}}, \bibinfo {author} {\bibfnamefont {D.~B.}\ \bibnamefont {Hume}}, \bibinfo {author} {\bibfnamefont {W.~M.}\ \bibnamefont {Itano}}, \bibinfo {author} {\bibfnamefont {D.}~\bibnamefont {Leibfried}}, \bibinfo {author} {\bibfnamefont {R.}~\bibnamefont {Reichle}}, \bibinfo {author} {\bibfnamefont {T.}~\bibnamefont {Rosenband}}, \bibinfo {author} {\bibfnamefont {T.}~\bibnamefont {Schaetz}}, \bibinfo {author} {\bibfnamefont {P.~O.}\ \bibnamefont {Schmidt}}, \ and\ \bibinfo {author}
  {\bibfnamefont {D.~J.}\ \bibnamefont {Wineland}},\ }\href {\doibase 10.1103/PhysRevLett.95.060502} {\bibfield  {journal} {\bibinfo  {journal} {Physical Review Letters}\ }\textbf {\bibinfo {volume} {95}},\ \bibinfo {pages} {060502} (\bibinfo {year} {2005})}\BibitemShut {NoStop}%
\bibitem [{\citenamefont {{Parra-Rodriguez}}\ \emph {et~al.}(2020)\citenamefont {{Parra-Rodriguez}}, \citenamefont {Lougovski}, \citenamefont {Lamata}, \citenamefont {Solano},\ and\ \citenamefont {Sanz}}]{parra-rodriguez2020}%
  \BibitemOpen
  \bibfield  {author} {\bibinfo {author} {\bibfnamefont {A.}~\bibnamefont {{Parra-Rodriguez}}}, \bibinfo {author} {\bibfnamefont {P.}~\bibnamefont {Lougovski}}, \bibinfo {author} {\bibfnamefont {L.}~\bibnamefont {Lamata}}, \bibinfo {author} {\bibfnamefont {E.}~\bibnamefont {Solano}}, \ and\ \bibinfo {author} {\bibfnamefont {M.}~\bibnamefont {Sanz}},\ }\href {\doibase 10.1103/PhysRevA.101.022305} {\bibfield  {journal} {\bibinfo  {journal} {Physical Review A}\ }\textbf {\bibinfo {volume} {101}},\ \bibinfo {pages} {022305} (\bibinfo {year} {2020})}\BibitemShut {NoStop}%
\bibitem [{\citenamefont {Foxen}\ \emph {et~al.}(2020)\citenamefont {Foxen}, \citenamefont {Neill}, \citenamefont {Dunsworth}, \citenamefont {Roushan}, \citenamefont {Chiaro}, \citenamefont {Megrant}, \citenamefont {Kelly}, \citenamefont {Chen}, \citenamefont {Satzinger}, \citenamefont {Barends}, \citenamefont {Arute}, \citenamefont {Arya}, \citenamefont {Babbush}, \citenamefont {Bacon}, \citenamefont {Bardin}, \citenamefont {Boixo}, \citenamefont {Buell}, \citenamefont {Burkett}, \citenamefont {Chen}, \citenamefont {Collins}, \citenamefont {Farhi}, \citenamefont {Fowler}, \citenamefont {Gidney}, \citenamefont {Giustina}, \citenamefont {Graff}, \citenamefont {Harrigan}, \citenamefont {Huang}, \citenamefont {Isakov}, \citenamefont {Jeffrey}, \citenamefont {Jiang}, \citenamefont {Kafri}, \citenamefont {Kechedzhi}, \citenamefont {Klimov}, \citenamefont {Korotkov}, \citenamefont {Kostritsa}, \citenamefont {Landhuis}, \citenamefont {Lucero}, \citenamefont {McClean}, \citenamefont {McEwen}, \citenamefont {Mi},
  \citenamefont {Mohseni}, \citenamefont {Mutus}, \citenamefont {Naaman}, \citenamefont {Neeley}, \citenamefont {Niu}, \citenamefont {Petukhov}, \citenamefont {Quintana}, \citenamefont {Rubin}, \citenamefont {Sank}, \citenamefont {Smelyanskiy}, \citenamefont {Vainsencher}, \citenamefont {White}, \citenamefont {Yao}, \citenamefont {Yeh}, \citenamefont {Zalcman}, \citenamefont {Neven},\ and\ \citenamefont {Martinis}}]{foxen2020}%
  \BibitemOpen
  \bibfield  {author} {\bibinfo {author} {\bibfnamefont {B.}~\bibnamefont {Foxen}}, \bibinfo {author} {\bibfnamefont {C.}~\bibnamefont {Neill}}, \bibinfo {author} {\bibfnamefont {A.}~\bibnamefont {Dunsworth}}, \bibinfo {author} {\bibfnamefont {P.}~\bibnamefont {Roushan}}, \bibinfo {author} {\bibfnamefont {B.}~\bibnamefont {Chiaro}}, \bibinfo {author} {\bibfnamefont {A.}~\bibnamefont {Megrant}}, \bibinfo {author} {\bibfnamefont {J.}~\bibnamefont {Kelly}}, \bibinfo {author} {\bibfnamefont {Z.}~\bibnamefont {Chen}}, \bibinfo {author} {\bibfnamefont {K.}~\bibnamefont {Satzinger}}, \bibinfo {author} {\bibfnamefont {R.}~\bibnamefont {Barends}}, \bibinfo {author} {\bibfnamefont {F.}~\bibnamefont {Arute}}, \bibinfo {author} {\bibfnamefont {K.}~\bibnamefont {Arya}}, \bibinfo {author} {\bibfnamefont {R.}~\bibnamefont {Babbush}}, \bibinfo {author} {\bibfnamefont {D.}~\bibnamefont {Bacon}}, \bibinfo {author} {\bibfnamefont {J.~C.}\ \bibnamefont {Bardin}}, \bibinfo {author} {\bibfnamefont {S.}~\bibnamefont {Boixo}},
  \bibinfo {author} {\bibfnamefont {D.}~\bibnamefont {Buell}}, \bibinfo {author} {\bibfnamefont {B.}~\bibnamefont {Burkett}}, \bibinfo {author} {\bibfnamefont {Y.}~\bibnamefont {Chen}}, \bibinfo {author} {\bibfnamefont {R.}~\bibnamefont {Collins}}, \bibinfo {author} {\bibfnamefont {E.}~\bibnamefont {Farhi}}, \bibinfo {author} {\bibfnamefont {A.}~\bibnamefont {Fowler}}, \bibinfo {author} {\bibfnamefont {C.}~\bibnamefont {Gidney}}, \bibinfo {author} {\bibfnamefont {M.}~\bibnamefont {Giustina}}, \bibinfo {author} {\bibfnamefont {R.}~\bibnamefont {Graff}}, \bibinfo {author} {\bibfnamefont {M.}~\bibnamefont {Harrigan}}, \bibinfo {author} {\bibfnamefont {T.}~\bibnamefont {Huang}}, \bibinfo {author} {\bibfnamefont {S.~V.}\ \bibnamefont {Isakov}}, \bibinfo {author} {\bibfnamefont {E.}~\bibnamefont {Jeffrey}}, \bibinfo {author} {\bibfnamefont {Z.}~\bibnamefont {Jiang}}, \bibinfo {author} {\bibfnamefont {D.}~\bibnamefont {Kafri}}, \bibinfo {author} {\bibfnamefont {K.}~\bibnamefont {Kechedzhi}}, \bibinfo {author}
  {\bibfnamefont {P.}~\bibnamefont {Klimov}}, \bibinfo {author} {\bibfnamefont {A.}~\bibnamefont {Korotkov}}, \bibinfo {author} {\bibfnamefont {F.}~\bibnamefont {Kostritsa}}, \bibinfo {author} {\bibfnamefont {D.}~\bibnamefont {Landhuis}}, \bibinfo {author} {\bibfnamefont {E.}~\bibnamefont {Lucero}}, \bibinfo {author} {\bibfnamefont {J.}~\bibnamefont {McClean}}, \bibinfo {author} {\bibfnamefont {M.}~\bibnamefont {McEwen}}, \bibinfo {author} {\bibfnamefont {X.}~\bibnamefont {Mi}}, \bibinfo {author} {\bibfnamefont {M.}~\bibnamefont {Mohseni}}, \bibinfo {author} {\bibfnamefont {J.~Y.}\ \bibnamefont {Mutus}}, \bibinfo {author} {\bibfnamefont {O.}~\bibnamefont {Naaman}}, \bibinfo {author} {\bibfnamefont {M.}~\bibnamefont {Neeley}}, \bibinfo {author} {\bibfnamefont {M.}~\bibnamefont {Niu}}, \bibinfo {author} {\bibfnamefont {A.}~\bibnamefont {Petukhov}}, \bibinfo {author} {\bibfnamefont {C.}~\bibnamefont {Quintana}}, \bibinfo {author} {\bibfnamefont {N.}~\bibnamefont {Rubin}}, \bibinfo {author} {\bibfnamefont
  {D.}~\bibnamefont {Sank}}, \bibinfo {author} {\bibfnamefont {V.}~\bibnamefont {Smelyanskiy}}, \bibinfo {author} {\bibfnamefont {A.}~\bibnamefont {Vainsencher}}, \bibinfo {author} {\bibfnamefont {T.~C.}\ \bibnamefont {White}}, \bibinfo {author} {\bibfnamefont {Z.}~\bibnamefont {Yao}}, \bibinfo {author} {\bibfnamefont {P.}~\bibnamefont {Yeh}}, \bibinfo {author} {\bibfnamefont {A.}~\bibnamefont {Zalcman}}, \bibinfo {author} {\bibfnamefont {H.}~\bibnamefont {Neven}}, \ and\ \bibinfo {author} {\bibfnamefont {J.~M.}\ \bibnamefont {Martinis}},\ }\href {\doibase 10.1103/PhysRevLett.125.120504} {\bibfield  {journal} {\bibinfo  {journal} {Physical Review Letters}\ }\textbf {\bibinfo {volume} {125}},\ \bibinfo {pages} {120504} (\bibinfo {year} {2020})},\ \Eprint {http://arxiv.org/abs/2001.08343} {arXiv:2001.08343 [quant-ph]} \BibitemShut {NoStop}%
\bibitem [{\citenamefont {Lacroix}\ \emph {et~al.}(2020)\citenamefont {Lacroix}, \citenamefont {Hellings}, \citenamefont {Andersen}, \citenamefont {Di~Paolo}, \citenamefont {Remm}, \citenamefont {Lazar}, \citenamefont {Krinner}, \citenamefont {Norris}, \citenamefont {Gabureac}, \citenamefont {Heinsoo}, \citenamefont {Blais}, \citenamefont {Eichler},\ and\ \citenamefont {Wallraff}}]{lacroix2020}%
  \BibitemOpen
  \bibfield  {author} {\bibinfo {author} {\bibfnamefont {N.}~\bibnamefont {Lacroix}}, \bibinfo {author} {\bibfnamefont {C.}~\bibnamefont {Hellings}}, \bibinfo {author} {\bibfnamefont {C.~K.}\ \bibnamefont {Andersen}}, \bibinfo {author} {\bibfnamefont {A.}~\bibnamefont {Di~Paolo}}, \bibinfo {author} {\bibfnamefont {A.}~\bibnamefont {Remm}}, \bibinfo {author} {\bibfnamefont {S.}~\bibnamefont {Lazar}}, \bibinfo {author} {\bibfnamefont {S.}~\bibnamefont {Krinner}}, \bibinfo {author} {\bibfnamefont {G.~J.}\ \bibnamefont {Norris}}, \bibinfo {author} {\bibfnamefont {M.}~\bibnamefont {Gabureac}}, \bibinfo {author} {\bibfnamefont {J.}~\bibnamefont {Heinsoo}}, \bibinfo {author} {\bibfnamefont {A.}~\bibnamefont {Blais}}, \bibinfo {author} {\bibfnamefont {C.}~\bibnamefont {Eichler}}, \ and\ \bibinfo {author} {\bibfnamefont {A.}~\bibnamefont {Wallraff}},\ }\href {\doibase 10.1103/PRXQuantum.1.020304} {\bibfield  {journal} {\bibinfo  {journal} {PRX Quantum}\ }\textbf {\bibinfo {volume} {1}},\ \bibinfo {pages} {020304}
  (\bibinfo {year} {2020})}\BibitemShut {NoStop}%
\bibitem [{\citenamefont {Clinton}\ \emph {et~al.}(2024)\citenamefont {Clinton}, \citenamefont {Cubitt}, \citenamefont {Flynn}, \citenamefont {Gambetta}, \citenamefont {Klassen}, \citenamefont {Montanaro}, \citenamefont {Piddock}, \citenamefont {Santos},\ and\ \citenamefont {Sheridan}}]{clinton2024}%
  \BibitemOpen
  \bibfield  {author} {\bibinfo {author} {\bibfnamefont {L.}~\bibnamefont {Clinton}}, \bibinfo {author} {\bibfnamefont {T.}~\bibnamefont {Cubitt}}, \bibinfo {author} {\bibfnamefont {B.}~\bibnamefont {Flynn}}, \bibinfo {author} {\bibfnamefont {F.~M.}\ \bibnamefont {Gambetta}}, \bibinfo {author} {\bibfnamefont {J.}~\bibnamefont {Klassen}}, \bibinfo {author} {\bibfnamefont {A.}~\bibnamefont {Montanaro}}, \bibinfo {author} {\bibfnamefont {S.}~\bibnamefont {Piddock}}, \bibinfo {author} {\bibfnamefont {R.~A.}\ \bibnamefont {Santos}}, \ and\ \bibinfo {author} {\bibfnamefont {E.}~\bibnamefont {Sheridan}},\ }\href {\doibase 10.1038/s41467-023-43479-6} {\bibfield  {journal} {\bibinfo  {journal} {Nature Communications}\ }\textbf {\bibinfo {volume} {15}},\ \bibinfo {pages} {211} (\bibinfo {year} {2024})}\BibitemShut {NoStop}%
\bibitem [{\citenamefont {Guise}\ \emph {et~al.}(2015)\citenamefont {Guise}, \citenamefont {Fallek}, \citenamefont {Stevens}, \citenamefont {Brown}, \citenamefont {Volin}, \citenamefont {Harter}, \citenamefont {Amini}, \citenamefont {Higashi}, \citenamefont {Lu}, \citenamefont {Chanhvongsak}, \citenamefont {Nguyen}, \citenamefont {Marcus}, \citenamefont {Ohnstein},\ and\ \citenamefont {Youngner}}]{guise2015}%
  \BibitemOpen
  \bibfield  {author} {\bibinfo {author} {\bibfnamefont {N.~D.}\ \bibnamefont {Guise}}, \bibinfo {author} {\bibfnamefont {S.~D.}\ \bibnamefont {Fallek}}, \bibinfo {author} {\bibfnamefont {K.~E.}\ \bibnamefont {Stevens}}, \bibinfo {author} {\bibfnamefont {K.~R.}\ \bibnamefont {Brown}}, \bibinfo {author} {\bibfnamefont {C.}~\bibnamefont {Volin}}, \bibinfo {author} {\bibfnamefont {A.~W.}\ \bibnamefont {Harter}}, \bibinfo {author} {\bibfnamefont {J.~M.}\ \bibnamefont {Amini}}, \bibinfo {author} {\bibfnamefont {R.~E.}\ \bibnamefont {Higashi}}, \bibinfo {author} {\bibfnamefont {S.~T.}\ \bibnamefont {Lu}}, \bibinfo {author} {\bibfnamefont {H.~M.}\ \bibnamefont {Chanhvongsak}}, \bibinfo {author} {\bibfnamefont {T.~A.}\ \bibnamefont {Nguyen}}, \bibinfo {author} {\bibfnamefont {M.~S.}\ \bibnamefont {Marcus}}, \bibinfo {author} {\bibfnamefont {T.~R.}\ \bibnamefont {Ohnstein}}, \ and\ \bibinfo {author} {\bibfnamefont {D.~W.}\ \bibnamefont {Youngner}},\ }\href {\doibase 10.1063/1.4917385} {\bibfield  {journal} {\bibinfo
  {journal} {Journal of Applied Physics}\ }\textbf {\bibinfo {volume} {117}},\ \bibinfo {pages} {174901} (\bibinfo {year} {2015})}\BibitemShut {NoStop}%
\bibitem [{\citenamefont {Burton}\ \emph {et~al.}(2023)\citenamefont {Burton}, \citenamefont {Estey}, \citenamefont {Hoffman}, \citenamefont {Perry}, \citenamefont {Volin},\ and\ \citenamefont {Price}}]{burton2023}%
  \BibitemOpen
  \bibfield  {author} {\bibinfo {author} {\bibfnamefont {W.~C.}\ \bibnamefont {Burton}}, \bibinfo {author} {\bibfnamefont {B.}~\bibnamefont {Estey}}, \bibinfo {author} {\bibfnamefont {I.~M.}\ \bibnamefont {Hoffman}}, \bibinfo {author} {\bibfnamefont {A.~R.}\ \bibnamefont {Perry}}, \bibinfo {author} {\bibfnamefont {C.}~\bibnamefont {Volin}}, \ and\ \bibinfo {author} {\bibfnamefont {G.}~\bibnamefont {Price}},\ }\href {\doibase 10.1103/PhysRevLett.130.173202} {\bibfield  {journal} {\bibinfo  {journal} {Physical Review Letters}\ }\textbf {\bibinfo {volume} {130}},\ \bibinfo {pages} {173202} (\bibinfo {year} {2023})}\BibitemShut {NoStop}%
\bibitem [{\citenamefont {Ivory}\ \emph {et~al.}(2021)\citenamefont {Ivory}, \citenamefont {Setzer}, \citenamefont {Karl}, \citenamefont {McGuinness}, \citenamefont {DeRose}, \citenamefont {Blain}, \citenamefont {Stick}, \citenamefont {Gehl},\ and\ \citenamefont {Parazzoli}}]{ivory2021}%
  \BibitemOpen
  \bibfield  {author} {\bibinfo {author} {\bibfnamefont {M.}~\bibnamefont {Ivory}}, \bibinfo {author} {\bibfnamefont {W.~J.}\ \bibnamefont {Setzer}}, \bibinfo {author} {\bibfnamefont {N.}~\bibnamefont {Karl}}, \bibinfo {author} {\bibfnamefont {H.}~\bibnamefont {McGuinness}}, \bibinfo {author} {\bibfnamefont {C.}~\bibnamefont {DeRose}}, \bibinfo {author} {\bibfnamefont {M.}~\bibnamefont {Blain}}, \bibinfo {author} {\bibfnamefont {D.}~\bibnamefont {Stick}}, \bibinfo {author} {\bibfnamefont {M.}~\bibnamefont {Gehl}}, \ and\ \bibinfo {author} {\bibfnamefont {L.~P.}\ \bibnamefont {Parazzoli}},\ }\href {\doibase 10.1103/PhysRevX.11.041033} {\bibfield  {journal} {\bibinfo  {journal} {Physical Review X}\ }\textbf {\bibinfo {volume} {11}},\ \bibinfo {pages} {041033} (\bibinfo {year} {2021})}\BibitemShut {NoStop}%
\bibitem [{\citenamefont {Wang}\ \emph {et~al.}(2020)\citenamefont {Wang}, \citenamefont {Crain}, \citenamefont {Fang}, \citenamefont {Zhang}, \citenamefont {Huang}, \citenamefont {Liang}, \citenamefont {Leung}, \citenamefont {Brown},\ and\ \citenamefont {Kim}}]{wang2020}%
  \BibitemOpen
  \bibfield  {author} {\bibinfo {author} {\bibfnamefont {Y.}~\bibnamefont {Wang}}, \bibinfo {author} {\bibfnamefont {S.}~\bibnamefont {Crain}}, \bibinfo {author} {\bibfnamefont {C.}~\bibnamefont {Fang}}, \bibinfo {author} {\bibfnamefont {B.}~\bibnamefont {Zhang}}, \bibinfo {author} {\bibfnamefont {S.}~\bibnamefont {Huang}}, \bibinfo {author} {\bibfnamefont {Q.}~\bibnamefont {Liang}}, \bibinfo {author} {\bibfnamefont {P.~H.}\ \bibnamefont {Leung}}, \bibinfo {author} {\bibfnamefont {K.~R.}\ \bibnamefont {Brown}}, \ and\ \bibinfo {author} {\bibfnamefont {J.}~\bibnamefont {Kim}},\ }\href {\doibase 10.1103/PhysRevLett.125.150505} {\bibfield  {journal} {\bibinfo  {journal} {Physical Review Letters}\ }\textbf {\bibinfo {volume} {125}},\ \bibinfo {pages} {150505} (\bibinfo {year} {2020})}\BibitemShut {NoStop}%
\bibitem [{\citenamefont {Shih}\ \emph {et~al.}(2021)\citenamefont {Shih}, \citenamefont {Motlakunta}, \citenamefont {Kotibhaskar}, \citenamefont {Sajjan}, \citenamefont {Habl{\"u}tzel},\ and\ \citenamefont {Islam}}]{shih2021}%
  \BibitemOpen
  \bibfield  {author} {\bibinfo {author} {\bibfnamefont {C.-Y.}\ \bibnamefont {Shih}}, \bibinfo {author} {\bibfnamefont {S.}~\bibnamefont {Motlakunta}}, \bibinfo {author} {\bibfnamefont {N.}~\bibnamefont {Kotibhaskar}}, \bibinfo {author} {\bibfnamefont {M.}~\bibnamefont {Sajjan}}, \bibinfo {author} {\bibfnamefont {R.}~\bibnamefont {Habl{\"u}tzel}}, \ and\ \bibinfo {author} {\bibfnamefont {R.}~\bibnamefont {Islam}},\ }\href {\doibase 10.1038/s41534-021-00396-0} {\bibfield  {journal} {\bibinfo  {journal} {npj Quantum Information}\ }\textbf {\bibinfo {volume} {7}} (\bibinfo {year} {2021}),\ 10.1038/s41534-021-00396-0}\BibitemShut {NoStop}%
\bibitem [{\citenamefont {Pogorelov}\ \emph {et~al.}(2021)\citenamefont {Pogorelov}, \citenamefont {Feldker}, \citenamefont {Marciniak}, \citenamefont {Postler}, \citenamefont {Jacob}, \citenamefont {Krieglsteiner}, \citenamefont {Podlesnic}, \citenamefont {Meth}, \citenamefont {Negnevitsky}, \citenamefont {Stadler}, \citenamefont {H{\"o}fer}, \citenamefont {W{\"a}chter}, \citenamefont {Lakhmanskiy}, \citenamefont {Blatt}, \citenamefont {Schindler},\ and\ \citenamefont {Monz}}]{pogorelov2021}%
  \BibitemOpen
  \bibfield  {author} {\bibinfo {author} {\bibfnamefont {I.}~\bibnamefont {Pogorelov}}, \bibinfo {author} {\bibfnamefont {T.}~\bibnamefont {Feldker}}, \bibinfo {author} {\bibfnamefont {{\relax Ch}.~D.}\ \bibnamefont {Marciniak}}, \bibinfo {author} {\bibfnamefont {L.}~\bibnamefont {Postler}}, \bibinfo {author} {\bibfnamefont {G.}~\bibnamefont {Jacob}}, \bibinfo {author} {\bibfnamefont {O.}~\bibnamefont {Krieglsteiner}}, \bibinfo {author} {\bibfnamefont {V.}~\bibnamefont {Podlesnic}}, \bibinfo {author} {\bibfnamefont {M.}~\bibnamefont {Meth}}, \bibinfo {author} {\bibfnamefont {V.}~\bibnamefont {Negnevitsky}}, \bibinfo {author} {\bibfnamefont {M.}~\bibnamefont {Stadler}}, \bibinfo {author} {\bibfnamefont {B.}~\bibnamefont {H{\"o}fer}}, \bibinfo {author} {\bibfnamefont {C.}~\bibnamefont {W{\"a}chter}}, \bibinfo {author} {\bibfnamefont {K.}~\bibnamefont {Lakhmanskiy}}, \bibinfo {author} {\bibfnamefont {R.}~\bibnamefont {Blatt}}, \bibinfo {author} {\bibfnamefont {P.}~\bibnamefont {Schindler}}, \ and\ \bibinfo
  {author} {\bibfnamefont {T.}~\bibnamefont {Monz}},\ }\href {\doibase 10.1103/PRXQuantum.2.020343} {\bibfield  {journal} {\bibinfo  {journal} {PRX Quantum}\ }\textbf {\bibinfo {volume} {2}},\ \bibinfo {pages} {020343} (\bibinfo {year} {2021})}\BibitemShut {NoStop}%
\bibitem [{\citenamefont {{Binai-Motlagh}}\ \emph {et~al.}(2023)\citenamefont {{Binai-Motlagh}}, \citenamefont {Day}, \citenamefont {Videnov}, \citenamefont {Greenberg}, \citenamefont {Senko},\ and\ \citenamefont {Islam}}]{binai-motlagh2023}%
  \BibitemOpen
  \bibfield  {author} {\bibinfo {author} {\bibfnamefont {A.}~\bibnamefont {{Binai-Motlagh}}}, \bibinfo {author} {\bibfnamefont {M.}~\bibnamefont {Day}}, \bibinfo {author} {\bibfnamefont {N.}~\bibnamefont {Videnov}}, \bibinfo {author} {\bibfnamefont {N.}~\bibnamefont {Greenberg}}, \bibinfo {author} {\bibfnamefont {C.}~\bibnamefont {Senko}}, \ and\ \bibinfo {author} {\bibfnamefont {R.}~\bibnamefont {Islam}},\ }\href {\doibase 10.48550/arXiv.2302.14711} {\enquote {\bibinfo {title} {A {{Guided Light System}} for {{Agile Individual Addressing}} of \{\vphantom\}{{B}}\vphantom\{\}a\${\textasciicircum}+\$ {{Qubits}} with \$10{\textasciicircum}\{-4\}\$ {{Level Intensity Crosstalk}}},}\ } (\bibinfo {year} {2023}),\ \Eprint {http://arxiv.org/abs/2302.14711} {arXiv:2302.14711 [physics, physics:quant-ph]} \BibitemShut {NoStop}%
\bibitem [{\citenamefont {Sotirova}\ \emph {et~al.}(2023)\citenamefont {Sotirova}, \citenamefont {Sun}, \citenamefont {Leppard}, \citenamefont {Wang}, \citenamefont {Wang}, \citenamefont {{Vazquez-Brennan}}, \citenamefont {Nadlinger}, \citenamefont {Moser}, \citenamefont {Jesacher}, \citenamefont {He}, \citenamefont {Pokorny}, \citenamefont {Booth},\ and\ \citenamefont {Ballance}}]{sotirova2023}%
  \BibitemOpen
  \bibfield  {author} {\bibinfo {author} {\bibfnamefont {A.~S.}\ \bibnamefont {Sotirova}}, \bibinfo {author} {\bibfnamefont {B.}~\bibnamefont {Sun}}, \bibinfo {author} {\bibfnamefont {J.~D.}\ \bibnamefont {Leppard}}, \bibinfo {author} {\bibfnamefont {A.}~\bibnamefont {Wang}}, \bibinfo {author} {\bibfnamefont {M.}~\bibnamefont {Wang}}, \bibinfo {author} {\bibfnamefont {A.}~\bibnamefont {{Vazquez-Brennan}}}, \bibinfo {author} {\bibfnamefont {D.~P.}\ \bibnamefont {Nadlinger}}, \bibinfo {author} {\bibfnamefont {S.}~\bibnamefont {Moser}}, \bibinfo {author} {\bibfnamefont {A.}~\bibnamefont {Jesacher}}, \bibinfo {author} {\bibfnamefont {C.}~\bibnamefont {He}}, \bibinfo {author} {\bibfnamefont {F.}~\bibnamefont {Pokorny}}, \bibinfo {author} {\bibfnamefont {M.~J.}\ \bibnamefont {Booth}}, \ and\ \bibinfo {author} {\bibfnamefont {C.~J.}\ \bibnamefont {Ballance}},\ }\href {\doibase 10.48550/arXiv.2310.13419} {\enquote {\bibinfo {title} {Low {{Cross-Talk Optical Addressing}} of {{Trapped-Ion Qubits Using}} a {{Novel
  Integrated Photonic Chip}}},}\ } (\bibinfo {year} {2023}),\ \Eprint {http://arxiv.org/abs/2310.13419} {arXiv:2310.13419 [physics, physics:quant-ph]} \BibitemShut {NoStop}%
\bibitem [{\citenamefont {Ospelkaus}\ \emph {et~al.}(2011)\citenamefont {Ospelkaus}, \citenamefont {Warring}, \citenamefont {Colombe}, \citenamefont {Brown}, \citenamefont {Amini}, \citenamefont {Leibfried},\ and\ \citenamefont {Wineland}}]{Ospelkaus2011}%
  \BibitemOpen
  \bibfield  {author} {\bibinfo {author} {\bibfnamefont {C.}~\bibnamefont {Ospelkaus}}, \bibinfo {author} {\bibfnamefont {U.}~\bibnamefont {Warring}}, \bibinfo {author} {\bibfnamefont {Y.}~\bibnamefont {Colombe}}, \bibinfo {author} {\bibfnamefont {K.~R.}\ \bibnamefont {Brown}}, \bibinfo {author} {\bibfnamefont {J.~M.}\ \bibnamefont {Amini}}, \bibinfo {author} {\bibfnamefont {D.}~\bibnamefont {Leibfried}}, \ and\ \bibinfo {author} {\bibfnamefont {D.~J.}\ \bibnamefont {Wineland}},\ }\href {\doibase 10.1038/nature10290} {\bibfield  {journal} {\bibinfo  {journal} {Nature}\ }\textbf {\bibinfo {volume} {476}},\ \bibinfo {pages} {181} (\bibinfo {year} {2011})}\BibitemShut {NoStop}%
\bibitem [{\citenamefont {Mehta}\ and\ \citenamefont {Ram}(2017)}]{mehta2017}%
  \BibitemOpen
  \bibfield  {author} {\bibinfo {author} {\bibfnamefont {K.~K.}\ \bibnamefont {Mehta}}\ and\ \bibinfo {author} {\bibfnamefont {R.~J.}\ \bibnamefont {Ram}},\ }\href {\doibase 10.1038/s41598-017-02169-2} {\bibfield  {journal} {\bibinfo  {journal} {Scientific Reports}\ }\textbf {\bibinfo {volume} {7}},\ \bibinfo {pages} {2019} (\bibinfo {year} {2017})}\BibitemShut {NoStop}%
\bibitem [{\citenamefont {West}\ \emph {et~al.}(2019)\citenamefont {West}, \citenamefont {Loh}, \citenamefont {Kharas}, \citenamefont {{Sorace-Agaskar}}, \citenamefont {Mehta}, \citenamefont {Sage}, \citenamefont {Chiaverini},\ and\ \citenamefont {Ram}}]{west2019}%
  \BibitemOpen
  \bibfield  {author} {\bibinfo {author} {\bibfnamefont {G.~N.}\ \bibnamefont {West}}, \bibinfo {author} {\bibfnamefont {W.}~\bibnamefont {Loh}}, \bibinfo {author} {\bibfnamefont {D.}~\bibnamefont {Kharas}}, \bibinfo {author} {\bibfnamefont {C.}~\bibnamefont {{Sorace-Agaskar}}}, \bibinfo {author} {\bibfnamefont {K.~K.}\ \bibnamefont {Mehta}}, \bibinfo {author} {\bibfnamefont {J.}~\bibnamefont {Sage}}, \bibinfo {author} {\bibfnamefont {J.}~\bibnamefont {Chiaverini}}, \ and\ \bibinfo {author} {\bibfnamefont {R.~J.}\ \bibnamefont {Ram}},\ }\href {\doibase 10.1063/1.5052502} {\bibfield  {journal} {\bibinfo  {journal} {APL Photonics}\ }\textbf {\bibinfo {volume} {4}},\ \bibinfo {pages} {026101} (\bibinfo {year} {2019})}\BibitemShut {NoStop}%
\bibitem [{\citenamefont {Sambles}\ \emph {et~al.}(1981)\citenamefont {Sambles}, \citenamefont {Elsom},\ and\ \citenamefont {{Sharp-Dent}}}]{sambles1981}%
  \BibitemOpen
  \bibfield  {author} {\bibinfo {author} {\bibfnamefont {J.~R.}\ \bibnamefont {Sambles}}, \bibinfo {author} {\bibfnamefont {K.~C.}\ \bibnamefont {Elsom}}, \ and\ \bibinfo {author} {\bibfnamefont {G.}~\bibnamefont {{Sharp-Dent}}},\ }\href {\doibase 10.1088/0305-4608/11/5/012} {\bibfield  {journal} {\bibinfo  {journal} {Journal of Physics F: Metal Physics}\ }\textbf {\bibinfo {volume} {11}},\ \bibinfo {pages} {1075} (\bibinfo {year} {1981})}\BibitemShut {NoStop}%
\bibitem [{\citenamefont {Delaney}\ \emph {et~al.}(2024)\citenamefont {Delaney}, \citenamefont {Sletten}, \citenamefont {Cich}, \citenamefont {Estey}, \citenamefont {Fabrikant}, \citenamefont {Hayes}, \citenamefont {Hoffman}, \citenamefont {Hostetter}, \citenamefont {Langer}, \citenamefont {Moses}, \citenamefont {Perry}, \citenamefont {Peterson}, \citenamefont {Schaffer}, \citenamefont {Volin}, \citenamefont {Vittorini},\ and\ \citenamefont {Burton}}]{delaney2024}%
  \BibitemOpen
  \bibfield  {author} {\bibinfo {author} {\bibfnamefont {R.~D.}\ \bibnamefont {Delaney}}, \bibinfo {author} {\bibfnamefont {L.~R.}\ \bibnamefont {Sletten}}, \bibinfo {author} {\bibfnamefont {M.~J.}\ \bibnamefont {Cich}}, \bibinfo {author} {\bibfnamefont {B.}~\bibnamefont {Estey}}, \bibinfo {author} {\bibfnamefont {M.}~\bibnamefont {Fabrikant}}, \bibinfo {author} {\bibfnamefont {D.}~\bibnamefont {Hayes}}, \bibinfo {author} {\bibfnamefont {I.~M.}\ \bibnamefont {Hoffman}}, \bibinfo {author} {\bibfnamefont {J.}~\bibnamefont {Hostetter}}, \bibinfo {author} {\bibfnamefont {C.}~\bibnamefont {Langer}}, \bibinfo {author} {\bibfnamefont {S.~A.}\ \bibnamefont {Moses}}, \bibinfo {author} {\bibfnamefont {A.~R.}\ \bibnamefont {Perry}}, \bibinfo {author} {\bibfnamefont {T.~A.}\ \bibnamefont {Peterson}}, \bibinfo {author} {\bibfnamefont {A.}~\bibnamefont {Schaffer}}, \bibinfo {author} {\bibfnamefont {C.}~\bibnamefont {Volin}}, \bibinfo {author} {\bibfnamefont {G.}~\bibnamefont {Vittorini}}, \ and\ \bibinfo {author}
  {\bibfnamefont {W.~C.}\ \bibnamefont {Burton}},\ }\href {\doibase 10.48550/arXiv.2403.00756} {\enquote {\bibinfo {title} {Scalable {{Multispecies Ion Transport}} in a {{Grid Based Surface-Electrode Trap}}},}\ } (\bibinfo {year} {2024}),\ \Eprint {http://arxiv.org/abs/2403.00756} {arXiv:2403.00756 [physics, physics:quant-ph]} \BibitemShut {NoStop}%
\bibitem [{\citenamefont {Hertzberg}\ \emph {et~al.}(2021)\citenamefont {Hertzberg}, \citenamefont {Zhang}, \citenamefont {Rosenblatt}, \citenamefont {Magesan}, \citenamefont {Smolin}, \citenamefont {Yau}, \citenamefont {Adiga}, \citenamefont {Sandberg}, \citenamefont {Brink}, \citenamefont {Chow},\ and\ \citenamefont {Orcutt}}]{hertzberg2021}%
  \BibitemOpen
  \bibfield  {author} {\bibinfo {author} {\bibfnamefont {J.~B.}\ \bibnamefont {Hertzberg}}, \bibinfo {author} {\bibfnamefont {E.~J.}\ \bibnamefont {Zhang}}, \bibinfo {author} {\bibfnamefont {S.}~\bibnamefont {Rosenblatt}}, \bibinfo {author} {\bibfnamefont {E.}~\bibnamefont {Magesan}}, \bibinfo {author} {\bibfnamefont {J.~A.}\ \bibnamefont {Smolin}}, \bibinfo {author} {\bibfnamefont {J.-B.}\ \bibnamefont {Yau}}, \bibinfo {author} {\bibfnamefont {V.~P.}\ \bibnamefont {Adiga}}, \bibinfo {author} {\bibfnamefont {M.}~\bibnamefont {Sandberg}}, \bibinfo {author} {\bibfnamefont {M.}~\bibnamefont {Brink}}, \bibinfo {author} {\bibfnamefont {J.~M.}\ \bibnamefont {Chow}}, \ and\ \bibinfo {author} {\bibfnamefont {J.~S.}\ \bibnamefont {Orcutt}},\ }\href {\doibase 10.1038/s41534-021-00464-5} {\bibfield  {journal} {\bibinfo  {journal} {npj Quantum Information}\ }\textbf {\bibinfo {volume} {7}} (\bibinfo {year} {2021}),\ 10.1038/s41534-021-00464-5}\BibitemShut {NoStop}%
\bibitem [{\citenamefont {Berke}\ \emph {et~al.}(2022)\citenamefont {Berke}, \citenamefont {Varvelis}, \citenamefont {Trebst}, \citenamefont {Altland},\ and\ \citenamefont {DiVincenzo}}]{berke2022}%
  \BibitemOpen
  \bibfield  {author} {\bibinfo {author} {\bibfnamefont {C.}~\bibnamefont {Berke}}, \bibinfo {author} {\bibfnamefont {E.}~\bibnamefont {Varvelis}}, \bibinfo {author} {\bibfnamefont {S.}~\bibnamefont {Trebst}}, \bibinfo {author} {\bibfnamefont {A.}~\bibnamefont {Altland}}, \ and\ \bibinfo {author} {\bibfnamefont {D.~P.}\ \bibnamefont {DiVincenzo}},\ }\href {\doibase 10.1038/s41467-022-29940-y} {\bibfield  {journal} {\bibinfo  {journal} {Nature Communications}\ }\textbf {\bibinfo {volume} {13}},\ \bibinfo {pages} {2495} (\bibinfo {year} {2022})}\BibitemShut {NoStop}%
\bibitem [{\citenamefont {{de Leon}}\ \emph {et~al.}(2021)\citenamefont {{de Leon}}, \citenamefont {Itoh}, \citenamefont {Kim}, \citenamefont {Mehta}, \citenamefont {Northup}, \citenamefont {Paik}, \citenamefont {Palmer}, \citenamefont {Samarth}, \citenamefont {Sangtawesin},\ and\ \citenamefont {Steuerman}}]{deleon2021}%
  \BibitemOpen
  \bibfield  {author} {\bibinfo {author} {\bibfnamefont {N.~P.}\ \bibnamefont {{de Leon}}}, \bibinfo {author} {\bibfnamefont {K.~M.}\ \bibnamefont {Itoh}}, \bibinfo {author} {\bibfnamefont {D.}~\bibnamefont {Kim}}, \bibinfo {author} {\bibfnamefont {K.~K.}\ \bibnamefont {Mehta}}, \bibinfo {author} {\bibfnamefont {T.~E.}\ \bibnamefont {Northup}}, \bibinfo {author} {\bibfnamefont {H.}~\bibnamefont {Paik}}, \bibinfo {author} {\bibfnamefont {B.~S.}\ \bibnamefont {Palmer}}, \bibinfo {author} {\bibfnamefont {N.}~\bibnamefont {Samarth}}, \bibinfo {author} {\bibfnamefont {S.}~\bibnamefont {Sangtawesin}}, \ and\ \bibinfo {author} {\bibfnamefont {D.~W.}\ \bibnamefont {Steuerman}},\ }\href {\doibase 10.1126/science.abb2823} {\bibfield  {journal} {\bibinfo  {journal} {Science}\ }\textbf {\bibinfo {volume} {372}},\ \bibinfo {pages} {eabb2823} (\bibinfo {year} {2021})}\BibitemShut {NoStop}%
\bibitem [{\citenamefont {Allcock}\ \emph {et~al.}(2011)\citenamefont {Allcock}, \citenamefont {Guidoni}, \citenamefont {Harty}, \citenamefont {Ballance}, \citenamefont {Blain}, \citenamefont {Steane},\ and\ \citenamefont {Lucas}}]{allcock2011}%
  \BibitemOpen
  \bibfield  {author} {\bibinfo {author} {\bibfnamefont {D.~T.~C.}\ \bibnamefont {Allcock}}, \bibinfo {author} {\bibfnamefont {L.}~\bibnamefont {Guidoni}}, \bibinfo {author} {\bibfnamefont {T.~P.}\ \bibnamefont {Harty}}, \bibinfo {author} {\bibfnamefont {C.~J.}\ \bibnamefont {Ballance}}, \bibinfo {author} {\bibfnamefont {M.~G.}\ \bibnamefont {Blain}}, \bibinfo {author} {\bibfnamefont {A.~M.}\ \bibnamefont {Steane}}, \ and\ \bibinfo {author} {\bibfnamefont {D.~M.}\ \bibnamefont {Lucas}},\ }\href {\doibase 10.1088/1367-2630/13/12/123023} {\bibfield  {journal} {\bibinfo  {journal} {New Journal of Physics}\ }\textbf {\bibinfo {volume} {13}},\ \bibinfo {pages} {123023} (\bibinfo {year} {2011})}\BibitemShut {NoStop}%
\bibitem [{\citenamefont {Hite}\ \emph {et~al.}(2012)\citenamefont {Hite}, \citenamefont {Colombe}, \citenamefont {Wilson}, \citenamefont {Brown}, \citenamefont {Warring}, \citenamefont {J{\"o}rdens}, \citenamefont {Jost}, \citenamefont {McKay}, \citenamefont {Pappas}, \citenamefont {Leibfried},\ and\ \citenamefont {Wineland}}]{hite2012}%
  \BibitemOpen
  \bibfield  {author} {\bibinfo {author} {\bibfnamefont {D.~A.}\ \bibnamefont {Hite}}, \bibinfo {author} {\bibfnamefont {Y.}~\bibnamefont {Colombe}}, \bibinfo {author} {\bibfnamefont {A.~C.}\ \bibnamefont {Wilson}}, \bibinfo {author} {\bibfnamefont {K.~R.}\ \bibnamefont {Brown}}, \bibinfo {author} {\bibfnamefont {U.}~\bibnamefont {Warring}}, \bibinfo {author} {\bibfnamefont {R.}~\bibnamefont {J{\"o}rdens}}, \bibinfo {author} {\bibfnamefont {J.~D.}\ \bibnamefont {Jost}}, \bibinfo {author} {\bibfnamefont {K.~S.}\ \bibnamefont {McKay}}, \bibinfo {author} {\bibfnamefont {D.~P.}\ \bibnamefont {Pappas}}, \bibinfo {author} {\bibfnamefont {D.}~\bibnamefont {Leibfried}}, \ and\ \bibinfo {author} {\bibfnamefont {D.~J.}\ \bibnamefont {Wineland}},\ }\href {\doibase 10.1103/PhysRevLett.109.103001} {\bibfield  {journal} {\bibinfo  {journal} {Physical Review Letters}\ }\textbf {\bibinfo {volume} {109}},\ \bibinfo {pages} {103001} (\bibinfo {year} {2012})}\BibitemShut {NoStop}%
\bibitem [{\citenamefont {Brown}\ \emph {et~al.}(2021)\citenamefont {Brown}, \citenamefont {Chiaverini}, \citenamefont {Sage},\ and\ \citenamefont {H{\"a}ffner}}]{brown2021}%
  \BibitemOpen
  \bibfield  {author} {\bibinfo {author} {\bibfnamefont {K.~R.}\ \bibnamefont {Brown}}, \bibinfo {author} {\bibfnamefont {J.}~\bibnamefont {Chiaverini}}, \bibinfo {author} {\bibfnamefont {J.~M.}\ \bibnamefont {Sage}}, \ and\ \bibinfo {author} {\bibfnamefont {H.}~\bibnamefont {H{\"a}ffner}},\ }\href {\doibase 10.1038/s41578-021-00292-1} {\bibfield  {journal} {\bibinfo  {journal} {Nature Reviews Materials}\ }\textbf {\bibinfo {volume} {6}},\ \bibinfo {pages} {892} (\bibinfo {year} {2021})}\BibitemShut {NoStop}%
\bibitem [{\citenamefont {{Berlin-Udi}}\ \emph {et~al.}(2022)\citenamefont {{Berlin-Udi}}, \citenamefont {Matthiesen}, \citenamefont {Lloyd}, \citenamefont {Alonso}, \citenamefont {Noel}, \citenamefont {Saarel}, \citenamefont {Orme}, \citenamefont {Kim}, \citenamefont {Nelson}, \citenamefont {Ray}, \citenamefont {Lordi},\ and\ \citenamefont {H{\"a}ffner}}]{berlin-udi2022}%
  \BibitemOpen
  \bibfield  {author} {\bibinfo {author} {\bibfnamefont {M.}~\bibnamefont {{Berlin-Udi}}}, \bibinfo {author} {\bibfnamefont {C.}~\bibnamefont {Matthiesen}}, \bibinfo {author} {\bibfnamefont {P.~N.~T.}\ \bibnamefont {Lloyd}}, \bibinfo {author} {\bibfnamefont {A.~M.}\ \bibnamefont {Alonso}}, \bibinfo {author} {\bibfnamefont {C.}~\bibnamefont {Noel}}, \bibinfo {author} {\bibfnamefont {B.}~\bibnamefont {Saarel}}, \bibinfo {author} {\bibfnamefont {C.~A.}\ \bibnamefont {Orme}}, \bibinfo {author} {\bibfnamefont {C.-E.}\ \bibnamefont {Kim}}, \bibinfo {author} {\bibfnamefont {A.~J.}\ \bibnamefont {Nelson}}, \bibinfo {author} {\bibfnamefont {K.~G.}\ \bibnamefont {Ray}}, \bibinfo {author} {\bibfnamefont {V.}~\bibnamefont {Lordi}}, \ and\ \bibinfo {author} {\bibfnamefont {H.}~\bibnamefont {H{\"a}ffner}},\ }\href {\doibase 10.1103/PhysRevB.106.035409} {\bibfield  {journal} {\bibinfo  {journal} {Physical Review B}\ }\textbf {\bibinfo {volume} {106}},\ \bibinfo {pages} {035409} (\bibinfo {year} {2022})}\BibitemShut
  {NoStop}%
\bibitem [{\citenamefont {Beck}\ \emph {et~al.}(2024)\citenamefont {Beck}, \citenamefont {Home},\ and\ \citenamefont {Mehta}}]{beck2024}%
  \BibitemOpen
  \bibfield  {author} {\bibinfo {author} {\bibfnamefont {G.~J.}\ \bibnamefont {Beck}}, \bibinfo {author} {\bibfnamefont {J.~P.}\ \bibnamefont {Home}}, \ and\ \bibinfo {author} {\bibfnamefont {K.~K.}\ \bibnamefont {Mehta}},\ }\href {\doibase 10.1109/JLT.2024.3381785} {\bibfield  {journal} {\bibinfo  {journal} {Journal of Lightwave Technology}\ } (\bibinfo {year} {2024}),\ 10.1109/JLT.2024.3381785}\BibitemShut {NoStop}%
\bibitem [{\citenamefont {Lin}\ \emph {et~al.}(2013)\citenamefont {Lin}, \citenamefont {Gaebler}, \citenamefont {Reiter}, \citenamefont {Tan}, \citenamefont {Bowler}, \citenamefont {S{\o}rensen}, \citenamefont {Leibfried},\ and\ \citenamefont {Wineland}}]{lin2013}%
  \BibitemOpen
  \bibfield  {author} {\bibinfo {author} {\bibfnamefont {Y.}~\bibnamefont {Lin}}, \bibinfo {author} {\bibfnamefont {J.~P.}\ \bibnamefont {Gaebler}}, \bibinfo {author} {\bibfnamefont {F.}~\bibnamefont {Reiter}}, \bibinfo {author} {\bibfnamefont {T.~R.}\ \bibnamefont {Tan}}, \bibinfo {author} {\bibfnamefont {R.}~\bibnamefont {Bowler}}, \bibinfo {author} {\bibfnamefont {A.~S.}\ \bibnamefont {S{\o}rensen}}, \bibinfo {author} {\bibfnamefont {D.}~\bibnamefont {Leibfried}}, \ and\ \bibinfo {author} {\bibfnamefont {D.~J.}\ \bibnamefont {Wineland}},\ }\href {\doibase 10.1038/nature12801} {\bibfield  {journal} {\bibinfo  {journal} {Nature}\ }\textbf {\bibinfo {volume} {504}},\ \bibinfo {pages} {415} (\bibinfo {year} {2013})}\BibitemShut {NoStop}%
\bibitem [{\citenamefont {Cole}\ \emph {et~al.}(2022)\citenamefont {Cole}, \citenamefont {Erickson}, \citenamefont {Zarantonello}, \citenamefont {Horn}, \citenamefont {Hou}, \citenamefont {Wu}, \citenamefont {Slichter}, \citenamefont {Reiter}, \citenamefont {Koch},\ and\ \citenamefont {Leibfried}}]{cole2022}%
  \BibitemOpen
  \bibfield  {author} {\bibinfo {author} {\bibfnamefont {D.~C.}\ \bibnamefont {Cole}}, \bibinfo {author} {\bibfnamefont {S.~D.}\ \bibnamefont {Erickson}}, \bibinfo {author} {\bibfnamefont {G.}~\bibnamefont {Zarantonello}}, \bibinfo {author} {\bibfnamefont {K.~P.}\ \bibnamefont {Horn}}, \bibinfo {author} {\bibfnamefont {P.-Y.}\ \bibnamefont {Hou}}, \bibinfo {author} {\bibfnamefont {J.~J.}\ \bibnamefont {Wu}}, \bibinfo {author} {\bibfnamefont {D.~H.}\ \bibnamefont {Slichter}}, \bibinfo {author} {\bibfnamefont {F.}~\bibnamefont {Reiter}}, \bibinfo {author} {\bibfnamefont {C.~P.}\ \bibnamefont {Koch}}, \ and\ \bibinfo {author} {\bibfnamefont {D.}~\bibnamefont {Leibfried}},\ }\href {\doibase 10.1103/PhysRevLett.128.080502} {\bibfield  {journal} {\bibinfo  {journal} {Physical Review Letters}\ }\textbf {\bibinfo {volume} {128}},\ \bibinfo {pages} {080502} (\bibinfo {year} {2022})}\BibitemShut {NoStop}%
\bibitem [{\citenamefont {Malinowski}\ \emph {et~al.}(2022)\citenamefont {Malinowski}, \citenamefont {Zhang}, \citenamefont {Negnevitsky}, \citenamefont {Rojkov}, \citenamefont {Reiter}, \citenamefont {Nguyen}, \citenamefont {Stadler}, \citenamefont {Kienzler}, \citenamefont {Mehta},\ and\ \citenamefont {Home}}]{malinowski2022}%
  \BibitemOpen
  \bibfield  {author} {\bibinfo {author} {\bibfnamefont {M.}~\bibnamefont {Malinowski}}, \bibinfo {author} {\bibfnamefont {C.}~\bibnamefont {Zhang}}, \bibinfo {author} {\bibfnamefont {V.}~\bibnamefont {Negnevitsky}}, \bibinfo {author} {\bibfnamefont {I.}~\bibnamefont {Rojkov}}, \bibinfo {author} {\bibfnamefont {F.}~\bibnamefont {Reiter}}, \bibinfo {author} {\bibfnamefont {T.-L.}\ \bibnamefont {Nguyen}}, \bibinfo {author} {\bibfnamefont {M.}~\bibnamefont {Stadler}}, \bibinfo {author} {\bibfnamefont {D.}~\bibnamefont {Kienzler}}, \bibinfo {author} {\bibfnamefont {K.~K.}\ \bibnamefont {Mehta}}, \ and\ \bibinfo {author} {\bibfnamefont {J.~P.}\ \bibnamefont {Home}},\ }\href {\doibase 10.1103/PhysRevLett.128.080503} {\bibfield  {journal} {\bibinfo  {journal} {Physical Review Letters}\ }\textbf {\bibinfo {volume} {128}},\ \bibinfo {pages} {080503} (\bibinfo {year} {2022})}\BibitemShut {NoStop}%
\bibitem [{\citenamefont {Sackett}\ \emph {et~al.}(2000)\citenamefont {Sackett}, \citenamefont {Kielpinski}, \citenamefont {King}, \citenamefont {Langer}, \citenamefont {Meyer}, \citenamefont {Myatt}, \citenamefont {Rowe}, \citenamefont {Turchette}, \citenamefont {Itano},\ and\ \citenamefont {Wineland}}]{sackett2000}%
  \BibitemOpen
  \bibfield  {author} {\bibinfo {author} {\bibfnamefont {C.~A.}\ \bibnamefont {Sackett}}, \bibinfo {author} {\bibfnamefont {D.}~\bibnamefont {Kielpinski}}, \bibinfo {author} {\bibfnamefont {B.~E.}\ \bibnamefont {King}}, \bibinfo {author} {\bibfnamefont {C.}~\bibnamefont {Langer}}, \bibinfo {author} {\bibfnamefont {V.}~\bibnamefont {Meyer}}, \bibinfo {author} {\bibfnamefont {C.~J.}\ \bibnamefont {Myatt}}, \bibinfo {author} {\bibfnamefont {M.}~\bibnamefont {Rowe}}, \bibinfo {author} {\bibfnamefont {{\relax QA}.}~\bibnamefont {Turchette}}, \bibinfo {author} {\bibfnamefont {W.~M.}\ \bibnamefont {Itano}}, \ and\ \bibinfo {author} {\bibfnamefont {D.~J.}\ \bibnamefont {Wineland}},\ }\href@noop {} {\bibfield  {journal} {\bibinfo  {journal} {Nature}\ }\textbf {\bibinfo {volume} {404}},\ \bibinfo {pages} {256} (\bibinfo {year} {2000})}\BibitemShut {NoStop}%
\end{thebibliography}%

\newpage

\begin{appendices}

\section{Single-qubit gate methods}\label{secA1}
\textbf{Single-qubit gate mechanism} Consider a shared trace placed along the $\hat{x}$ direction as in section~\ref{sec:experiments}. An oscillating current $I \cos(\omega_0 t + \phi)$ applied to the shared trace generates a magnetic field $\vec{B} \cos(\omega_0 t + \phi)$ at the ion location. For simplicity, we assume that the wire is very long along the x-direction, such that the oscillating magnetic field $\vec{B} = B_y \hat{y} + B_z \hat{z}$ is polarized in the y-z plane.

For a $\sigma$-polarized qubit placed in a magnetic field oriented along $\hat{y}$ such as the one discussed in section~\ref{sec:experiments}, the Hamiltonian can then be written as $H = H_0 + H_1$, where the bare qubit Hamiltonian $H_0 = \frac{1}{2} \hbar \omega_0 \sigma_z$ and the interaction Hamiltonian $H_1 = \frac{1}{2} \hbar \frac{\partial \omega_0}{\partial B} (B_z \sigma_x + B_y \sigma_z) \cos(\omega_0 t + \phi)$. Transforming $H$ into the interaction picture with respect to $H_0$ and taking the rotating wave approximation to drop all oscillating terms results in a time-independent Hamiltonian $H' = \frac{1}{2} \Omega_1 \sigma_\phi$, where $\Omega_1 = \frac{1}{2} \frac{\partial \omega_0}{\partial B} B_z$. Integrating $H'$ from $t=0$ to $t=t_1$ results in a single-qubit rotation unitary as in the main text.

For the geometry described in Fig.~\ref{fig:1q}, we expect that at $y = 0$ (i.e. directly above the shared drive) $B_z = 0$ and $\partial B_z / \partial z = 0$. Thus, in the absence of any imperfections, the single-qubit Rabi frequency $\Omega_1 = 0$ at $y=0$ for any value of $z$. At the same time, $\partial B_z / \partial y \neq 0$, which allows $\Omega_1$ to be tuned by translating the ion along $y$. The experimental data in Fig.~\ref{fig:1q} B) differs from this simplified model in two ways. First, the Rabi frequency minimum is shifted from $y_0 = 0$ to $y_0 \approx -0.2 \mathrm{\mu}$m. We believe this to be caused by a small tilt ($\approx 0.3$ deg) of the static magnetic field in the y-z plane. Second, the Rabi frequency at $y = y_0$ does not drop to zero, and we instead record a minimum $\Omega_1 \approx 2 \pi \times 17$ kHz when applying current $I \approx 70$ mA. We believe this to be caused by a small tilt ($\approx 0.2$ deg) of the static magnetic field in the x-y plane. We expect both of these effects can be compensated to a high degree using additional magnetic field coils.

\textbf{Single-qubit operation benchmarking}. The quality of single-qubit operations is measured using single-qubit Clifford RB, which is implemented by decomposing Clifford gates onto single-qubit gates with $\theta_1$ in the set $\{\pi/2, \pi\}$ and $\phi$ selected from the set $\{0, \pi/2, \pi, 3 \pi/2\}$. Each single-qubit gate, in turn, is implemented as a sequence of three square pulses: the target rotation followed by a pair of rotations with $\theta_1 = 2 \pi$ and phase $\phi' = \phi \pm \arccos{(-\frac{\theta_0}{4\pi})}$, which implements a Solovay-Kitaev-1 (SK1) composite pulse \cite{merrill2012}. The longest SK1 sequence ($\theta_1 = \pi$) takes $8 \ \mathrm{\mu}$s, including $7.5\ \mathrm{\mu}$s of coherent evolution and $0.5\ \mathrm{\mu}$s dead time. All single-qubit RB includes Pauli randomization to eliminate possible bias from asymmetric readout errors.

\section{Two-qubit gate methods}\label{secA2}
\textbf{Two-qubit gate mechanism}. The two-qubit gate mechanism can be analyzed by extending the discussion in section~\ref{secA1} to include ion motion and spin-motion coupling. Consider a qubit at frequency $\omega_0$ placed directly above a narrow and long shared trace, coupled to a single radial mode of motion at frequency $\omega_m$ oriented along $\hat{y}$. A bichromatic oscillating current $I \cos((\omega_0 \pm (\omega_m \pm \delta)) t + \phi)$ generates a bichromatic magnetic field $\vec{B} \cos((\omega_0 \pm \omega_m \pm \delta) t + \phi)$ at the ion location. Due to the trace geometry, we find that $B_z \approx 0$ and $\partial B_y / \partial y \approx 0$, allowing us to approximate the system Hamiltonian as $H = H_0 + H_1 + H_2$, where the bare spin-motion Hamiltonian $H_0 = \frac{1}{2} \hbar \omega_0 \sigma_z + \frac{1}{2} \hbar \omega_m \hat{a}^\dagger \hat{a}$, the single-qubit coupling Hamiltonian $H_1 = \frac{1}{2} \hbar \frac{\partial \omega_0}{\partial B} B_y \sigma_z f(t)$ , and the spin-motion coupling Hamiltonian

\begin{align*}
    H_2 = \frac{1}{2} \hbar \frac{\partial \omega_0}{\partial B} \frac{\partial B_z}{\partial y} \sigma_x f(t) y_0 (\hat{a}^\dagger e^{i \omega_m t} + \hat{a} e^{-i \omega_m t})
\end{align*}
where $\hat{a}$ is the motion lowering operator, $y_0 = \sqrt{\hbar/(2 m \omega_m)}$ is the size of the ground-state ion wavepacket, with $m$ denoting ion mass, and 
\begin{align*}
f(t) = \cos((\omega_0 + \omega_m + \delta) t + \phi) + \cos((\omega_0 - \omega_m - \delta) t + \phi).
\end{align*}
Once again transforming $H$ into the interaction picture with respect to $H_0$, and performing the rotating wave approximation to drop all terms oscillating at frequencies above $\delta$, we arrive at $H' = \frac{1}{2} \hbar \Omega_2 \sigma_x (\hat{a}^\dagger e^{i \delta t} + \hat{a} e^{-i \delta t})$ with $\Omega_2 = \frac{1}{2} \frac{\partial \omega_0}{\partial B} \frac{\partial B_z}{\partial y} y_0$, which is a standard Hamiltonian for a state-dependent force that can generate two-qubit M{\o}lmer-S{\o}rensen unitaries as discussed in the main text.

\textbf{SPAM error estimation}. The estimate of entangled-state fidelity includes a correction for state preparation and measurement (SPAM) errors. We first describe how we characterize the SPAM error.

We denote the probability for ion $i$ to be prepared in $\ket{\downarrow}$ instead of $\ket{\uparrow}$ as $\epsilon_{\mathrm{sp}, i}$. The average per-ion state preparation error $\epsilon_{\mathrm{sp}} = \left(\epsilon_{\mathrm{sp}, 1} + \epsilon_{\mathrm{sp}, 2}\right) / 2$ is measured by applying the same experimental sequence as used for entangled state generation, but with no current applied to the shared drive, and recording the average probability $P(\downarrow)$ to find an ion in $\ket{\downarrow}$ instead of $\ket{\uparrow}$. The duration of fluorescence detection is selected to ensure negligible error ($\ll 1 \times 10^{-4}$) from $\ket{\uparrow}$ and $\ket{\downarrow}$ photon count overlap, and the Rabi frequency of the shelving pulse at 729 nm is selected to ensure negligible probability to off-resonantly excite $\ket{\uparrow}$. Thus, we attribute all measurements of $\ket{\downarrow}$ to imperfect state preparation, and set $\epsilon_{\mathrm{sp}} = P(\downarrow)$. Over the full 60 hours of data acquisition, we record a stable state preparation error of $\epsilon_{\mathrm{sp}} = 1.4(2) \times 10^{-4}$.

For perfect readout, an ion in state $\ket{\downarrow}$ is recorded as ``dark'', while an ion in state $\ket{\uparrow}$ is recorded as ``bright''. We denote the probability for ion $i$ in $\ket{\uparrow}$ to be erroneously recorded as ``dark'' as $\epsilon_{\mathrm{d,i}}$, and for an ion in $\ket{\downarrow}$ to be erroneously recorded as ``bright'' as $\epsilon_{\mathrm{b,i}}$. We also denote the average per-ion readout error as ${\bar{\epsilon}_{\mathrm{r}} = \left(\epsilon_{\mathrm{b}} + \epsilon_{\mathrm{d}}\right) / 2}$. As $\epsilon_{\mathrm{b}} \ll 1 \times 10^{-4}$, it can be neglected in the analysis. To estimate $\epsilon_{\mathrm{d}} = \left(\epsilon_{\mathrm{d}, 1} + \epsilon_{\mathrm{d}, 2}\right) / 2$, we apply the same experimental sequence as used for entangled state generation, but with the two-qubit entangling pulse replaced with a single-qubit $\pi$ rotation which ideally prepares $\ket{\downarrow \downarrow}$. We then record the probability $P(\downarrow \uparrow + \uparrow \downarrow)$ of measuring one ion bright, associating all such events to imperfect state preparation and measurement, i.e. $P(\downarrow \uparrow + \uparrow \downarrow) = \epsilon_{\mathrm{sp},1} + \epsilon_{\mathrm{sp},2} + \epsilon_{\mathrm{d},1} + \epsilon_{\mathrm{d},2} = 2 \left(\epsilon_{\mathrm{sp}} + \epsilon_{\mathrm{d}}\right)$. Using the estimate of $\epsilon_{\mathrm{sp}}$ presented earlier, we thus infer $\epsilon_{\mathrm{d}} = 6.0(5) \times 10^{-4}$, and an average readout error of $\bar{\epsilon}_{\mathrm{r}} = 3.0(2) \times 10^{-4}$, stable throughout the data acquisition period. The readout error is consistent with the expected probability of spontaneous emission from the shelve state during the readout window.

\textbf{Fidelity estimate of two-qubit operation}. The quality of two-qubit operations is measured by the fidelity with which they transform the initial state $\ket{\uparrow\uparrow}$ into a maximally entangled state of the form  $\left(\ket{\uparrow\uparrow} + e^{i\phi_0}\ket{\downarrow\downarrow}\right) / \sqrt{2}$ using regularly spaced measurements over 12 hours (target data acquisition period) and 60 hours (full data acquisition period). The entangling operation fidelity estimation experiments are interleaved with calibration of the motional mode frequency (approximately every 2 minutes) and SPAM error monitoring (approximately every 20 minutes). No other parameters are calibrated or manually adjusted over the measurement period.

The entangled state fidelity $F$ is estimated using SPAM-corrected partial state tomography~\cite{sackett2000}. To that end, we first measure the population $P = P(\downarrow \downarrow) + P(\uparrow \uparrow)$. Interleaved with that we apply a $\pi/2$ ``analysis'' pulse to the final state with a variable phase $\phi$, and estimate the contrast $C$ by fitting the amplitude of $P(\uparrow \downarrow) + P (\downarrow \uparrow)$ assuming a functional form $(C/2)\cos (2 \phi - \phi_0)+C_0$ using a maximum-likelihood estimator with a binomial fitter. These results allow us to estimate the ``raw'' fidelity as $F' = (P + C)/2$. 

In the presence of state preparation and measurement (SPAM) errors, $F'$ is known to be a biased estimator for the true Bell-state fidelity $F$. For uncorrelated SPAM errors, assuming $F \approx 1$, it can be shown that the true fidelity is given by $F = F' + 2 \, \epsilon_{\mathrm{sp}} + 3 \, \bar{\epsilon}_{\mathrm{r}}$ up to the leading order in $\epsilon_{\mathrm{sp}}$ and $\epsilon_{\mathrm{r}}$. We use this formula, with $\epsilon_{\mathrm{sp}} = 1.4(2) \times 10^{-4}$ and $\bar{\epsilon}_{\mathrm{r}} = 3.0(2) \times 10^{-4}$, to estimate the fidelities in the main text. To obtain the confidence intervals, we bootstrap~\cite{efron1994} the raw population, parity, and SPAM data with 10,000 resamples that are run through the same analysis procedure.

\end{appendices}

\end{document}